\documentclass[a4paper,11pt]{article}
\pdfoutput=1 

\usepackage{jheppub}
\usepackage{upgreek}

\newcommand{\be}{\begin{equation}}
\newcommand{\ee}{\end{equation}}
\newcommand{\beq}{\begin{equation}}
\newcommand{\eeq}{\end{equation}}
\newcommand{\beqa}{\begin{eqnarray}}
\newcommand{\eeqa}{\end{eqnarray}}

\newcommand{\bear}{\begin{eqnarray}}
\newcommand{\eear}{\end{eqnarray}}

\newcommand{\D}{{\mathcal D}}

\title{\boldmath Transseries for causal diffusive systems}

\author[1,2]{Michal P.\ Heller,}
\author[2]{Alexandre Serantes,}
\author[2,3]{Micha\l\ Spali\'nski,}
\author[2,1]{Viktor Svensson}
\author[4]{and Benjamin Withers}

\affiliation[1]{Max Planck Institute for Gravitational Physics (Albert Einstein Institute),
14476 Potsdam-Golm, Germany}
\affiliation[2]{National Centre for Nuclear Research, 02-093 Warsaw, Poland}
\affiliation[3]{Physics Department, University of Bia{\l}ystok, 15-245 Bia\l ystok, Poland}
\affiliation[4]{Mathematical Sciences and STAG Research Centre, University of Southampton, Highfield, Southampton SO17 1BJ, UK}

\emailAdd{michal.p.heller@aei.mpg.de}
\emailAdd{alexandre.serantesrubianes@ncbj.gov.pl}
\emailAdd{michal.spalinski@ncbj.gov.pl}
\emailAdd{viktor.svensson@aei.mpg.de}
\emailAdd{b.s.withers@soton.ac.uk}

\abstract{The large proper-time behaviour of expanding boost-invariant fluids has provided many crucial insights into quark-gluon plasma dynamics. 
Here we formulate and
explore the late-time behaviour of nonequilibrium dynamics at the level of
linearized perturbations of equilibrium, but without any special symmetry assumptions. 
We introduce a useful quantitative approximation scheme in which hydrodynamic modes appear as perturbative contributions while transients are nonperturbative. In this way, solutions are naturally organized into transseries as they are in the case of boost-invariant flows.
We focus our attention on the ubiquitous telegrapher's equation, the simplest example of a causal theory with a hydrodynamic sector. In position space we uncover novel transient contributions as well as Stokes phenomena which change the structure of the transseries based on the spacetime region or the choice of initial data.
}

\begin{document} 
\maketitle

\newpage

\section{Introduction\label{introduction}}

Understanding nonequilibrium phenomena with hydrodynamic tails has been a very active research direction of the past two decades. One motivation for this quest has been ultra-relativistic heavy-ion collisions at RHIC and LHC and the success of hydrodynamic modelling there~\cite{Florkowski:2017olj,Romatschke:2017ejr,Busza:2018rrf,Shen:2020mgh}. Another set of motivations came from condensed matter and quantum-many body physics~\cite{Hartnoll:2016apf}. There have also been more foundational questions driving this field, such as existence of bounds on transport~\cite{Kovtun:2004de}, or the character of the hydrodynamic gradient expansion~\cite{Heller:2013fn}.

One of the driving forces in this endeavour was holography~\cite{Maldacena:1997re, Witten:1998qj, Gubser:1998bc}, in which the nonequilibrium dynamics of a certain class of quantum field theories is represented by time-dependent geometries involving black holes~\cite{Hubeny:2010ry}. 
Results in this context are often based on numerical solutions of the equations of motion, with analytic or semi-analytic insights limited to basically two cases: linear response theory around equilibrium or highly symmetric dynamics. The former cases are based on Fourier transform techniques
while the latter concern situations  which can be effectively captured as comoving flows known from cosmology, or from heavy-ion collisions.

Our paper is directly motivated by such a description of ultrarelativistic heavy-ion collisions in terms of a Bjorken flow -- one-dimensional expansion of matter, which looks the same in any coordinate frame boosted in the direction of expansion~\cite{Bjorken:1982qr}. In the case of conformal models, such dynamics can be 
expressed as 
a single function $\mathcal{A}(w)$ which measures deviations away from local equilibrium. $w$ is a dimensionless clock variable and the key motivating point for our paper is that $\cal A$ has an asymptotic late-time form of a transseries,
i.e. a double expansion in powers of $1/w$ and an exponential suppression factor:
\begin{equation}
\label{eq.Atransseries}
{\cal A} = \sum_{n = 1}^{\infty} \mu_{n} \, w^{-n} + e^{-\Omega \, w}w^\gamma \sum_{n = 0}^{\infty} \nu_{n} \, w^{-n}  + \ldots \,.
\end{equation}
The ellipsis in the above equation stands for other exponentially suppressed contributions to the sum. 

The first sum in~\eqref{eq.Atransseries} is the onshell hydrodynamic gradient expansion of the energy-momentum tensor,
\begin{subequations}
\label{Pi_constitutive}
\beqa
&T^{a b} = \mathcal{E} \, U^a \, U^b + \frac{\mathcal{E}}{d-1} (\eta^{ab} + U^a \, U^b) + \Pi^{a b}, \\
&\Pi^{a b} = -\eta \, \sigma^{ab} + \eta\, \tau\,  \D\sigma^{ab} + \lambda_1 {\sigma^{\langle a}}_c \sigma^{b\rangle c} + \ldots,
\eeqa
\end{subequations}
where we focus on conformal theories and the ellipsis denotes three additional nonlinear terms with two derivatives of velocity $U^{a}$ which are irrelevant for our discussion, as well as higher order contributions, see, for example,~\cite{Florkowski:2017olj} for details. In~\eqref{eq.Atransseries}, the $\mu_{1}/w$ term represents the shear viscosity~$\eta$ contribution and the $\mu_{2}/w^2$ term represents the combined effect of $\tau$ and $\lambda_{1}$~\cite{Baier:2007ix}. More generally, each term $\mu_{n}/w^{n}$ comes from the $n^{\mathrm{th}}$ order of the hydrodynamic gradient expansion. The key aspect of~\cite{Heller:2013fn,Heller:2015dha,Basar:2015ava,Aniceto:2015mto,Florkowski:2016zsi,Heller:2016rtz,Florkowski:2016zsi,Heller:2018qvh,Casalderrey-Solana:2017zyh,Aniceto:2018uik,Behtash:2020vqk} was the ability to explicitly calculate higher order contributions to the sum~\eqref{eq.Atransseries} with the conclusion that the gradient expansion evaluated on shell for the Bjorken flow diverges, i.e. $\mu_{n} \sim n!$ at sufficiently large~$n$ (see, however,~\cite{Denicol:2019lio}).

The second contribution in~\eqref{eq.Atransseries} is associated with transient, exponentially decaying phenomena known from linear response theory and dressed in the hydrodynamic variables~\cite{Janik:2006gp, Heller:2014wfa}, hence the other, also divergent gradient expansion with $\nu_{n}/w^{n}$ terms. As is the case in the mainstream application area of resurgence in theoretical high-energy physics~\cite{Dorigoni:2014hea, Aniceto:2018bis}, i.e. coupling constant expansions in interacting quantum mechanical systems, the hydrodynamic sector carries information about transient phenomena through the phenomenon of resurgence. 

It is an important open problem to what extent the picture uncovered for Bjorken flow, i.e. divergent hydrodynamic expansion and spatiotemporal dependence as transseries, survives when symmetry assumptions are lifted. New light on this question was shed by our recent article~\cite{Heller:2020uuy}, in which we combined linear response theory with the 
Fourier transform to investigate convergence of hydrodynamic gradient expansion in linearized conformal hydrodynamics in complete generality. 

In linear response theory, a component of conserved currents~$\rho(t,\textbf{x})$ acquires spatiotemporal dependence given by
\beq
\rho(t,\textbf{x}) = \int_{\mathbb R^d}\, d^d \textbf{k}\, \hat\rho(t,\textbf{k}) e^{i\,\textbf{k}\cdot \textbf{x}}, \quad \hat\rho(t,\textbf{k}) = \sum_{q= 0}^N f_q(\textbf{k}) e^{- i\,\omega_q(|\textbf{k}|)\,t},
\label{spectral_decomposition}
\eeq
where $\omega_{q}(k)$ are dispersion relations for different modes in the system.\footnote{This form of $\rho(t,\textbf{x})$ is appropriate, for example, for holographic systems or in hydrodynamic models. In kinetic theory, the situation is more intricate, see, for example,~\cite{Romatschke:2015gic}.} Hydrodynamic modes are those for which
\beq
\omega_q(k) = \kappa \, k^z + \ldots, \quad \kappa \in \mathbb C, \label{lifdef}
\eeq
with Lifshitz exponent $z > 0$, which guarantees that dissipation accounted for by the imaginary part of $\kappa$ can be made arbitrary small by giving initial data support at arbitrarily small $k$. The ellipsis in~\eqref{lifdef} denote terms with higher powers of $k$, which are a counterpart of the hydrodynamic gradient expansion~\eqref{Pi_constitutive}. The aforementioned transients are simply contributions to the sum in~\eqref{spectral_decomposition} which do not share the property~\eqref{lifdef}.

The aim of the present article is to build on~\cite{Heller:2020uuy} to construct a transseries solution describing nonequilibrium processes going beyond the class of comoving flows represented by Bjorken dynamics. Our guiding principle will be to have hydrodynamic phenomena captured by the perturbative part of the transseries
with nonperturbative  transient phenomena captured by higher transseries sectors -- in analogy with Bjorken flow, where the perturbative part is given by the $1/w$ expansion while the transient effects are expressed by terms exponentially suppressed in $w$. In any given model, there may be many different ways to construct such an expansion, e.g. by treating some microscopic parameter as small. The expansion we use here can be applied to any model.

At the technical level, the key idea is to introduce a formal expansion parameter for the transseries by rescaling space and time coordinates in Minkowski space where a nonequilibrium phenomenon of interest takes place. 
Since we expect hydrodynamics to be a late time phenomenon,   
we introduce a formal parameter $\epsilon$ through the rescaling
\be
t \to \frac{t}{\epsilon^\alpha}, \qquad x \to \frac{x}{\epsilon}, \label{rescaling}
\ee
with $\alpha > 0$, and treat $\epsilon$ as small.
The resultant effect on \eqref{lifdef}, through the corresponding scaling $\omega \to \epsilon^\alpha\,\omega, k \to \epsilon \, k$, is given by,
\beq
\omega_q  = \kappa \, \epsilon^{z - \alpha}\, k^z + \ldots.
\eeq
For a given hydrodynamic sector parameterized by some Lifshitz exponent $z$, the natural choice is therefore the marginal one, $\alpha = z$, preserving the hydrodynamic scaling~\eqref{lifdef}. This choice focuses attention on the sector of interest, whilst not scaling as far as to render it trivial. It also has the desired effect of ensuring that all nonhydrodynamic modes appear nonperturbatively in a small $\epsilon$ expansion, since they scale as $\omega = O(\epsilon)^{-z}$. The outcome is that the spectral decomposition \eqref{spectral_decomposition} becomes a transseries in the parameter~$\epsilon$ with perturbative sectors corresponding to hydrodynamic mode contributions, and nonperturbative sectors corresponding to nonhydrodynamic ones. We emphasise that $\epsilon$ is only a formal parameter, and it takes the value $\epsilon = 1$ at the end of the calculation.

For definiteness, in the present work we focus on nonequilibrium phenomena described by the telegrapher's equation
\begin{equation}
\tau \partial_{t}^2 \rho + \partial_t \rho - D \partial_{x}^2 \rho = 0, \label{eq.telegraphers}
\end{equation}
which at large distances and long times describes diffusion, or $z=2$ hydrodynamic scaling~\eqref{lifdef}. Later, unless we keep explicitly $\tau$ and $D$, we use their following numerical values
\begin{equation}
\label{eq.paramschoice}
\tau = D = \frac{1}{2}.
\end{equation}
The linear partial differential equation~\eqref{eq.telegraphers} is well-known in the literature and, as we review in appendix~\ref{MIS}, it arises as the description of shear channel perturbations in the M{\"u}ller-Israel-Stewart (MIS) formulation of relativistic hydrodynamics~\cite{Romatschke:2009im}.\footnote{In the AdS/CFT context, the natural counterpart of this problem would be a shear channel fluctuation in all-order hydrodynamics as discussed in \cite{Bu:2014sia,Bu:2014ena}.} From a broader perspective, the telegrapher's equation features prominently in the context of quasihydrodynamics~\cite{Grozdanov:2018fic}, where it provides the simplest example of a diffusion-to-sound crossover. Quasihydrodynamics is the natural generalization of standard hydrodynamics in the presence of weakly broken symmetries. Apart from MIS itself, examples of theories featuring a diffusion-to-sound crossover described\footnote{It is worth remarking that the telegrapher's equation might only emerge in a suitable parametric limit.} by the telegrapher's equation include quantum fluctuating superconductors \cite{Davison:2016hno}, systems breaking spatial translations spontaneously in the presence of phase relaxation \cite{Delacretaz:2017zxd} and, in the AdS/CFT context, probe branes at finite temperature and large baryon density \cite{Kaminski:2009dh,Davison:2011ek,Chen:2017dsy}, models of momentum relaxation \cite{Davison:2014lua}, higher-derivative gravity \cite{Grozdanov:2016vgg,Grozdanov:2016fkt}, and constructions based on generalized global symmetries that describe dynamical electromagnetism in the boundary QFT \cite{Grozdanov:2017kyl,Hofman:2017vwr} or viscoelastic media \cite{Grozdanov:2018ewh}.\footnote{See also \cite{Baggioli:2020whu} for an embedding of the telegrapher's equation into a field-theoretic context.} Furthermore, with a straightforward modification, our methods also apply to the chiral magnetic waves in the presence of axial charge relaxation discussed in ~\cite{Jimenez-Alba:2014iia,Stephanov:2014dma}. These observations suggest that the results we will derive in this work are potentially relevant for a wide range of distinct physical systems.

\begin{figure}[h!]
\begin{center}
\includegraphics[width=7.2cm]{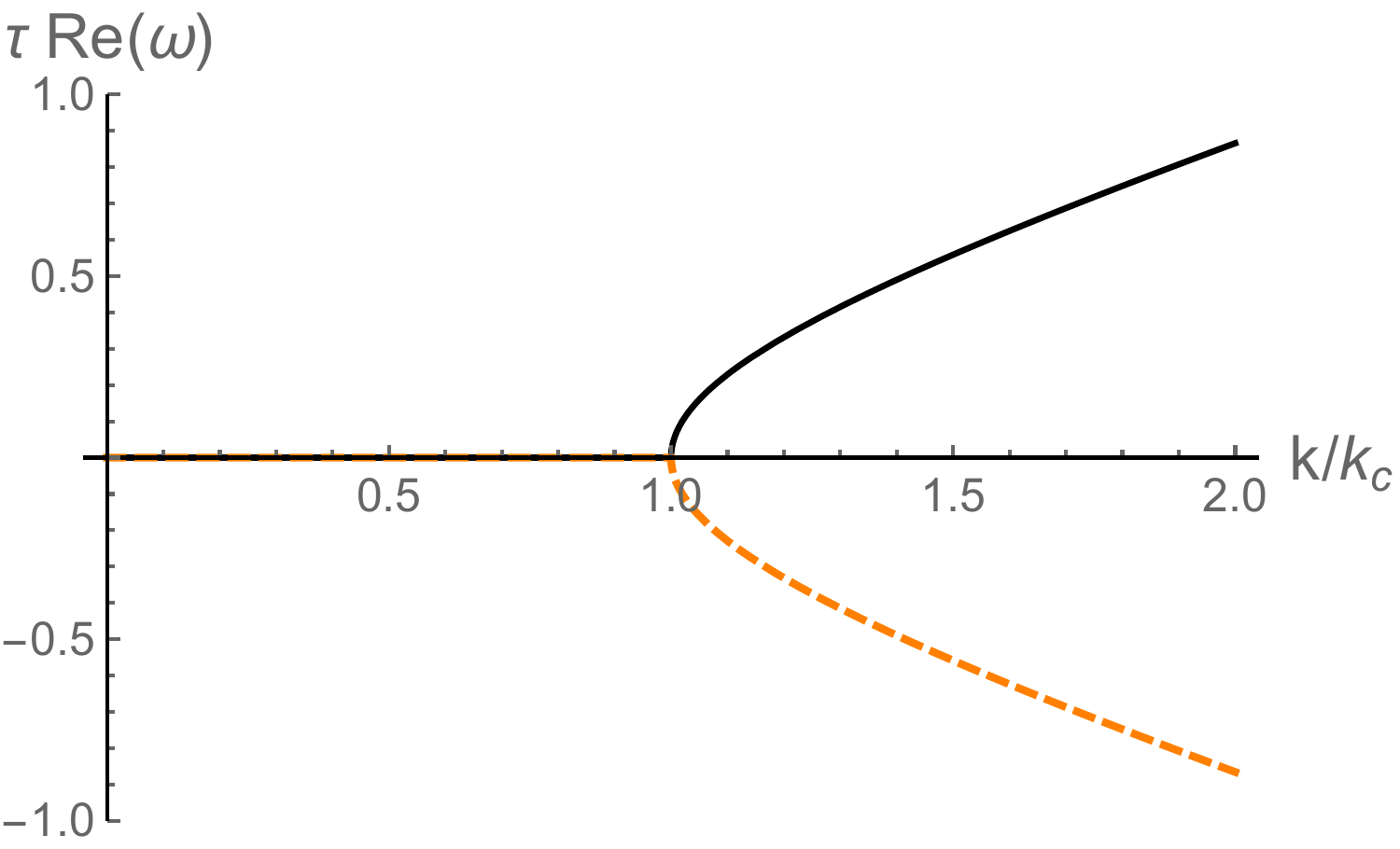}\,\,\,\includegraphics[width=7.2cm]{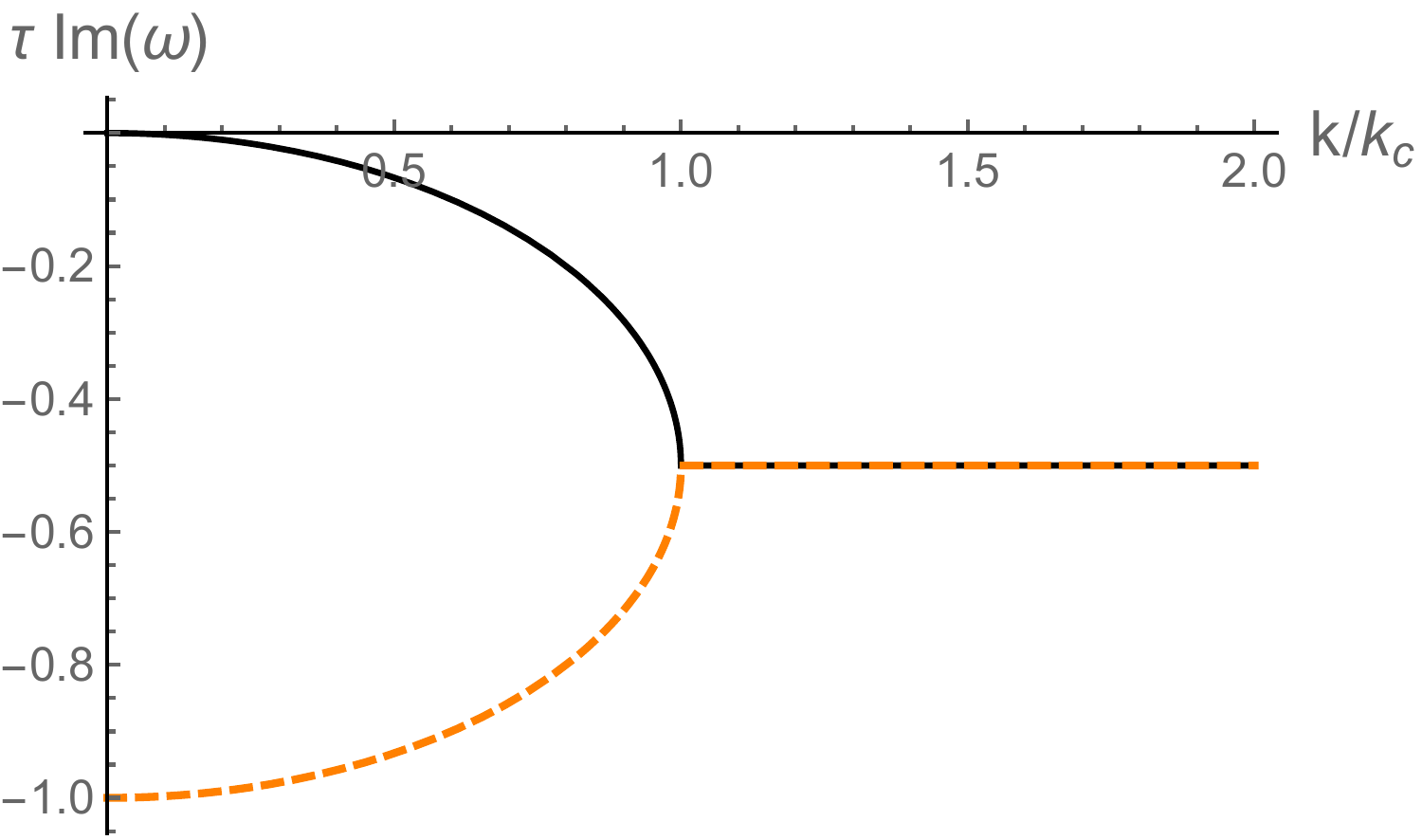}
\caption{\small Real (left) and imaginary (right) parts of the hydrodynamic (solid black) and nonhydrodynamic (dashed orange) modes in the shear channel of the M\"uller-Israel-Stewart theory, as functions of momentum. The modes are given by \eqref{modes}.}
\label{fig:modes}
\end{center}
\end{figure}

After these considerations, let us comment briefly on the mode structure of the telegrapher's equation. We have two modes in the sense of~\eqref{spectral_decomposition}:
\begin{equation}
\omega_H(k) = \frac{- i + i \Delta(k)}{2 \tau}, \quad \omega_{NH}(k) = \frac{- i - i \Delta(k)}{2 \tau}, \quad  \Delta(k) = \sqrt{1 - 4 D \tau k^2}. \label{modes}
\end{equation}
Among these two modes, $\omega_H$ is a hydrodynamic diffusion mode 
\begin{equation}
\omega_H(k) \to - i D k^2,\,\,\,k \to 0, 
\end{equation}
while $\omega_{NH}$ remains gapped in the same limit. As figure \ref{fig:modes} illustrates, both modes are nonpropagating and purely decaying below a critical momentum 
\beq 
k_c^2 = \frac{1}{4 D \tau}, \label{criticalk}
\eeq 
which corresponds to a branch point where the hydrodynamic and the nonhydrodynamic modes collide in the sense introduced in~\cite{Withers:2018srf} and developed in~\cite{Grozdanov:2019kge,Grozdanov:2019uhi,Abbasi:2020ykq,Jansen:2020hfd,Choi:2020tdj,Arean:2020eus}. For $\tau$ and $D$ given by~\eqref{eq.paramschoice}, $k_{c}^2 = 1$. Past the critical momentum, the modes acquire a propagating component. For asymptotically large $|k|$, both modes become purely propagating, with a linear dispersion relation $\omega \to \pm \sqrt{D/\tau}\, k$. This should not come as a surprise, since in the end the telegrapher's equation is nothing but a dissipative wave equation. If one wants to impose relativistic causality with the speed of light set to unity, this requires~$D \leq \tau$.\footnote{In Appendix \ref{causal} we discuss the causality properties of the telegrapher's equation in more detail.}

\noindent Another illustrative perspective is to view the mode collision represented by the telegrapher's equation as the simplest incarnation of the so-called $k$-gap phenomenon, which features widely across physics (see \cite{Baggioli:2019jcm} for a review). Apart from the examples mentioned before, the existence of a $k$-gap in the dispersion relation of the transverse collective excitations is a crucial feature distinguishing liquids and solids \cite{Baggioli:2018vfc,Baggioli:2018nnp,Baggioli:2020loj}. Other instances of the $k$-gap phenomenon in the AdS/CFT context include $p$-wave superfluids \cite{Arias:2014msa} or plasmons \cite{Gran:2018vdn,Baggioli:2019aqf,Baggioli:2019sio}.  

In the special case of the telegrapher's equation, the expansion in powers of $\epsilon$ can equally well be regarded as an expansion in powers of the relaxation time $\tau$. In consequence of our choice to seek a perturbative scheme in which hydrodynamic modes appear at leading order while nonhydrodynamic degrees of freedom enter as nonperturbative corrections, we observe that we are expanding around the acausal Navier-Stokes limit, and nonperturbative contributions are necessary to ensure causality. Likewise, physical effects of the propagating modes mentioned above enter only at the nonperturbative level. These observations are illustrated further in appendix \ref{comparison}, where we apply our perturbative scheme to a particular example.

From a more mathematical perspective, we will be studying solutions of a partial differential equation~\eqref{eq.telegraphers} in the form of a transseries
\begin{equation}
\hat{\rho}(t,k) = \sum_{n=0}^\infty \hat{a}_{n}(t,k)\epsilon^{2n} + e^{-\frac{t}{\tau \epsilon^2}} \sum_{n=0}^{\infty} \hat{b}_{n}(t,k) \epsilon^{2n}, \label{rho_transseries}
\end{equation}
where the hats indicate momentum space quantities. One of the crucial points of our work will be to make sense of this transseries expression in position space. Similar analyses of other partial differential equations have been carried out before in various contexts, for example in~\cite{chapman2005exponential, chapman2007shock, Howls2004Aug}. In holography, the large-$w$ expansion of Bjorken flow~\cite{Heller:2013fn, Casalderrey-Solana:2017zyh, Aniceto:2018uik} is also governed by partial differential equations, the fully nonlinear Einstein equations with negative cosmological constant.\footnote{When the microscopics is given by kinetic theory models, see~\cite{Heller:2016rtz, Heller:2018qvh}, the situation is even richer, as the collisional kernel in the Boltzmann equation generically involves integration over momenta.} One of the interesting conclusions in~\cite{chapman2005exponential, chapman2007shock, Howls2004Aug} is that the structure of the transseries may be different in different parts of space and time. As we will see, this is also the case in our studies. Furthermore, in our study we want to stress the dependence of the transseries on initial conditions.

Transseries have also been studied in the context of attractors in Bjorken flow \cite{Heller:2015dha}. The non-perturbative contributions quickly decay, leaving a universal perturbative piece identified as the attractor. Similarly, in the present context there are many different solutions that differ only by the behaviour of the transients. For that class of solutions, the perturbative part acts as an attractor in the same sense.\footnote{The series is generically divergent, so it must be properly resummed or optimally truncated.} However, the Bjorken flow attractor arises from far from equilibrium behaviour involving the fast expansion at early times \cite{kurkela2019attracts} while in this paper, because of linearity, all dynamics can be understood as small perturbations away from equilibrium. On the other hand, we are able to study more general flows, but leave open the question of what happens when non-linearities are introduced. Note on this front that less symmetric, non-linear flows have been studied numerically in~\cite{Romatschke:2017acs,kurkela2019attracts,Kurkela2020Dec,Ambrus:2021sjg}.

Finally, let us come back to our initial motivation, i.e. the (conformal) Bjorken flow, and discuss the $\epsilon$ expansion and~\eqref{rho_transseries} in relation to $w$-transseries in~\eqref{eq.Atransseries}. While the Bjorken flow is a genuinely nonlinear one-dimensional expansion, nothing stops us from applying an $\epsilon$-rescaling~\eqref{rescaling} also to this case. Choosing $x$ to be the expansion direction, the boost-invariance forces the dynamics to depend on proper time $\uptau = \sqrt{t^2 - x^2}$ only (not to be confused with the relaxation time $\tau$). As a result, the natural rescaling preserving the character of the proper time is the homogeneous one, i.e. $\alpha = 1$, which simply takes $\uptau \rightarrow \uptau/\epsilon$. The clock variable is defined as \begin{equation}
w \equiv \uptau \, T(\uptau),
\end{equation}
where $T(\uptau)$ is the local effective temperature associated with the local energy density. The homogeneous $\epsilon$ expansion forces large proper time expansion of $w$ in powers of $\epsilon^{2/3}/\uptau^{2/3}$ starting with the term $\left(\epsilon^{2/3}/\uptau^{2/3}\right)^{-1}$. 
This reorganizes the transseries~\eqref{eq.Atransseries} from a transseries in $w$, whose each perturbative contribution corresponds to a given order of hydrodynamic gradient expansion evaluated on-shell, to a transseries in $\epsilon^{2/3}/\uptau^{2/3}$. In the present work we will be working with the analogue of the latter expansion, whereas the former was discussed in full generality in linearized hydrodynamics in our recent paper~\cite{Heller:2020uuy}. In appendix~\ref{sec.relationtogradexp} we discuss the link between the hydrodynamic gradient expansion and the perturbative part of the $\epsilon$ transseries.

\section{The momentum-space transseries and its Fourier transform\label{epsilon_exp}}

\subsection{Constructing the transseries}

In this section we construct transseries solutions to the telegrapher's equation \eqref{eq.telegraphers} in the small formal parameter $\epsilon$ as defined in \eqref{rescaling}. Our starting point is thus the following $\epsilon$-rescaled telegrapher's equation, 
\beq
\epsilon^2 \tau \partial_t^2 \rho(t,x) + \partial_t \rho(t,x) - D \partial_x^2 \rho(t,x)=0, \label{eq_telegrapher_rescaled_real_space}
\eeq
with \eqref{eq.telegraphers} and its actual solution recovered by setting $\epsilon = 1$. The point of introducing $\epsilon$ is that this is the approach that one can adopt generally, for example in holography or kinetic theory, while the $\tau$-expansion is specific to the telegrapher's equation and related systems.

Due to the spatial translational invariance of \eqref{eq_telegrapher_rescaled_real_space}, it is convenient to work in momentum space, where \eqref{eq_telegrapher_rescaled_real_space} reduces to a linear ODE,\footnote{Momentum space quantities are denoted by a hat.}
\beq
\epsilon^2 \tau \partial_t^2 \hat{\rho}(t,k) + \partial_t \hat{\rho}(t,k) + D k^2 \hat{\rho}(t,k) =0. \label{eq_telegrapher_rescaled}
\eeq
One then immediately recognises an appropriate transseries ansatz for  $\hat{\rho}(t,k)$ as $\epsilon \to 0$ as given by~\eqref{rho_transseries}. Plugging it into \eqref{eq_telegrapher_rescaled} and considering terms order-by-order in $\epsilon$, reveals the following pair of nested ODE systems
\begin{subequations}
\label{eq_nested_full}
\begin{align}
&\partial_t \hat{a}_{n+1}(t,k) + D k^2 \hat{a}_{n+1}(t,k) + \tau \partial_t^2 \hat{a}_{n}(t,k) = 0, \label{eq_nested_a}\\
&\partial_t \hat{b}_{n+1}(t,k) - D k^2 \hat{b}_{n+1}(t,k) - \tau \partial_t^2 \hat{b}_{n}(t,k) = 0. \label{eq_nested_b}
\end{align}
\end{subequations}
where we take $\hat{a}_{-1} = \hat{b}_{-1} = 0$. We see that $\hat{a}_{n+1}(t,k)$ obeys a heat equation sourced by the previous order, while $\hat{b}_{n+1}(t,k)$ is governed by the time-reversed equation. At this stage, the two equations are decoupled from one another. In order to solve \eqref{eq_nested_full}, we supplement the equations with initial data at $t=0$,
\beq
\rho(0,x) = u(x),\quad \partial_t \rho(0,x) = v(x) \label{uv_def}
\eeq
and denote the corresponding Fourier-transformed functions as $\hat{u}(k), \hat{v}(k)$. For the sake of simplicity of presentation in what follows, we are going to focus on the $u(x) = 0$ case. At the level of the expansion coefficients, this initial condition reduces to $\hat{a}_0(t,k) = \hat{b}_0(t,k) = 0$, $\hat{b}_1(0,k) = -\tau \hat{v}(k)$ and 
\begin{subequations}
\label{bc_full}
\begin{align}
&\hat{a}_{n+1}(0,k) + \hat{b}_{n+1}(0,k) = 0, \label{bc_1} \\
&\hat{b}_{n+1}(0,k) - \tau (\partial_t \hat{a}_{n}(0,k) + \partial_t \hat{b}_{n}(0,k)) = 0, \label{bc_2} 
\end{align}
\end{subequations}
where the last equality applies only for $n > 0$. The initial conditions couple the coefficients arising in the perturbative series to those in the nonperturbative series. Finally, it is possible to find closed-form expressions for $\hat{a}_n(t,k)$ and $\hat{b}_n(t,k)$ that solve \eqref{eq_nested_full} and obey \eqref{bc_full}. They are given by  
\begin{subequations}
\begin{align}
&\hat{a}_{n+1}(t,k) = \frac{2^{2n}\Gamma\left(n+\frac{1}{2} \right)}{\sqrt{\pi} \Gamma(n+1)}D^n \tau^{n+1}  k^{2n}  {}_1F_1\left(2n+1,n+1,- D k^2 t\right)\hat{v}(k), \label{a_n} \\
&\hat{b}_{n+1}(t,k) = -\hat{a}_{n+1}(-t,k). \label{b_n}
\end{align}
\end{subequations}

Let us now turn our attention to position space. Our guiding principle here will be to define the expansion coefficients of a given sector of the position space transseries as the inverse Fourier transform of the corresponding momentum space coefficients, as long as this inverse Fourier transform exists for positive real $t$. This is always the case for the perturbative sector of \eqref{rho_transseries} (i.e. the $\hat{a}_n$ coefficients). In fact, the perturbative sector  proceeds straightforwardly, and we can immediately obtain closed-form results which we present in the remainder of this section. Nonperturbative contributions in position space are both subtle and interesting, and later sections of this paper are devoted to this topic.

The position space coefficient,
\beq
a_{n+1}(t,x) \equiv \int_{\mathbb R} dk \, \hat{a}_{n+1}(t,k)e^{ikx}  
\eeq
can be computed in closed form, with the result
\begin{equation}
a_{n+1}(t,x) = \frac{\tau^{n+1} \Gamma\left(n+\frac{1}{2}\right)^2}{2 \pi^\frac{3}{2}\sqrt{D} \Gamma(n+1) t^{n + \frac{1}{2}}} (K_{n+1} * v)(x), \label{rho_n_real_space_full}
\end{equation}
where $*$ represents a convolution in the spatial coordinate, and the $(n+1)$-th kernel $K_{n+1}(x)$ is given by 
\begin{equation}
K_{n+1}(x) = {}_1 F_1\left(\frac{1}{2} + n, \frac{1}{2} - n, - \frac{x^2}{4 D t} \right). \label{kernel}
\end{equation}
An alternative representation of the same result is given by
\begin{equation}
a_{n+1}(t,x) = (-1)^n D^n \tau^{n+1}\sum_{q=0}^n {{2n}\choose{n-q}}\frac{D^q t^q}{q!} \partial_x^{2 (n+q)} (G_0 * v),\label{rho_n_real_space_gradient}
\end{equation}
where $G_0(t,x)$ is the propagator for the heat equation, 
\begin{equation}
G_0(t,x) = \frac{e^{-\frac{x^2}{4Dt}}}{2 \sqrt{\pi} \sqrt{D t}}. \label{heat_kernel}
\end{equation}

With $\hat{a}_{n+1}(t,k)$ and $a_{n+1}(t,x)$ now computed, we can arrange them to produce a piece of the full solution  for $\hat{\rho}$ and $\rho$ respectively. 
These we label with a superscript `$H$',
\beq
\hat{\rho}_\epsilon^{(H)}(t,k) \equiv \sum_{n=0}^\infty \hat{a}_{n+1}(t,k) \epsilon^{2n+2}, \quad \rho_\epsilon^{(H)}(t,x) \equiv \sum_{n=0}^\infty a_{n+1}(t,x) \epsilon^{2n+2}, \label{rhoHepsilonT}
\eeq
and since we obtained them by inverse Fourier transform of a series in $\epsilon$ we have added a subscript $\epsilon$ to denote the fact that they are not exact in $\epsilon$. 
In the light of \eqref{rho_n_real_space_gradient}, each term in $\rho^{(H)}_\epsilon(t,x)$ corresponds to a gradient series in $\partial_x^2$ acting upon a solution of the heat equation that depends on the initial data. In this sense, as advocated in the Introduction and expanded upon in Appendix~\ref{sec.relationtogradexp}, $\rho^{(H)}_\epsilon(t,x)$ provides a~particular reorganization of a gradient expansion construction of the contribution of the hydrodynamic mode to $\rho(t,x)$, hence the label~`$H$'. The convergence of $\rho^{(H)}_\epsilon(t,x)$ will be the focus of the next section, and 
we refer the reader to Appendix~\ref{comparison} for an analysis of how well $\rho^{(H)}_\epsilon(t,x)$ -- when we set $\epsilon=1$ -- reproduces the exact microscopic $\rho(t,x)$ in a particular example.  

\subsection{Large-order behavior}
\label{convergence}

In this section we analyse the large-order behavior of $\rho^{(H)}_{\epsilon}(t,x)\big|_{\epsilon = 1}$. We first provide a fully general, model-independent condition for the convergence of this object, which relies only on the support of the initial data in momentum space. Then, we discuss the large-order behavior of the series for initial data where this condition fails. 

We begin by splitting the full microscopic $\rho(t,x)$ into hydrodynamic and nonhydrodynamic mode contributions, $\rho(t,x) = \rho^{(H)}(t,x) + \rho^{(NH)}(t,x)$, as defined by individual contributions to the Fourier integral along a path $\gamma$,\footnote{Note that the splitting is only unequivocally defined once a particular integration path $\gamma$ is specified. While, for entire initial data $\hat{u}(k)$ and $\hat{v}(k)$, $\rho(t,x)$ is the Fourier transform of an entire function, this is not the case for the individual hydrodynamic and nonhydrodynamic contributions as defined in the text, due to the branch points of the mode frequencies at $k^2 = k_c^2$. Each individual contribution is well-defined only after a particular $\gamma$ to go around the corresponding branch cuts has been provided.} where
\begin{equation}
\rho^{(H)}(t,x) \equiv \int_{\gamma} dk\,\hat{\rho}^{(H)}(t,k)\, e^{i k x}, \quad \hat{\rho}^{(H)}(t,k) = f_H(k)  e^{-i \frac{\omega_H(\epsilon k)}{\epsilon^2} t}, \label{rho_H_general}
\end{equation}
with analogous expressions for the nonhydrodynamic mode, $NH$. As written, these expressions are exact in $\epsilon$, and notationally this is indicated by the lack of an $\epsilon$ subscript. For our choice of initial data (\eqref{uv_def} with $u=0$), we have
\beq
f_H(k) = - f_{NH}(k) = \frac{\epsilon^2 \tau}{\sqrt{1- 4 D \tau \epsilon^2 k^2}}\hat{v}(k). 
\eeq
By series expanding the exact $\hat{\rho}^{(H)}(t,k)$ around $\epsilon =0$ we recover the perturbative sector of the momentum space transseries, $\hat{\rho}^{(H)}_\epsilon(t,k)$, computed earlier \eqref{rhoHepsilonT}. The existence of two branch points in $\omega_H(\epsilon k)$ and $f_H(k)$ at $\epsilon k = \pm |k_c|$ implies that, for $\epsilon |k| > |k_c|$, $\hat{\rho}_\epsilon^{(H)}$ is a divergent series. On the other hand, in defining $\rho_\epsilon^{(H)}$ we assumed that the momentum space integral commutes  with the infinite sum in $\hat{\rho}_\epsilon^{(H)}$, in such a way that the individual series expansion coefficients, $\hat{a}_n$ and $a_n$, were directly related by the Fourier transform. Hence, 
\beq
\rho^{(H)}_\epsilon(t,x) = \int_{\mathbb R} dk\, \hat{\rho}^{(H)}_\epsilon(t,k) e^{ikx}. 
\eeq
From the expression above it is immediate to see that, unless the momentum space support of the initial data $\hat{v}$ is restricted to\footnote{Recall that one should set the formal parameter $\epsilon=1$ at the end of the analysis.}
\beq
|k| \leq \frac{|k_c|}{\epsilon}, \label{convergence_condition}
\eeq
the integral contains contributions from the momentum space region where $\hat{\rho}^{(H)}_\epsilon$ diverges. Modulo fine-tuned cancellations, the natural expectation to draw from this analysis is that the series~$\rho^{(H)}_\epsilon$ does not converge if the initial data does not satisfy the condition \eqref{convergence_condition}. 

It is important to note that, even if we have obtained \eqref{convergence_condition} for a particular class of initial data for the telegrapher's equation, its applicability is not restricted to this example. As long as the hydrodynamic mode frequency $\omega_H(k)$ has a complex singularity at $|k| = |k_c|$ and $f_H(k)$ is analytic for $|k| \leq |k_c|$, equation \eqref{convergence_condition} will hold. It can be argued that the existence of this complex singularity in $\omega_H(k)$ is a necessary condition for a microscopic theory to behave causally \cite{Heller:2020uuy}. From this viewpoint, $\rho_\epsilon^{(H)}$ would be a divergent series in any theory that respects relativistic causality for generic initial data with unrestricted momentum space support. 
\\\\
\noindent In order to illustrate how the convergence condition \eqref{convergence_condition} applies to our case, we consider compactly supported initial data in momentum space of the form 
\begin{equation}
\hat{v}(k) = \frac{1}{2\pi} \Theta(A^2 - k^2), \label{id_compact_support}
\end{equation}
where $\Theta$ denotes the Heaviside step function. These initial data correspond to a regularized $\delta$-function in position space. For $x = 0$, $a_n(t,0)$ can be explicitly computed
\small
\begin{equation}
a_{n+1}(t,0) = \frac{2^{4n} D^n \tau^{n+1}A^{2n+1} \Gamma\left(n+\frac{1}{2} \right)^2}{\pi^2 \Gamma(2n+2)}{}_2 F_2\left(2n+1, r+\frac{1}{2}; n+1, n + \frac{3}{2}, - A^2 D t\right). \label{rho_n_compact_support}
\end{equation}
\normalsize
\begin{figure}[h!]
\begin{center}
\includegraphics[width=7.7cm]{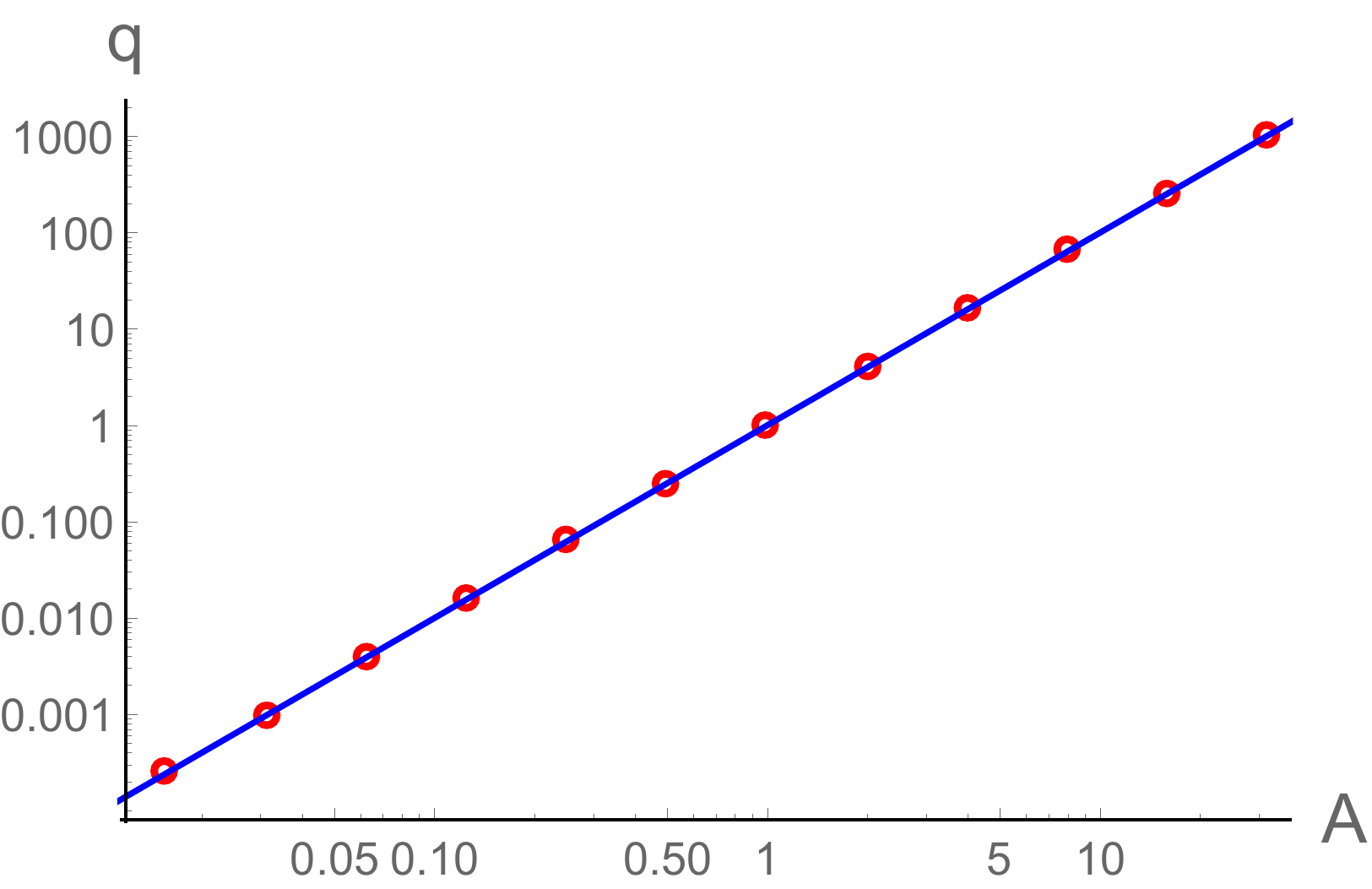}\includegraphics[width=7.7cm]{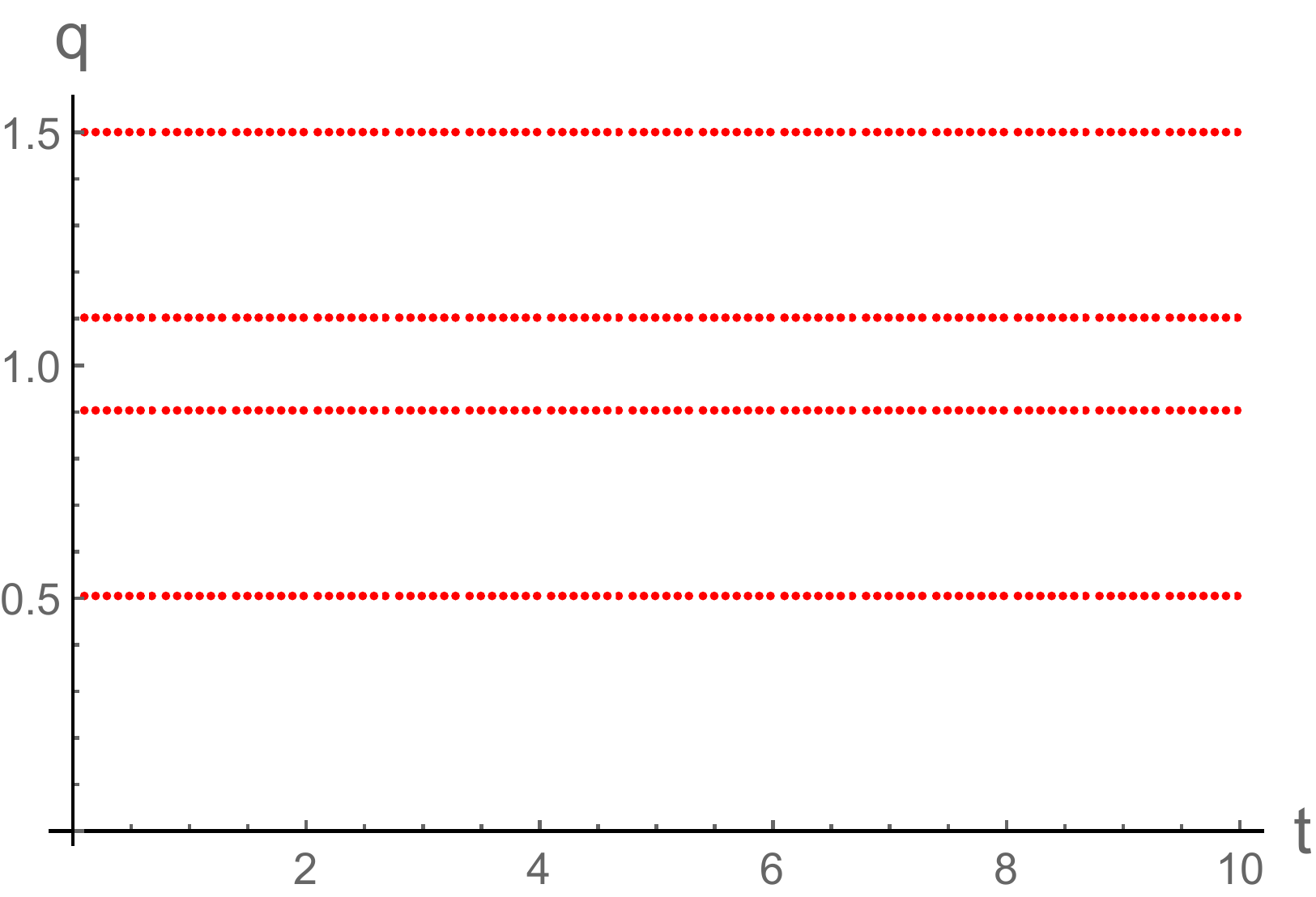}
\caption{\small The ratio test for the perturbative sector in position space for initial data \eqref{id_compact_support}. Left: The ratio $q$ defined in \eqref{ratiotest}  vs. maximum support of the initial data in momentum space, $A$, at $t = 1$. Right: $q$ vs. $t$ for different values of $A$; from top to bottom, $A^2 = 1.5, 1.1, 0.9$ and $0.5$.}
\label{fig:ratio_test_compact_support}
\end{center}
\end{figure}

To check whether the series expansion defined by the coefficients above is convergent, we use the ratio test and compute numerically
\beq
q =\lim_{n\to\infty} q_n, \quad q_n = |a_{n+1}/a_n|. \label{ratiotest}
\eeq
A sufficient condition for convergence is that $q < 1$. In figure \ref{fig:ratio_test_compact_support} (left), we plot $q$ as a function of $A$ at $t=1$. We always find that $q = A^2$ to exceedingly good accuracy. This implies that $\rho^{(H)}_{\epsilon}(t,x)\big|_{\epsilon = 1}$ only converges for $A < 1$, i.e., for initial data with no support for $|k| > |k_c| = 1$ in $k$-space. If the support exceeds this bound, we obtain a divergent asymptotic series. This statement is time-independent, as figure \ref{fig:ratio_test_compact_support} (right) shows. 

It is worth mentioning that, as long as $|k_c| < A < \infty$, the series divergence is just a geometric one. A factorial growth of the $a_n(t,0)$ coefficients only appears in the $A \to \infty$ limit. In this case, $v(x) = \delta(x)$,  
\begin{equation}
a_{n+1}(t,0) = \frac{\tau^{n+1}\Gamma\left( n + \frac{1}{2}\right)^2}{2 \pi^\frac{3}{2} \sqrt{D}\Gamma(n+1) t^{n+\frac{1}{2}}},  \label{rho_n_delta}
\end{equation}
and it follows that the ratio between successive coefficients increases linearly with $n$, 
\beq
q_n \sim \frac{\tau}{t} n,
\eeq
implying a factorial divergence.    

We have found that the correlation between the infinite support of the initial data in $k$-space and the factorially divergent character of $\rho^{(H)}_{\epsilon}(t,x)\big|_{\epsilon = 1}$ always extends beyond the particular case of $\delta$-function initial data discussed above. This is the behavior to expect for generic initial data: initial data with a sharp cutoff in momentum space represent fine-tuned states from a position space perspective, in the sense that any modification of the initial data which is localized in position space -- for instance, by any Gaussian -- would make the momentum space support unbounded.   

\begin{figure}[h!]
\begin{center}
\includegraphics[width=16cm]{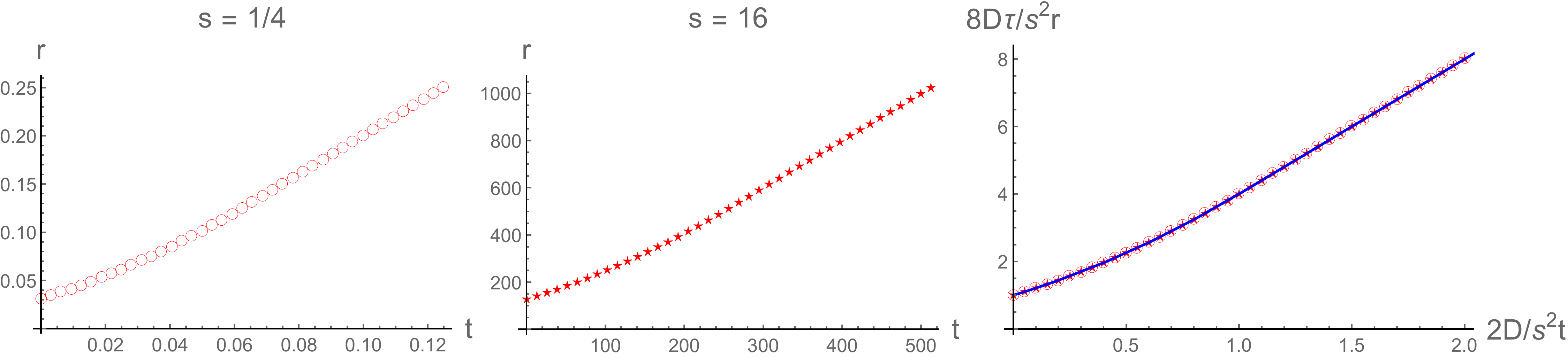}
\caption{Factorial growth prefactor \small $r$ defined by~\eqref{qn_GID}~(left) for Gaussian initial data obtained numerically from~\eqref{an_GID}, as a function of time, for $s = 1/4$ (left) and $s=16$ (middle). Both sets of data collapse on a single universal curve as shown in the right plot, which confirms~\eqref{qn_GID}~(right).}
\label{fig:GID_convergence}
\end{center}
\end{figure}

Let us illustrate further the generic factorial divergence of the perturbative series by focusing on the example provided by Gaussian initial data of the form\footnote{See appendix \ref{Lorentzian} for initial data of the Lorentzian form, i.e.~$v(x) = \frac{\alpha}{\pi (x^2 + \alpha^2)}$.}
\begin{equation}
v(x) = \frac{e^{-\frac{x^2}{2s^2}}}{\sqrt{2 \pi s^2}},\,\,\,\hat v(k) = \frac{1}{2 \pi}e^{-\frac{1}{2}s^2 k^2},  \label{GID}
\end{equation}
As it happened with our compactly-supported initial data, this case is also analytically solvable at $x = 0$. We obtain
\begin{equation}
a_{n+1}(t,0) = \frac{2^{3n}\tau^{n+1}D^n \Gamma\left(n+\frac{1}{2} \right)^2}{\sqrt{2}\pi^\frac{3}{2}s^{2n+1}\Gamma(n+1)} {}_2F_1\left(n + \frac{1}{2}, 2n+1, n+1, - \frac{2 D t}{s^2} \right).\label{an_GID}
\end{equation}
These coefficients, which diverge factorially as $n \to \infty$, show a more intricate behavior than their  $\delta$-function initial data counterparts. As $n \to \infty$,
\beq
q_n \sim r^{-1} n, \quad r = \frac{s^2}{8 D \tau} \upmu\left(\frac{2Dt}{s^2} \right), \label{qn_GID}
\eeq
where the function $\upmu$ depends only the scaling variable
\beq
\label{eq.tcdef}
\upeta = \frac{t}{t_c}, \quad  t_c = \frac{s^2}{2 D}, 
\eeq
and is given by 
\begin{equation}
\upmu(\upeta) = \left\{
\begin{array}{ll}
      (1+\upeta)^2, & \upeta < 1 \\
      4 \upeta, & \upeta > 1.
\end{array} 
\right.
\end{equation}
This implies that, for Gaussian initial data, $q_n$ only shows the behavior of $\delta$-function initial data for times greater than a critical time. In figure \ref{fig:GID_convergence}, we plot how $r$ evolves with $t$ for $s =1/4$ (left) and $s = 16$ (center). These two curves, when re-scaled by $8 D \tau/s^2$  and expressed as functions of $2Dt/s^2$, collapse to a universal function, which agrees with $\upmu(\upeta)$ (right, in blue). 

Let us close this section by mentioning that the $r$ parameter can be equivalently extracted by means of Borel transforms. 
Given a factorially divergent asymptotic series 
\begin{equation}
f\left(\epsilon^2\right) = \sum_{n=0}^{\infty} c_n \left( \epsilon^2 \right)^{n},  
\end{equation}
its Borel transform $f_B(z)$ given by 
\begin{equation}
f_B(z) = \sum_{n=0}^\infty \frac{c_n}{n!} z^n,  
\end{equation}
and, by definition, has a finite convergence radius in $z$. This convergence radius is set by the singularity which is closest to the origin in the complex $z$ plane. In general, the analytical continuation of $f_B(z)$ to complex $z$ cannot be computed in closed form and some approximation technique needs to be invoked. A common choice is to use Padé approximants. The $(m,n)$-Padé approximant of $f_B(z)$ is the unique rational function 
\beq
\label{eq.pade}
\mathcal{P}(z) = \frac{\sum_{p=0}^m a_p z^p}{1+\sum_{p=1}^n b_p z^p}
\eeq
such that
\beq
f_B(z) - \mathcal{P}(z) = O(z^{m+n+1}). 
\eeq
A well-known property of Padé approximants is that branch cuts in $f_B(z)$ manifest themselves as lines of pole condensation in $\mathcal{P}(z)$. For the Gaussian initial data~\eqref{GID}, including their $\delta$-function limit $s \rightarrow 0$, we always find a line of pole condensation along the positive $z$-axis starting at 
\beq
z_c = r,
\eeq
where $r$ is given by~\eqref{qn_GID}. Therefore, we conclude that the exact analytically-continued Borel transform has a branch point at this location, as expected on the basis of the large order behaviour of the series. Note, however, that for Gaussian initial data the location of the branch point depends on $t/t_c$, and is given by $t/\tau$ only for $t>t_c$. For $t<t_c$ the branch point location also depends on the initial data through its dependence on $s$. Note that a condensation of poles can hide more than a single branch cut,\footnote{For example, for $f_{B}(z) = \sqrt{1-z} + \sqrt{2-z}$, the Padé approximation~\eqref{eq.pade} is going to display a~condensation of poles emanating from $1$ towards larger values of $z$ on the real axis. The poles with $1 \leq z < 2$ will correspond to the branch cut associated with $\sqrt{1-z}$ and the poles with $z \geq 2$ will include also contributions from the branch cut associated with $\sqrt{2-z}$.\label{footnote.padecoincidentcuts}} as happened, for example, in~\cite{Heller:2015dha}. We will come back to this point in the next section, where we discuss the physical origin of these singularities and their dependence on $t$.

\section{Saddle point analysis}

In this section we take a different perspective and address the $\epsilon \to 0$ limit of the hydrodynamic mode contribution $\rho^{(H)}(t,x)$ by means of a saddle point analysis. As we will show, and as expected on general grounds, this procedure allows us to understand the physical origin of the branch points in the Borel plane. 

To start, let us consider a schematic integral 
\beq
I(\lambda) = \int_\gamma du \, G(u) \, e^{\lambda S(u)}, 
\eeq
and focus on its behavior for $|\lambda| \to \infty$. For us, the parameter $\lambda$ will be simply $\frac{1}{\epsilon^2}$ and we introduce it purely for notational convenience. This behavior can be obtained from a saddle point analysis. 
The relevant analysis can decomposed intro three subsequent steps. First, one finds the stationary points $u_s$ of the `action' $S(u)$, \beq
\frac{d}{du} S(u) \Bigr|_{u=u_s} = 0.
\eeq
Second, the steepest descent contours emanating from these saddle points are determined. The steepest descent contour $\gamma_s$ associated to the $u_s$ saddle point is the path emanating from $u_s$ along which $\textrm{Re}\,S(u)$ decreases the fastest. It obeys 
\beq
\textrm{Im}\,\lambda S(u(\xi)) = \textrm{Im}\,\lambda S(u_s). 
\eeq
Finally, one decomposes the original integration path $\gamma$ into steepest descent contours $\gamma_s$ and calculates the corresponding integrals in a $|\lambda| \to \infty$ expansion. 

The steepest descent path associated to a saddle $u_0$ can collide with another saddle~$u_1$ when $\arg \lambda$ is such that $\textrm{Im}\, \lambda S(u_0) = \textrm{Im}\,\lambda S(u_1)$. These saddles are known as adjacent saddles \cite{berry1991hyperasymptotics} and play a prominent role in controlling the large-order behavior of the $|\lambda| \to \infty $ expansion around the original $u_0$ saddle. In particular, when considering the Borel transform of this $|\lambda| \to \infty$ expansion, adjacent saddles manifest themselves as branch points in the Borel plane \cite{Serone:2017nmd}, located at
\beq
z_c = - (S_1 - S_0). 
\eeq
Let us apply this line of reasoning to understand the large-order behavior of $\rho^{(H)}_\epsilon(t,0)$ we reported in the previous section for the case of Gaussian initial data~\eqref{GID}. In this case, we have that
\beq
\rho^{(H)}(t,x) = \int_\gamma dk \, G_H( k) \, e^{\frac{S_H(t,x)}{\epsilon^2}}, \quad G_H( k) = \frac{\epsilon^2 \tau}{2 \pi\sqrt{1- 4 D \tau \epsilon^2 k^2}}, 
\eeq
with action 
\begin{equation}
\label{eq.SHdef}
S_H(t,x) = - \frac{1}{2}s^2 \epsilon^2 k^2 + \frac{(-1+\sqrt{1- 4 D \tau \epsilon^2 k^2})t}{2\tau} + i \epsilon^2 k x.  
\end{equation}
The analogous expressions for $\rho^{(NH)}, G_{NH}$ and $S_{NH}$ are obtained by flipping the sign of the square root in~\eqref{eq.SHdef}.
The most important point to draw from the expression above is that the initial data contribute nontrivially to the relevant action and, as a result, also to its saddle points. Up to this point, our expressions are exact in $\epsilon$. 
\begin{figure}[h!]
\begin{center}
\includegraphics[width=7.5cm]{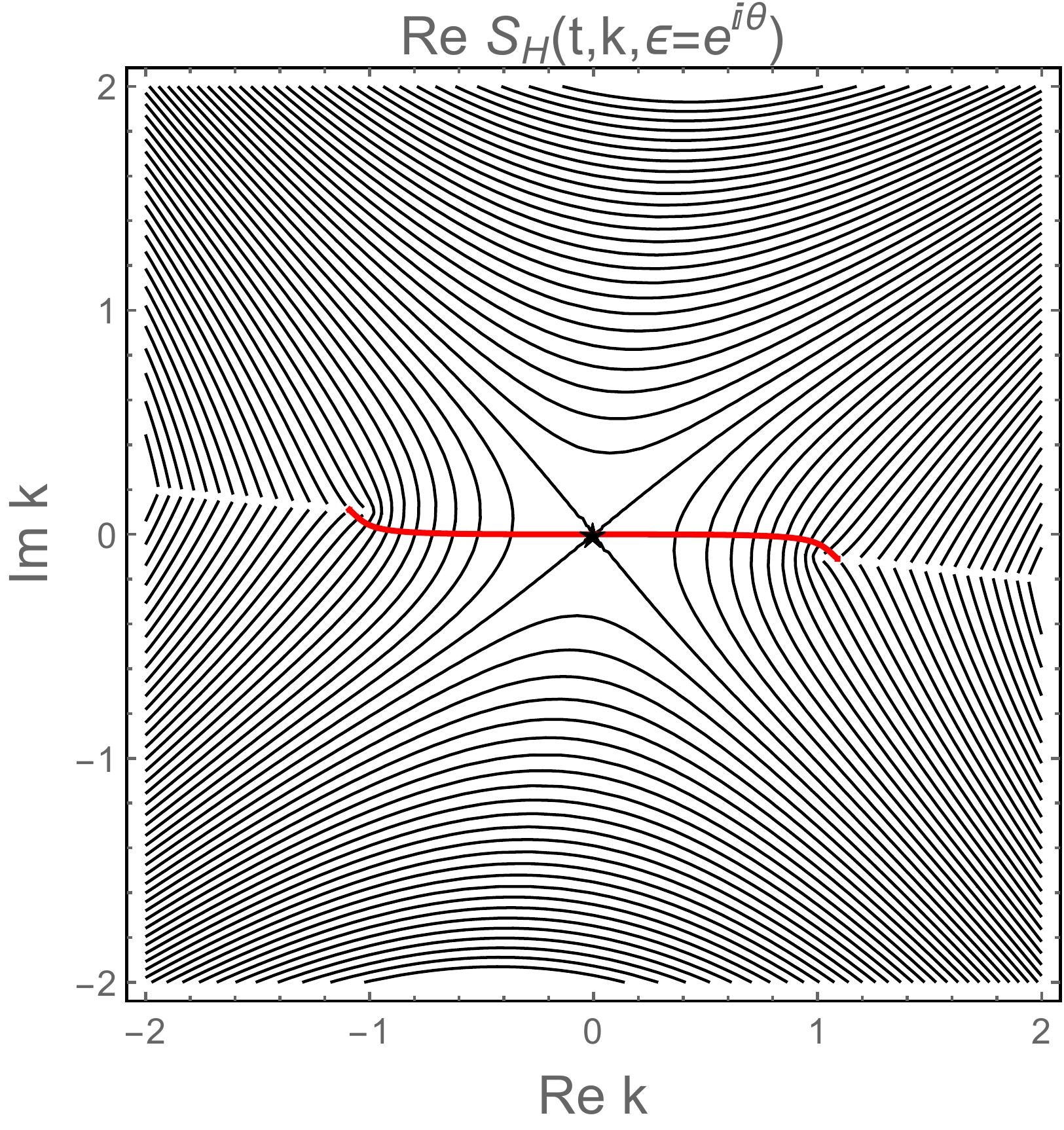}\includegraphics[width=7.5cm]{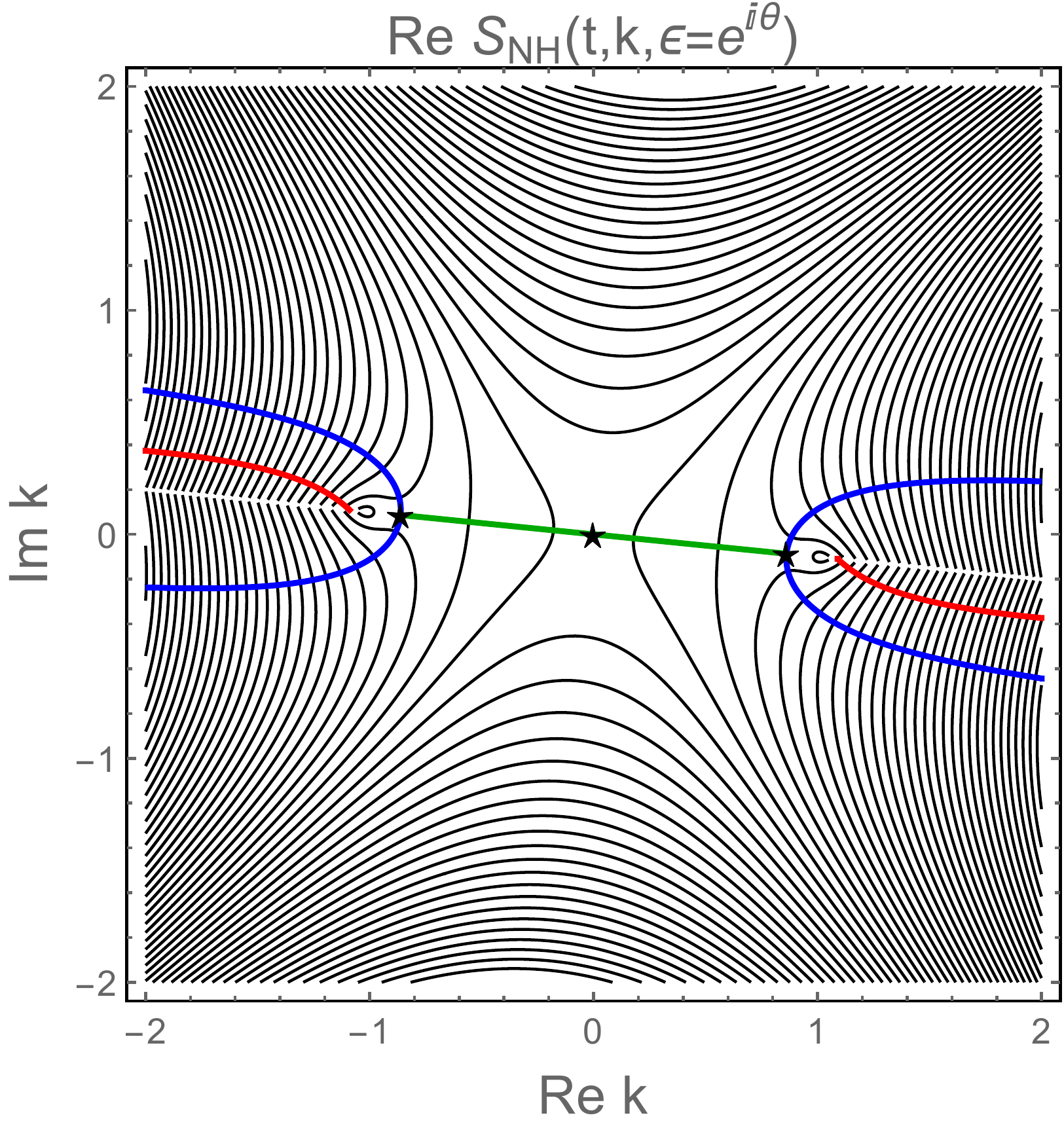}\\
\caption{\small Level plots of the saddle point actions $\textrm{Re}\,S_H(t,k)$ (left) and $\textrm{Re}\,S_{NH}(t,k)$ (right) for $\epsilon=e^{i\theta}$ for Gaussian initial data~\eqref{GID} with $s = 1$, $t = 0.5$ and $\theta = 0.1$. The saddle points are denoted by the black stars and associated steepest descent paths are shown in red ($k = 0$, hydrodynamic), green ($k_0$, nonhydrodynamic) and blue ($k_\pm$). The red path crosses the branch cuts (visible as white discontinuities in the plots) and continues on the other sheet, while the green path ends at the $k_\pm$ saddles.}
\label{fig:sd_paths}
\end{center}
\end{figure}

When $x = 0$, the saddle points of $S_H,\,S_{NH}$ are as follows. For $S_H$, we have a single saddle point located at $k=0$ with vanishing action. On the other hand, $S_{NH}$ has three saddle points. The first one is located at $k = k_0 = 0$ at all times and has action 
\beq
S_0 = -\frac{t}{\tau}. 
\eeq
The positions of the two remaining ones are time dependent
\beq
k = k_\pm = \pm \sqrt{1 - \left(\frac{t}{t_c}\right)^2} k_c
\eeq
with $t_{c}$ given by~\eqref{eq.tcdef} and lead to actions 
\beq
S_\pm = - \frac{1}{2}s^2 \left(1 + \frac{t}{t_c}\right)^2 k_c^2. 
\eeq
At $t = 0$, the $k_\pm$ saddles start at the critical momenta $\pm k_c$. As $t$ grows, they approach each other along the real $k$-axis, until colliding with the $k_0$ saddle at $t = t_c$. Past this time, they recede from the origin in opposite directions along the imaginary $k$-axis.

The perturbative sector of our transseries, $\rho^{(H)}_\epsilon(t,x)$, corresponds to the saddle point expansion around the hydrodynamic saddle.\footnote{See Appendix \ref{sp_details} for further technical details.} Crucially, for complex
\begin{equation}
\epsilon = |\epsilon| e^{i\theta},
\end{equation}
the steepest descent path emanating from this saddle crosses the branch cuts in the hydrodynamic dispersion relation and continues on the nonhydrodynamic sheet. Once there, this path can collide with adjacent nonhydrodynamic saddle points at specific values of~$\theta$. An example of this behavior is provided in figure \ref{fig:sd_paths}, where we also show the steepest descent contours associated to the three nonhydrodynamical saddles. 
\begin{figure}[h!]
\begin{center}
\includegraphics[width=7.5cm]{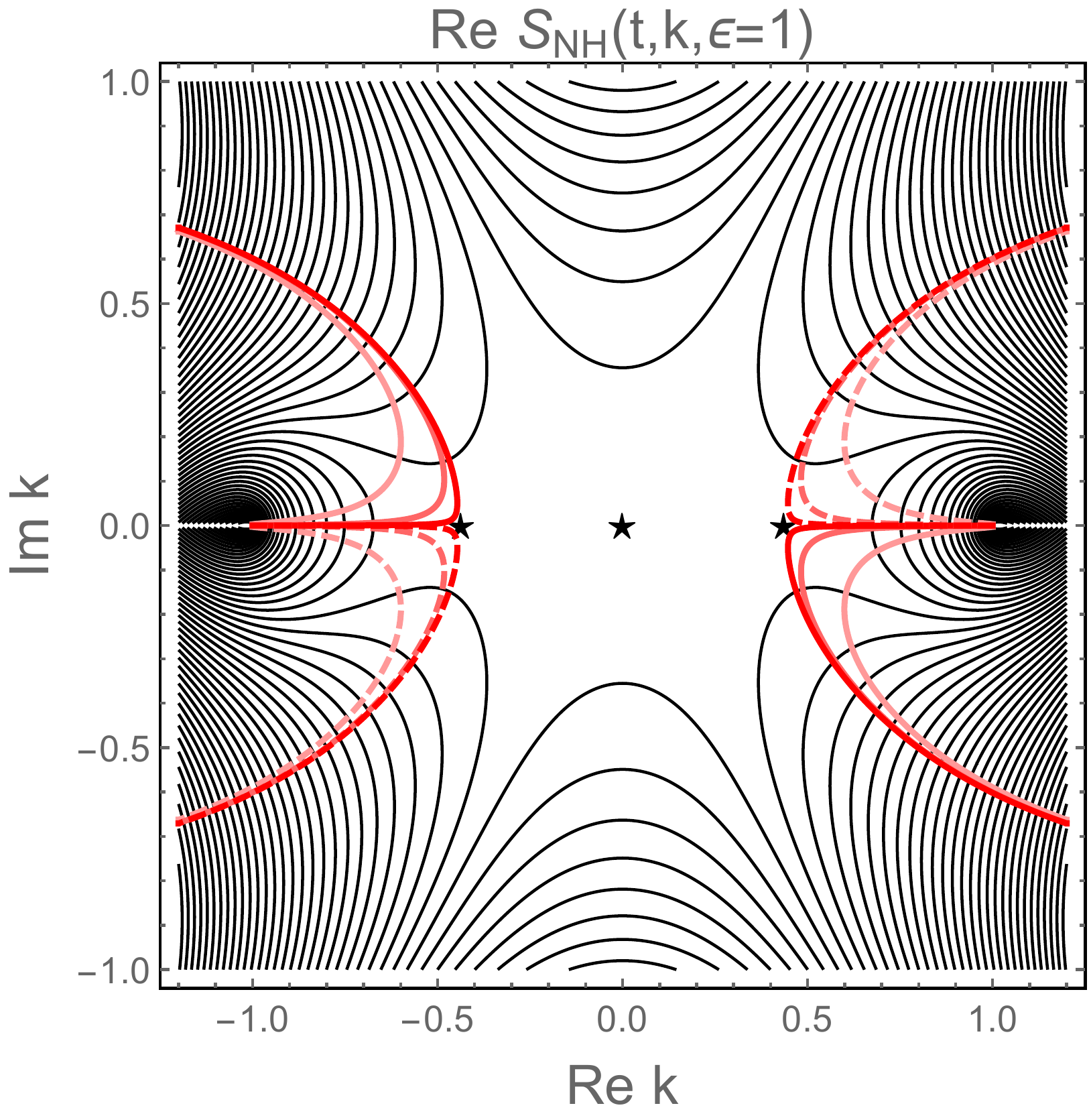}\includegraphics[width=7.5cm]{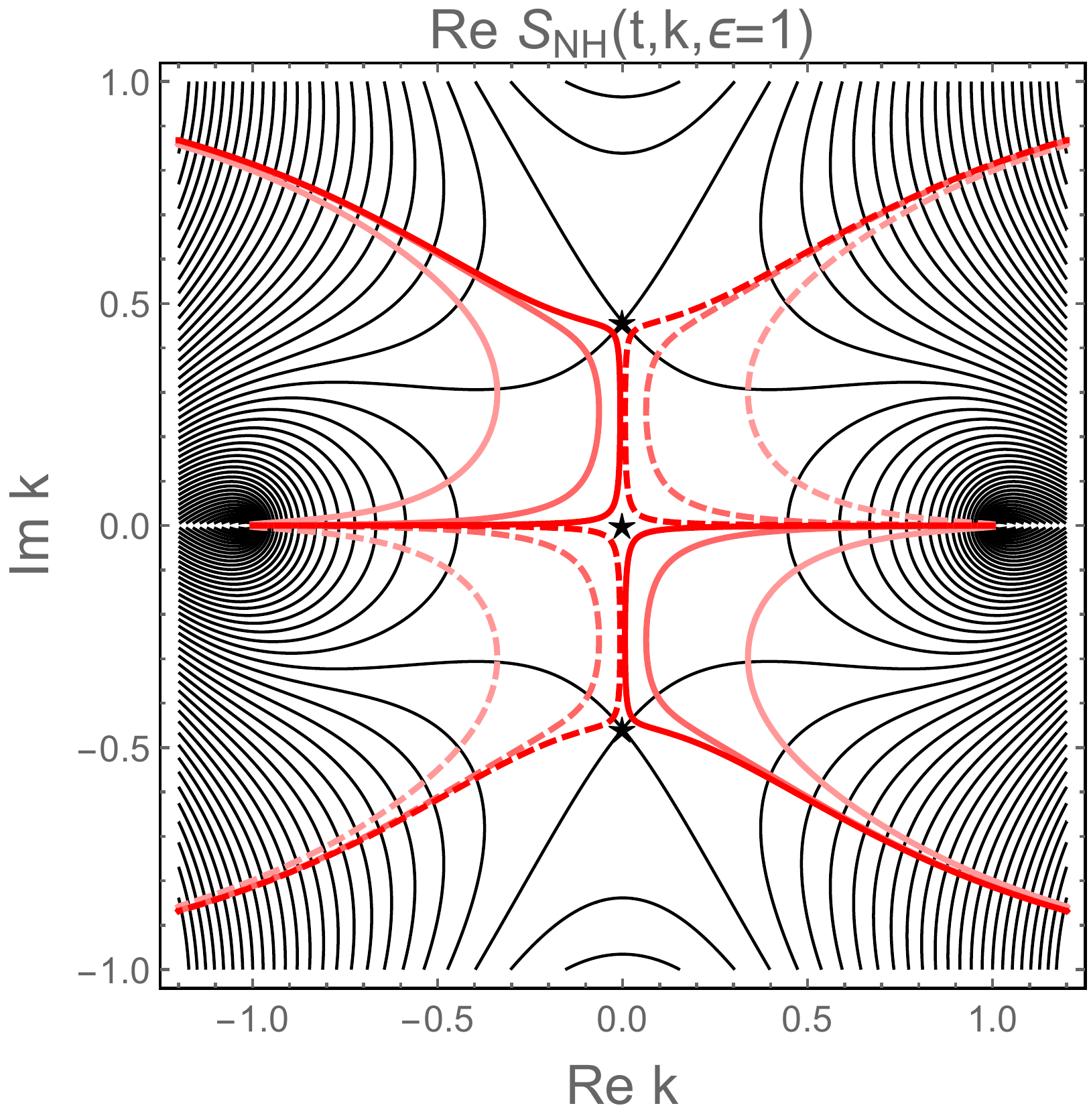}
\caption{\small For $s=1$, level plot of $\textrm{Re}\,S_{NH}(t,k)$ for $\epsilon=1$, superimposed with the steepest descent contours emanating from the $k = 0$ hydrodynamic saddle for different values of $\epsilon = e^{i \theta}.$ Solid red lines, from lighter to darker, correspond to $\theta = 2.5\times 10^{-3}, 2.5\times 10^{-4}, 2.5\times 10^{-5}$. Dashed lines represent the negative values of $\theta$, with the same color code. Black stars correspond to saddle point locations. Left: $t = 0.9$. Right: $t = 1.1$}
\label{fig:GID_sd}
\end{center}
\end{figure}

Figure \ref{fig:GID_sd} (left) further illustrates the behavior of the steepest descent path for the hydrodynamic saddle on the nonhydrodynamic sheet as $\theta \to 0$ for $0 < t \leq t_c$. Solid red lines, from lighter to darker, represent the values $\theta = 2.5\times 10^{-3}, 2.5\times 10^{-4}$ and $2.5\times 10^{-5}$, while dashed red lines represent their negatives. As $\theta$ approaches zero, the steepest descent path gets progressively closer to the $k_\pm$ saddles before veering off to infinity. Furthermore, $\theta = 0$ marks a discontinuous change in the path behavior, which is the reason why we considered $\theta$ as a parameter to vary; for instance, the quadrant in which the red curve extends to infinity undergoes a sudden change. These observations are compatible with the hypothesis that $k_\pm$ are the adjacent saddles controlling the divergence of the perturbative series expansion. On the other hand, as illustrated in figure \ref{fig:GID_sd} (right), the collision of nonperturbative saddle points at $t=t_c$ causes the nature of the adjacent saddles to change: past $t_c$, $k_0$ becomes a new adjacent saddle for the hydrodynamic steepest descent contour. 

We can now relate this saddle point analysis to the singularities we observed in the Borel plane in section \ref{convergence}. Since for $t > t_c$ we have that $|S_0| < |S_\pm|$, the above observations entail that the branch point of the Borel transform which is closest to the origin should go from being located at $z_c = - S_\pm$ for $t < t_c$ to being located at $z_c = -S_0$ for $t>t_c$. This prediction matches precisely the behavior of $r$ reported at the end of section \ref{convergence}. 

Let us emphasize that for $t > t_c$ the branch point associated to the $S_\pm$ adjacent saddles, $z_\pm = - S_\pm$, does not disappear. What actually happens is that, since both $z_0 = - S_0$ and $z_\pm$ are real positive quantities with $z_\pm > z_0$, the branch cut associated to $z_0$ superimposes with the branch cuts associated to $z_\pm$, causing the latter to be superficially hidden in the Borel plane (as we illustrated using a simple example in footnote~\ref{footnote.padecoincidentcuts}).

To expose the hidden $z_\pm$ branch points, we proceed along the lines of \cite{Costin:2019xql}. We take our original Padé approximant $\mathcal{P}(z)$ and introduce the variable change $z=z(w)$, with $z(w)$ analytic at $w=0$. Then, we series expand $\mathcal{P}(z = z(w))$ around $w = 0$. Finally, we compute the Padé approximant of the resulting series, $\mathcal{P}(\omega)$. By a suitable choice of $z(w)$, the images of our original branch points lead to non-superimposing branch cuts in the $w$ Borel plane.  
A convenient choice of variable change is as follows. Define 
\beq
z(w) = z_c \frac{2w}{1+w^2} \label{z_w}
\eeq
with $z_c$ being, as before, the branch point closest to the origin in the $z$ Borel plane. The variable change \eqref{z_w} maps a point $z \in (z_c, \infty)$ to two complex conjugated images on the right half of the $|w|=1$ unit circle, with $z_c$ being mapped to $w=1$. 
\begin{figure}[h!]
\begin{center}
\includegraphics[width=7.5cm]{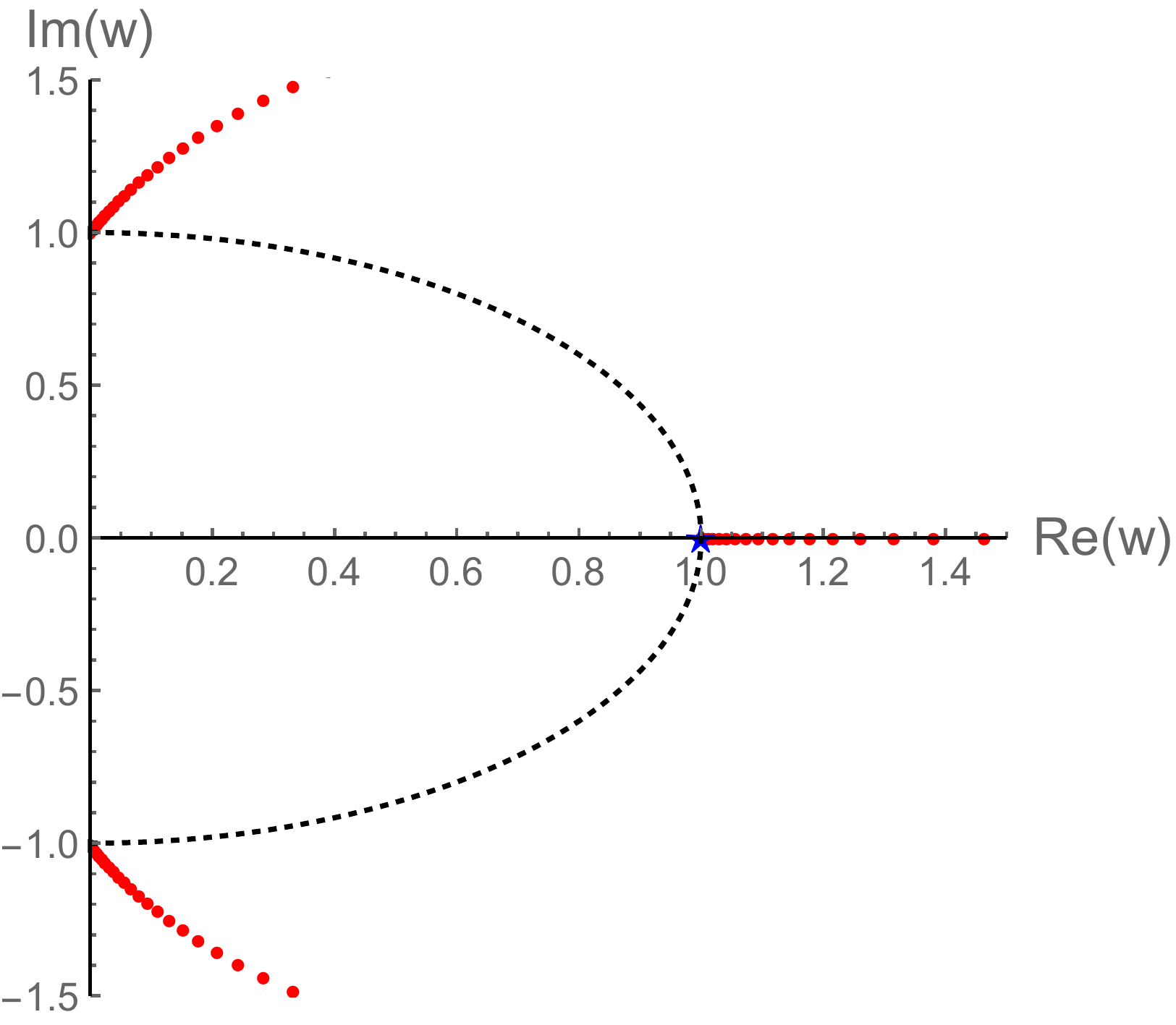}\includegraphics[width=7.5cm]{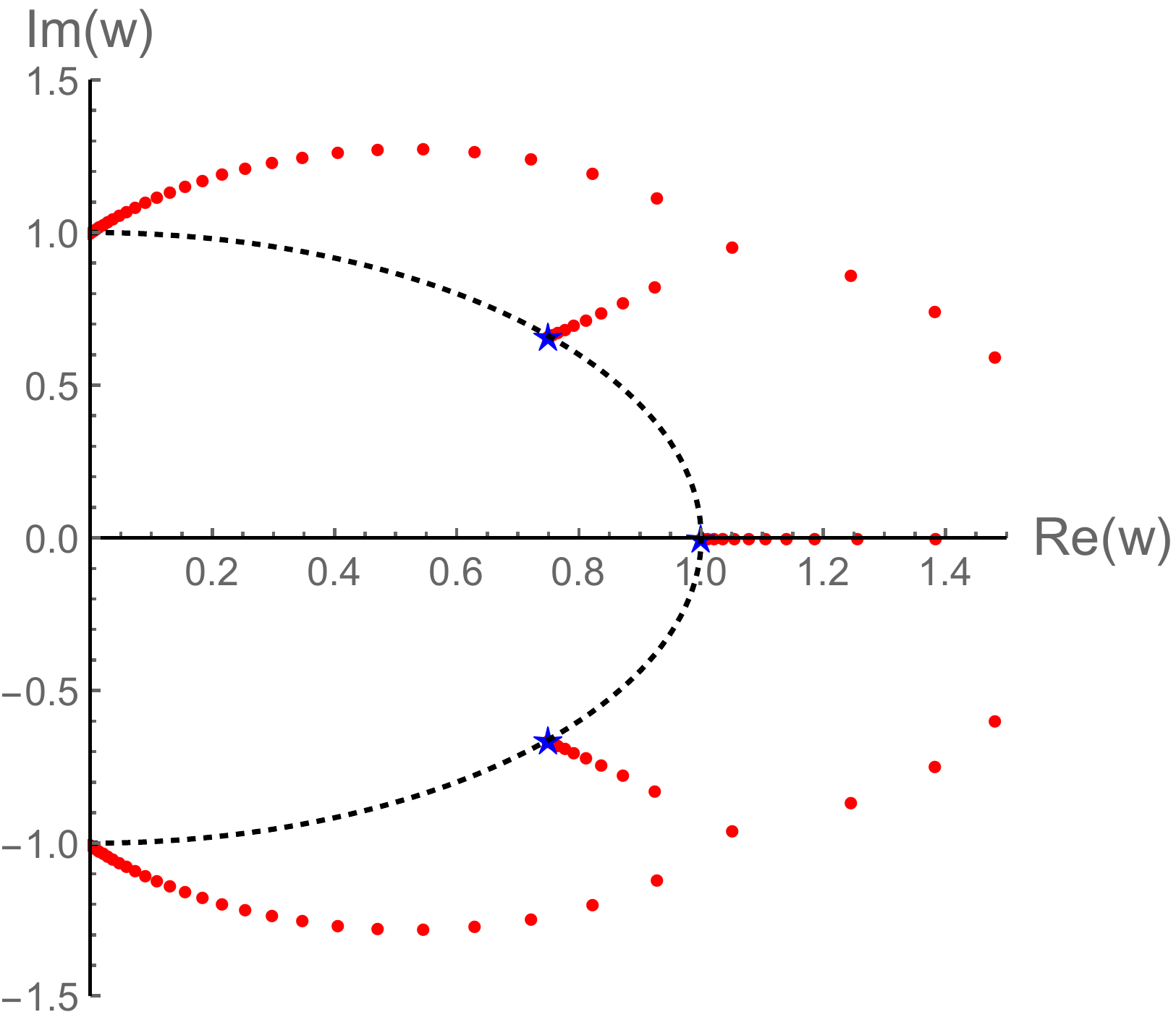}
\caption{\small Poles of the Padé approximant $\mathcal{P}(w)$ for $t = 1/2$ (left) and $t = 3$ (right) for Gaussian initial data with $s=1$. The blue stars signal the images of $z_0$ and $z_\pm$ under the map \eqref{z_w}, while the dashed black line corresponds to the $|w|=1$ circle.} 
\label{fig:poles_pade_w}
\end{center}
\end{figure}

For $t < t_c$, the pole structure of $\mathcal{P}(w)$ is as shown in figure \ref{fig:poles_pade_w} (left). We see three lines of pole condensation: two of them are complex conjugated and emanate from the singular points of the map \eqref{z_w}, located at $w= \pm i$, and are therefore unphysical; the third one starts at $w=1$ and runs along the positive real axis. This latter line corresponds to the original branch cut starting at $z_c=z_\pm$. For $t<t_c$ we see no trace of additional branch points associated to $z_0$ in the $w$ plane, confirming that for $t < t_c$ $k_0$ is not an adjacent saddle. 

On the other hand, for $t > t_c$, this state of affairs changes. As figure \ref{fig:poles_pade_w} (right) shows, besides the branch point at $w=1$ corresponding now to $z_c = z_0$, we also find two additional, complex conjugated lines of pole condensation emanating from the unit circle. It is immediate to check that these points are nothing but the images of $z_\pm$ under the map \eqref{z_w}. These images are represented by the upper and lower blue stars in the figure.

\section{Superexponential decay of the nonhydrodynamic saddle points}

From a physical standpoint, the existence of the $k_+$, $k_-$ saddles bears crucial consequences for the late-time behavior of the nonhydrodynamic mode contribution $\rho^{(NH)}$ for the Gaussian initial data, \eqref{GID}. To expose these consequences, for the rest of this section we focus on the original $\epsilon = 1$ telegrapher's equation. We take the integration path $\gamma$, used to compute $\rho^{(NH)}$, to lie strictly below the real axis in the complex $k$-plane. With this choice of path, $\rho^{(NH)}$ is real. Linearity of the telegrapher's equation means that $\rho^{(NH)}$ is now a fully-fledged solution to the telegrapher's equation on its own, associated to the initial data,
\beq
\hat{u}(k) = -\frac{\tau e^{-\frac{1}{2}s^2 k^2}}{2\pi \Delta(k)}, \quad \hat{v}(k) = \frac{e^{-\frac{1}{2}s^2 k^2} (1+\Delta(k))}{4\pi \Delta(k)}, \label{id_new} 
\eeq
in such a way that $f_{H}(k) = 0$.\footnote{Since we integrate along a path $\gamma$ that goes strictly below the real axis, the singularities of $\Delta$ do not require attention.} In the remaining part of this section, we will denote this solution simply as $\rho$. 

\begin{figure}[h!]
\begin{center}
\includegraphics[width=7.5cm]{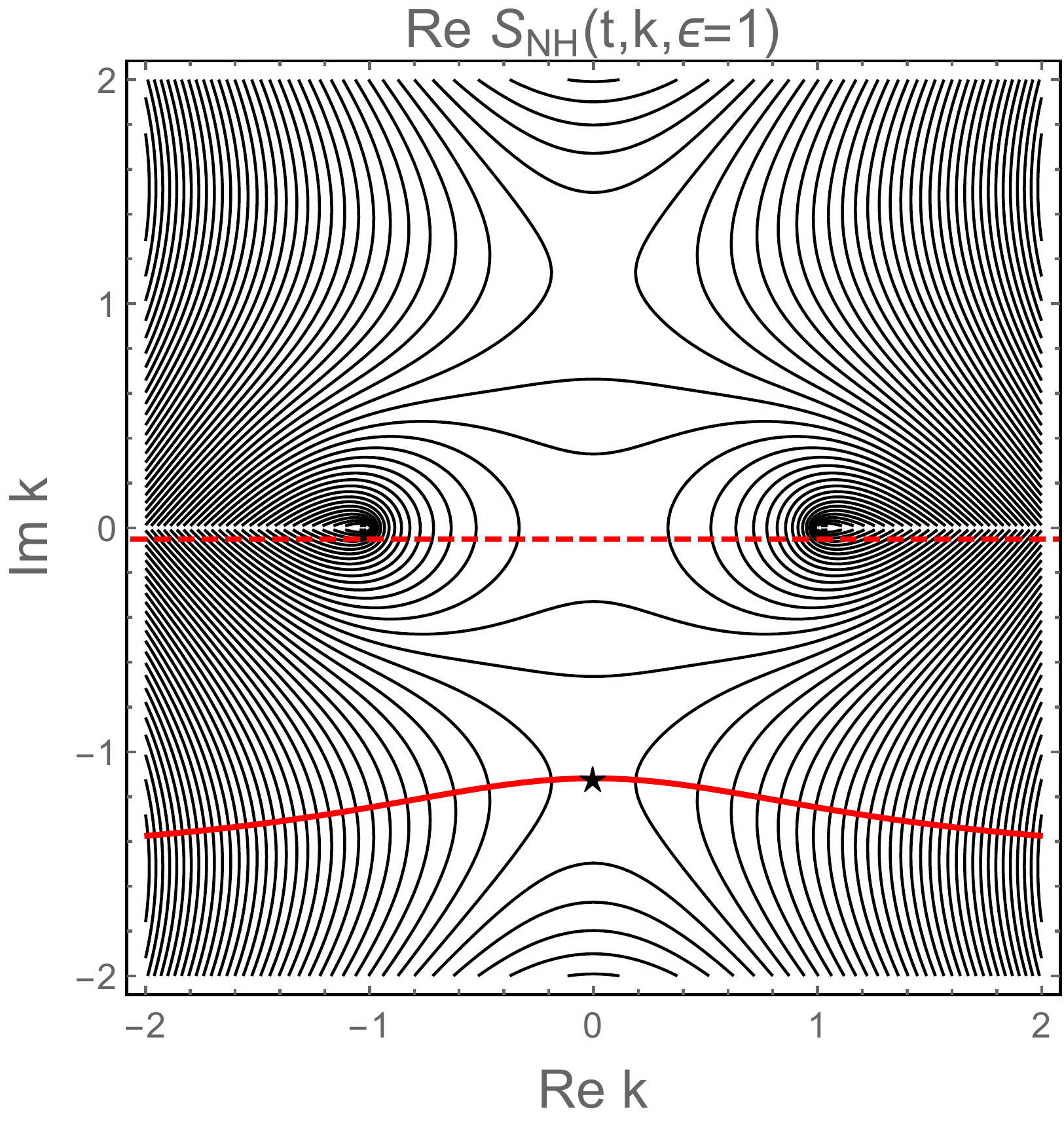}
\caption{\small Level plot of $\textrm{Re}\,S_{NH}(t,k)$ for Gaussian initial data with $s=1$ at $t = 3/2$. The dashed red line shows the $\gamma$ integration contour; the solid red line is the steepest descent path emanating from the $k_-$ saddle.}
\label{fig:NH_GID_sd}
\end{center}
\end{figure}

The chosen integration path $\gamma$ can be deformed into the steepest descent contour associated with the $k_-$ saddle-point (see figure \ref{fig:NH_GID_sd}). Therefore, the natural expectation is that the late-time behavior of the initial data \eqref{id_new} at $x=0$ is controlled by the $S_-$ action, which predicts a late-time decay \textit{faster than $e^{-\frac{t}{\tau}}$}.

Two questions arise naturally at this point. The first one is whether the faster-than-exponential decay we have uncovered extends to finite $x$. The second one is whether the existence of these additional saddle points extends to generic initial data with Gaussian asymptotic behavior as $k \to \infty$.  

Both questions can be answered affirmatively by the following argument. For initial data with Gaussian asymptotic behavior, the relevant action to consider in the nonhydrodynamic sheet at finite $x$ is (we take $\epsilon = 1$ from now on) 
\beq
S_{NH}(t,k) = - \frac{1}{2}s^2 k^2 - \frac{1}{2\tau} \left(1+\sqrt{1- 4 D \tau k^2} \right) + i k x. \label{NH_GID_action}
\eeq
By solving the saddle point equation in a late-time expansion, we find a saddle $k_*$ located in the lower half complex $k$-plane given by\footnote{We assume that $x > 0$ in what follows.}
\beq
k_* = - i \frac{\sqrt{D}}{s^2\sqrt{\tau}}t + i \frac{x}{s^2} + i\frac{s^2}{8 D^\frac{3}{2}\sqrt{\tau}}\frac{1}{t} + \ldots,  
\eeq
with action 
\beq
S_* = S_{NH}(t,k_*) = - \frac{D}{2 \tau s^2} t^2 - \frac{1}{\tau}\left(\frac{1}{2} - \frac{\sqrt{D \tau}}{s^2} x \right) t - \frac{s^4 + 4 D \tau x^2}{8 D \tau s^2} + \ldots. \label{GID_late_time_NH_action} 
\eeq
As the expression above shows, the finiteness of $x$ does not change the leading order late-time behavior of $S_*$ we encountered before for $x = 0$. Moreover, only the asymptotic behavior of the initial data at large $k$ matters in reaching this conclusion. 

\begin{figure}[h!]
\begin{center}
\includegraphics[width=7.5cm]{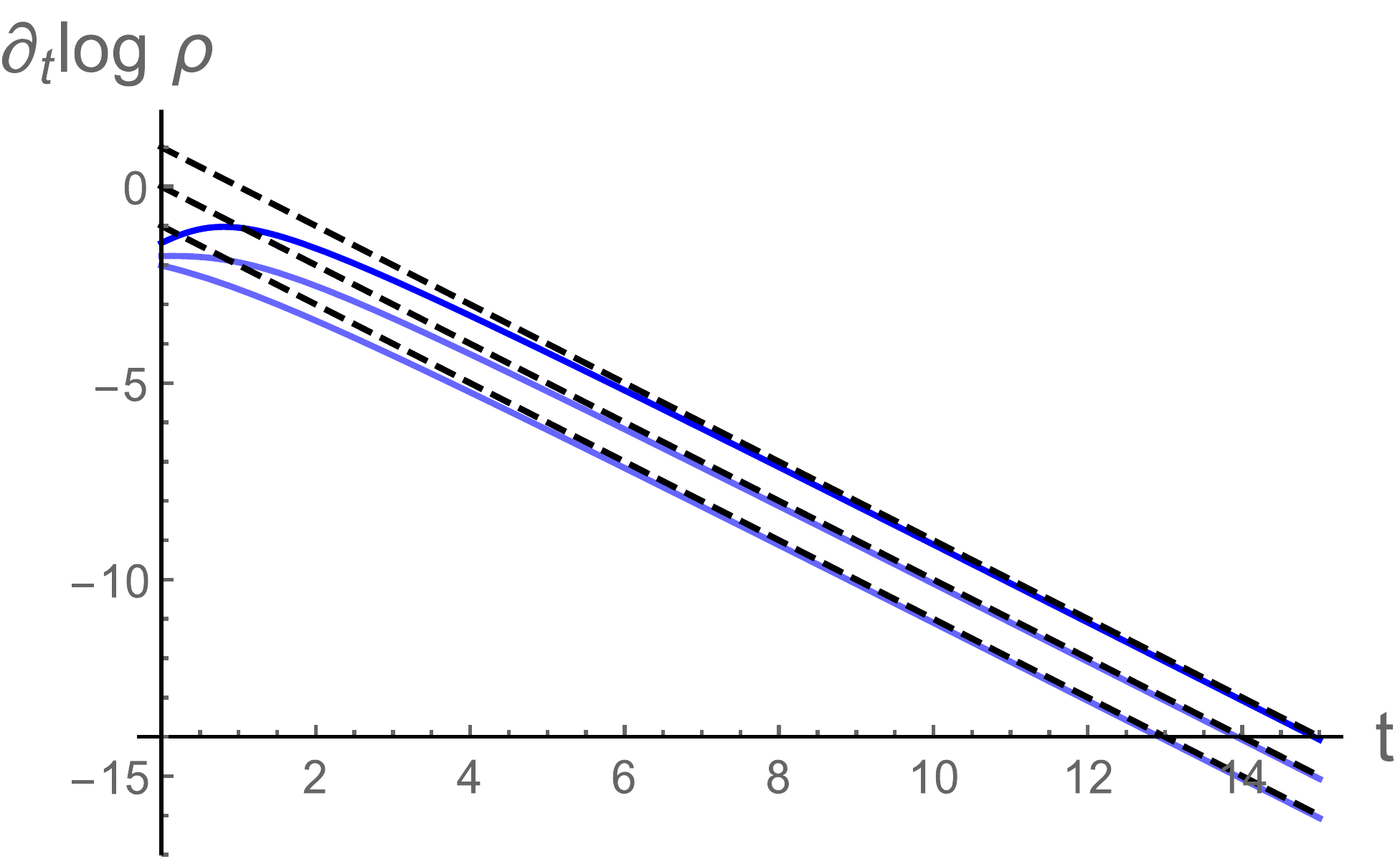}\includegraphics[width=7.5cm]{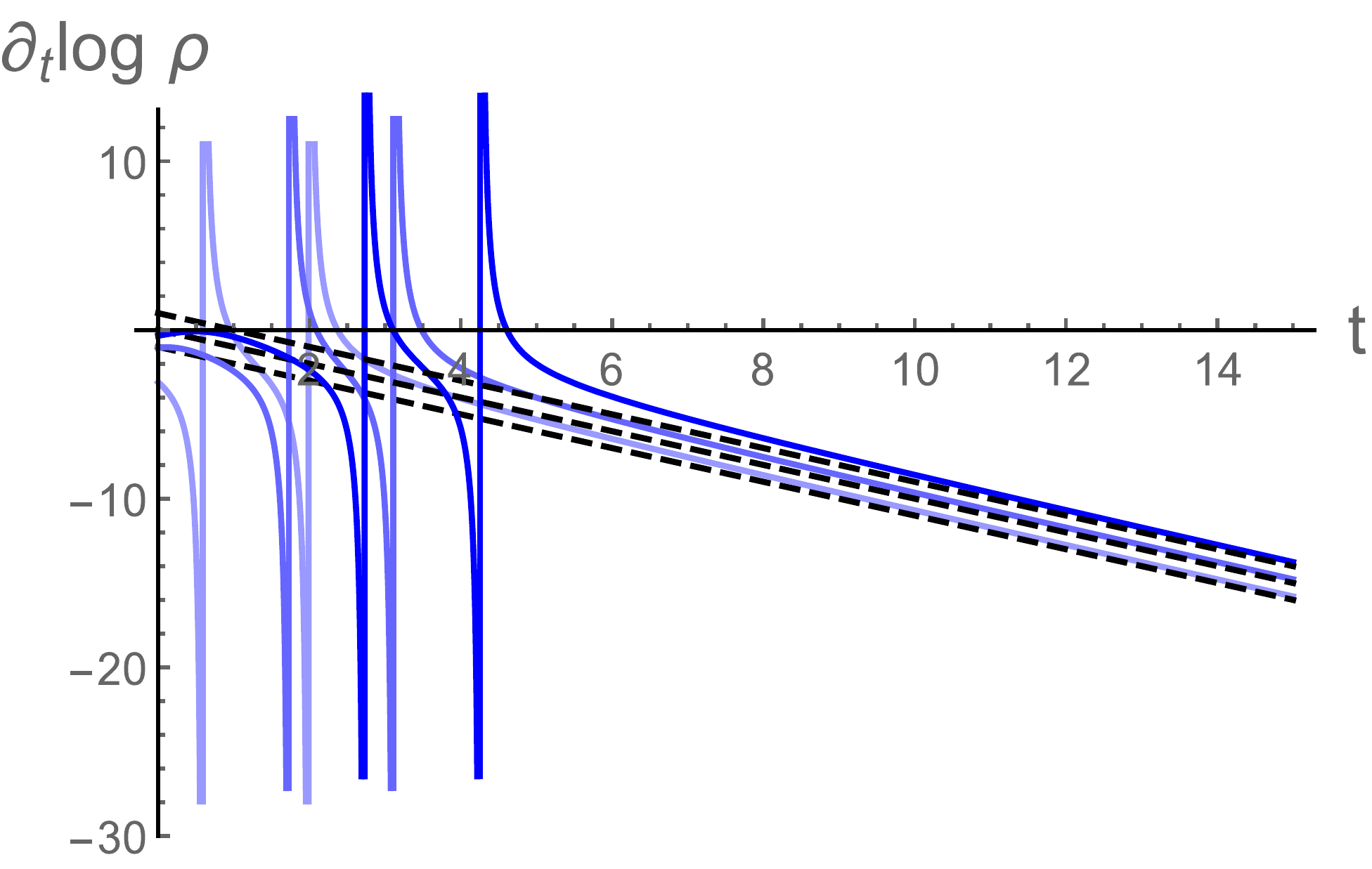}
\caption{\small Left: $\partial_t \log \rho$ for $s = 1$ and $P(k) = 1$. Solid blue lines correspond to the numerical integration of the exact expression, dashed black lines to the late-time behavior prediction \eqref{GID_log_derivative_prediction}. From bottom to top, $x = 0$, $1$ and $2$. Right: same as left, now with $P(k)$ given by a even, fourth-order polynomial in $k$ with generated once and for all random coefficients.}
\label{fig:GID_log_derivative_check}
\end{center}
\end{figure}

To test whether the late-time behavior predicted by \eqref{GID_late_time_NH_action} is actually realized, we select the logarithmic derivative $\partial_t \log \rho(t,x)$ as our probe. According to \eqref{GID_late_time_NH_action}, we must have that 
\beq
\partial_t \log \rho(t,x) = - \frac{D}{\tau s^2} t - \left(\frac{1}{2\tau} - \frac{\sqrt{D}}{\sqrt{\tau}s^2} x \right) + ...
\label{GID_log_derivative_prediction}
\eeq
In order to check whether equation \eqref{GID_log_derivative_prediction} holds, we restrict ourselves to initial data 
in which we multiply the Gaussian appearing in \eqref{id_new} by a polynomial $P(k)$ (taken to be a function of $k^2$ with real coefficients), fix $s$, and compute numerically $\partial_t \log \rho(t,x)$ for a range of spatial positions. In figure~\ref{fig:GID_log_derivative_check}, we show results for $P(k) = 1$ (left) and a fourth-order $P(k)$ with random coefficients drawn from the interval $[-1,1]$ (right), for $s=1$ and at $x=0,1$ and $2$. These results are in agreement with the hypothesis that the nonhydrodynamic contribution displays a faster-than-exponential decay. 

\begin{figure}[h!]
\begin{center}
\includegraphics[width=10cm]{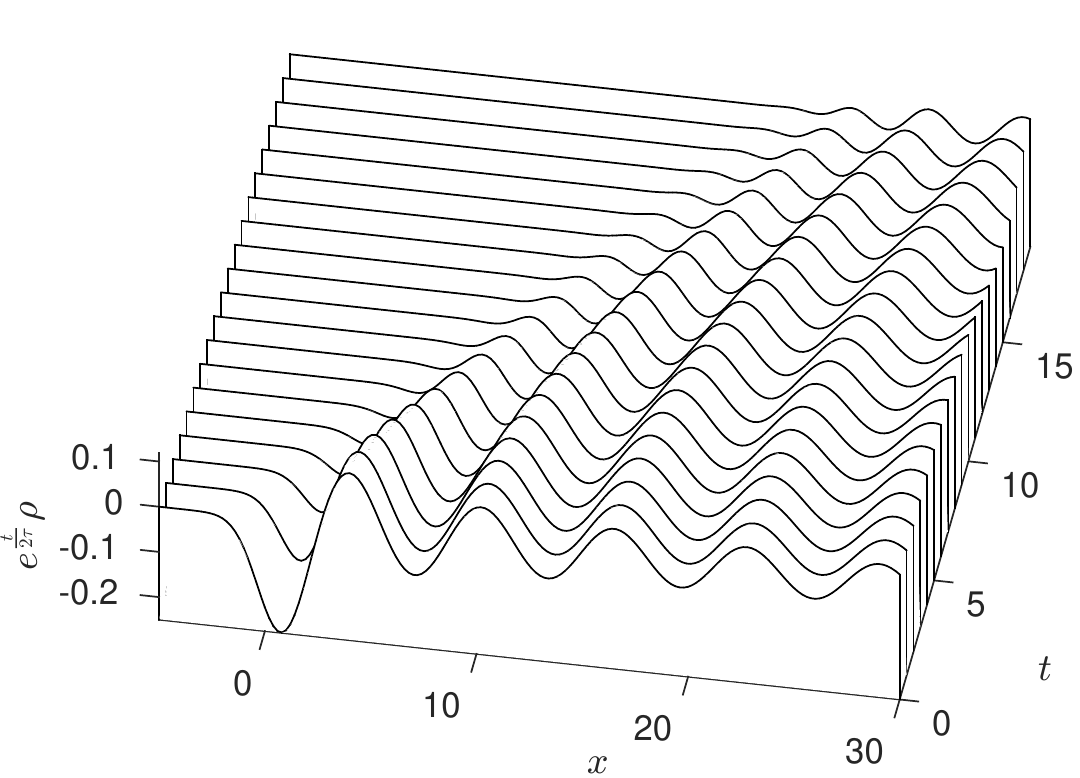}
\caption{\small Spacetime evolution of $e^{\frac{t}{2\tau}} \rho(t,x)$ for the initial data \eqref{id_new} with $s=1$. We have multiplied by $e^{\frac{t}{2\tau}}$ to factor out the leading late time decay of the wave (recall that $D=\tau=1/2$ in our numerical computations).}
\label{fig:trail}
\end{center}
\end{figure}

The physical origin of the faster-than-exponential decay is propagation of the data with a Gaussian tail rather than an effect of dissipation governed by the imaginary part of the mode frequency. To see this, note that  \eqref{GID_late_time_NH_action} can be suggestively rewritten as 
\begin{equation}
S_* = -\frac{s^2}{8 D \tau }-\frac{t}{2 \tau } -\frac{\left(x-\sqrt{\frac{D}{\tau }} t\right)^2}{2 s^2} + \dots.
\end{equation}
It is (the tail of) a Gaussian peak propagating in the direction of the positive $x$ with velocity $\sqrt{\frac{D}{\tau }}$ and decay rate $\frac{1}{2 \tau }$. These values match the propagating modes at large $k$ seen in figure \ref{fig:modes}. Figure \ref{fig:trail} shows the utility of this line of reasoning. In this figure we plot a dense set of snapshots of $\rho$ -- determined by direct numerical integration -- as a function of $x$. We observe that the initial data \eqref{id_new} indeed evolve as a propagating wave that recedes from $x=0$ as time grows. Furthermore, as the argument above suggested this wave gets progressively damped, but just exponentially so. 

\section{Stokes phenomena in spacetime}

For Gaussian initial data, figures \ref{fig:GID_sd} and \ref{fig:poles_pade_w} show that both the saddle point structure and the Borel plane structure take a different character depending on whether $t$ is larger or smaller than the critical time $t_c = s^2/2D$. The transseries also undergoes an abrupt change, known as a Stokes phenomenon \cite{Aniceto:2018bis}. It occurs when crossing a point where the imaginary part of the action is the same, i.e.
\begin{equation}
	\textrm{Im}\,(S_0 - S_\pm) = 0.
\end{equation}
If we consider the action as a function of complex $t$, with all other parameters real\footnote{There is no obstruction against considering complex values of $s$, which give rise to oscillating solutions.} and $x=0$, this condition is realized when
\begin{subequations}
\begin{align}
	\textrm{Im}\,t &= 0 \quad \text{or} \\
	\textrm{Re}\,t &= \frac{s^2}{2D}.
\end{align}
\end{subequations}
As the Stokes line at $\frac{s^2}{2D}$ is crossed along the real axis, the contribution of the $S_0$ saddle is turned off and the contributions of $S_-$ and $S_+$ are turned on. 

In terms of the coefficients $b_n(t,x)$, this is seen as follows. Recall that for Gaussian initial data of the form \eqref{GID}, $\hat{b}_n(t,k) = -\hat{a}_n(-t,k)$, where $\hat{a}_n(t,k)$ is given by~\eqref{a_n}. For $k \in \mathbb R$ and as $|k| \to \infty$, we have that $\hat{b}_{n+1}(t,k)$ behaves as 
\begin{equation}
\hat{b}_{n+1} = d_n t^n k^{4n} e^{-D (t_c - t) k^2} + \ldots, \quad d_n \in {\mathbb R}
\end{equation}
Therefore, the Fourier integral that computes $b_n(t,x)$ exists for positive real $t$ as long as $t < t_c$. The result is $b_n(t,0) = -a_n(-t,0)$ with the latter given by \eqref{an_GID}. This changes dramatically for $t > t_c$, where the Fourier integral diverges along the standard Fourier contour. The contour must be deformed to the $k_-$ saddle point. 

\begin{figure}[h!]
\begin{center}
\includegraphics[width=7.5cm]{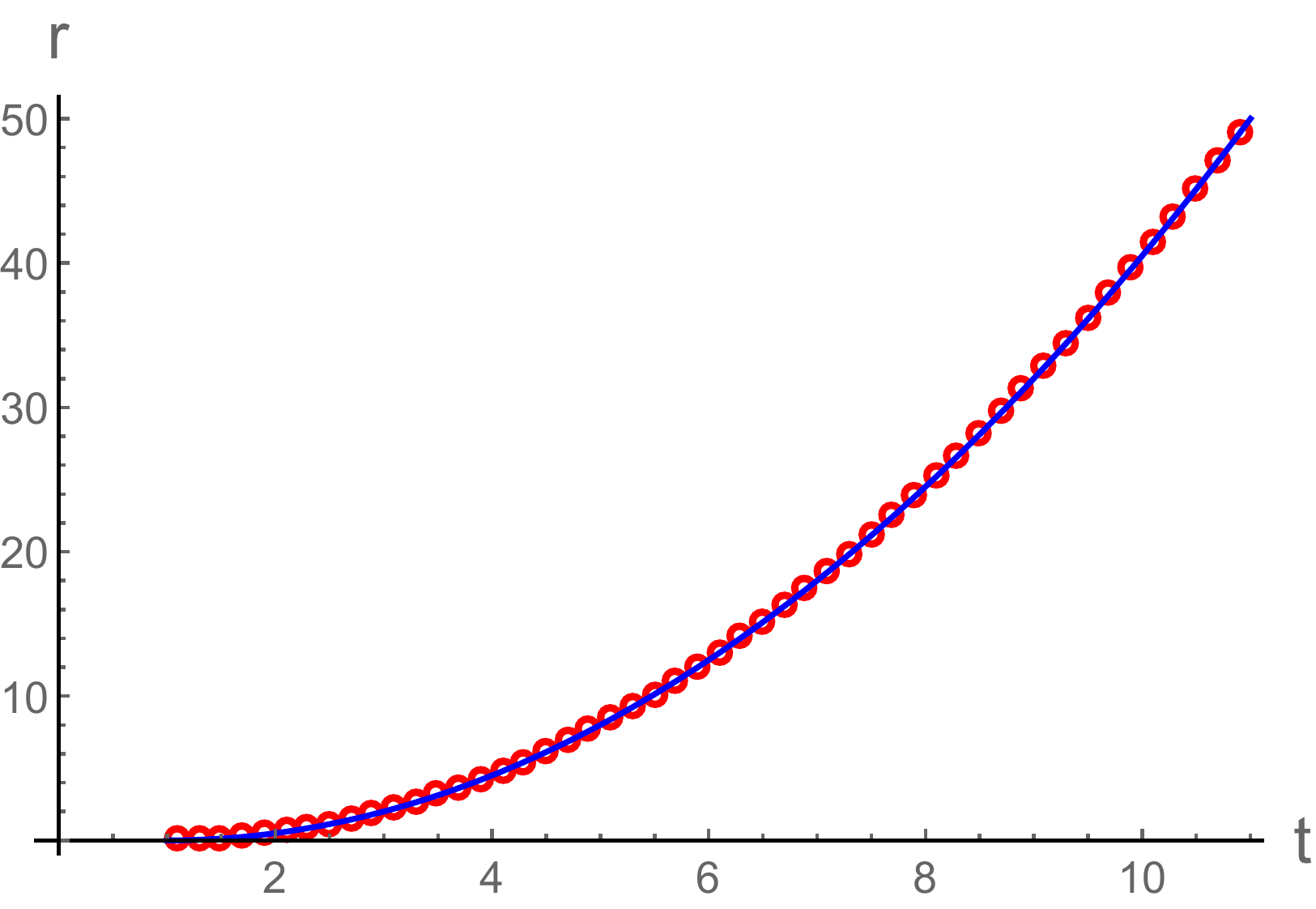}\includegraphics[width=7.5cm]{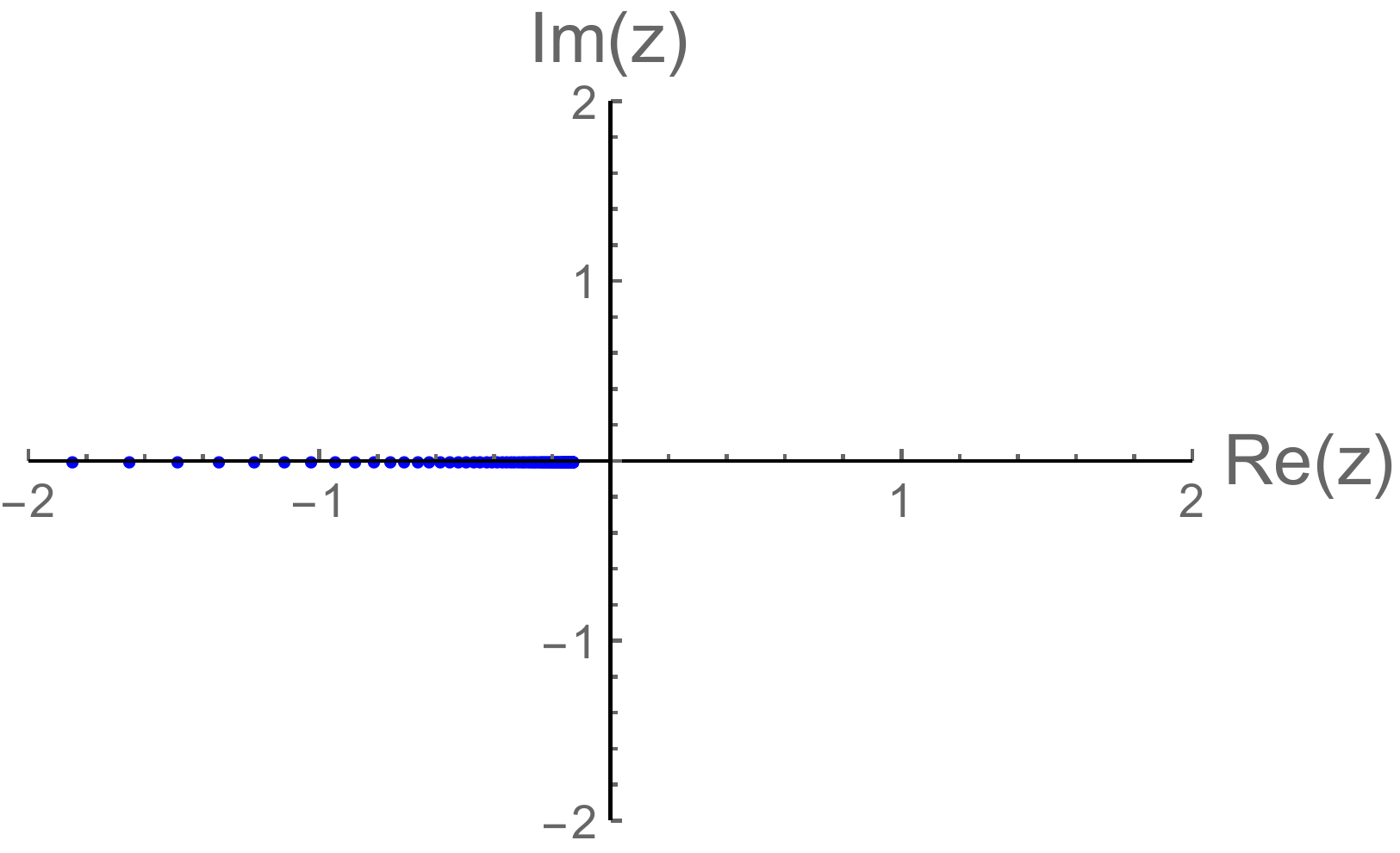}
\caption{\small Left: for $s = 1$, $r$ as a function of $t$ as determined directly from the closed-form expression for $c_n(t,0)$ (open red circles) vs. the function $S_0- S_-$ (solid blue line). Right: poles of the Padé approximant of $\rho^{(NH)}_\epsilon(t,0)$ at $t = 3/2$ for $s = 1$.}
\label{fig:GID_asymptotic_expansion_past_r}
\end{center}
\end{figure}

For $x=0$ this series expansion can be computed in closed form, with the end result that
\begin{subequations}
\begin{align}
&\rho^{(NH)}_\epsilon(t,0) = e^{\frac{S_-}{\epsilon^2}} \sum_{n=0}^\infty c_{n+1}(t,0)\epsilon^{2n+2}, \label{GID_asymptotic_expansion_past_full}\\
&c_{n+1}(t,0) = \frac{(-1)^{n+1} {}_2 F_1\left(\frac{1}{2}, -2 n, \frac{1}{2} - 2n, \frac{\upeta+1}{\upeta-1} \right)}{2^{n+\frac{3}{2}} t_c^{n+\frac{1}{2}} (\upeta+1)^{2n} \sqrt{\upeta^2-1} \Gamma\left(n+1\right) \Gamma\left(\frac{1}{2}-2n\right)}, \label{GID_asymptotic_expansion_past_coeffs}
\end{align}
\end{subequations}
where $\upeta = t/t_c > 1$. This series is factorially divergent. We find that the ratio test behaves as 
\begin{equation}
\Bigl|\frac{c_{n+1}(t,0)}{c_n(t,0)}\Bigr| \sim \frac{n}{r}, \quad n \to \infty
\end{equation}
where $r$ is a function given by the difference between the $k_0$ and $k_-$ saddle point actions, 
\begin{equation}
r(t,s) = S_0 - S_{-} = - \frac{t}{\tau} + \frac{(s^2 + 2 D t)^2}{8 D \tau s^2}. 
\end{equation}
We provide an example of this behavior in figure \ref{fig:GID_asymptotic_expansion_past_r} (left). In line with this result, the Padé approximant to the Borel transform of the asymptotic expansion \eqref{GID_asymptotic_expansion_past_full} displays a line of pole condensation along the negative real axis, starting at $z_c = S_- - S_0$ (see the right plot in figure \ref{fig:GID_asymptotic_expansion_past_r}). This location of the branch cut follows from the fact that $k = 0$ is an adjacent saddle for the steepest descent contour emanating from $k_-$ when $\arg \epsilon^2 = \pi$. 

 To summarize, for these initial data, we take our nonperturbative transseries sector to be defined by 
\begin{equation}
\rho^{(NH)}_\epsilon(t,0) = \left\{
\begin{array}{ll}
e^{\frac{S_0}{\epsilon^2}} \sum_{n=0}^\infty b_{n+1}(t,0)\epsilon^{2n+2}, & t < t_c \\
e^{\frac{S_-}{\epsilon^2}} \sum_{n=0}^\infty c_{n+1}(t,0)\epsilon^{2n+2}, & t > t_c.
\end{array} 
\right. 
\end{equation}
We see that the form of the nonperturbative transseries sector depends on the spacetime location. This is the reason why we started with the transseries in momentum space, see~\eqref{rho_transseries}, which is uniquely defined in terms of the modes in the system.

This kind of Stokes phenomena in spacetime is not unique to this model. It has been investigated previously in, for example, references~\cite{chapman2005exponential, chapman2007shock, Howls2004Aug}. Those studies show that nonlinearity can also be handled and that a more intricate higher-order Stokes phenomenon can occur. 

To close this section, let us emphasize the role that the initial data plays in the spacetime picture. Rather than the Stokes phenomenon occurring as $t$ is varied, one can equivalently regard it as occurring as the initial data is varied. It arises here because the transseries coefficients depend on the initial data, which we expect is generic and holds also in nonlinear models. If a richer set of initial conditions were considered, a richer set of transseries sectors could be the result. The takeaway lesson here is that both the spacetime location as well as the initial data must be taken into account when formulating the transseries.

Finally, we refer the reader to Appendix \ref{resummation}, where we employ Borel resummation to illustrate how the exact $\rho^{(H)}(t,0)$, $\rho^{(NH)}(t,0)$ -- as determined by a numerical integration -- can be recovered from the asymptotic expansions $\rho^{(H)}_\epsilon(t,0)$, $\rho^{(NH)}_\epsilon(t,0)$. 

\section{Discussion}

The main motivation for our work is understanding nonequilibrium phenomena with a hydrodynamic tail by expressing them as transseries with resurgent relations connecting their various sectors. This point of view originates from the studies of expanding matter in ultrarelativistic heavy-ion collisions described by the paradigmatic example of Bjorken flow. Our work proposes and explores an approach that allows one to frame more general examples of dynamics in the same kind of language. We focus on the linearized regime but, reminiscent of~\cite{Bhattacharyya:2008jc}, transseries techniques can in principle be applied when deviations from equilibrium are large. It would very interesting to study this question in detail, and to make contact with far from equilibrium attractors, which we leave for future research. 

To describe a nonequilibrium process with a transseries one needs to define a small parameter. Guided by results from Bjorken flow, we introduce a formal parameter~$\epsilon$ based on rescalings of spacetime coordinates~\eqref{rescaling}, that organizes the hydrodynamic and nonhydrodynamic  contributions into different transseries sectors. While the momentum space picture is straightforward, when passing to coordinate space we see new features which are as yet \emph{unseen} in Bjorken flow and other expanding plasma systems.

In particular, we find that the initial conditions affect the form of nonperturbative contributions in $\epsilon$ in the spacetime picture. In our work we focused on a particular simple yet rich and widely encountered equation of motion -- the telegrapher's equation~\eqref{eq.telegraphers} -- and a few classes of initial conditions. The nonhydrodynamic sector of the transseries took on two different forms. One is as a nonpropagating transient mode evaluated at zero momentum, similar to what was found in the Bjorken flow. However, when the initial conditions produce propagating wave packets, the receding tails of these wave packets gave rise to new transseries sectors. We have seen then that the decay of the nonhydrodynamic data is not only governed by the transient mode at zero momentum, but also by the form of the initial data and the dispersion relations at finite momentum. 

While we observed this phenomenon for the telegrapher's equation, its ingredients seem to originate from the underlying causality of the system. This is necessarily shared by all the models of relativistic matter, in particular holography, which suggests that it is ubiquitous.\footnote{Perhaps similar phenomena can even occur in the holographic Bjorken flow. One can view the gravity dual to the Bjorken flow as a set of nonlinear wave equations with constraints in two variables. This is not too dissimilar from what we considered here if the transseries analysis is extended into the bulk.} In studies of the transition to hydrodynamics in relativistic heavy-ion collisions using holography, the dominant theme has been the decay of transient modes as the mechanism governing it. Here we see that contributions from the nonhydrodynamic sectors can propagate away from a given spatial location, which effectively may render them zero. It would be very interesting to understand possible phenomenological implications of this observation.

Finally, let us comment on the utility of transseries solutions. The transseries allows one to organize hydrodynamic and nonhydrodynamic phenomena using a unified mathematical language. This is crucial when the hydrodynamic gradient expansion diverges and requires resummation. The transseries provides a framework to resum it yielding a unique answer for a nonequilibrium solution. Furthermore, transseries provides a way to encode different asymptotic behavior in different spacetime regions, or for different initial data. Transitions between these behaviors are described by the Stokes phenomena. In our case, these considerations allowed us to uncover a new interesting physical effect in the context of hydrodynamization.

\acknowledgments

We would like to acknowledge useful exchanges with In\^es Aniceto, Matteo Baggioli, Gerald Dunne, Blaise Gout\'eraux and Daniel Hasenbichler. The Gravity, Quantum Fields and Information group at the Max Planck Institute for Gravitational Physics (Albert Einstein Institute) is supported by the Alexander von Humboldt Foundation and the Federal Ministry for Education and Research through the Sofja Kovalevskaja Award.
AS and MS are supported by the Polish National Science Centre grant 2018/29/B/ST2/02457. BW is supported by a Royal Society University Research Fellowship.

\appendix 

\section{M\"uller-Israel-Stewart in the shear channel} 
\label{MIS}

MIS theory is the simplest phenomenological model of stress-energy tensor equilibration that agrees with relativistic Navier-Stokes hydrodynamics at leading order in the gradient expansion and, at the same time, respects causality. 

For a conformal fluid, the construction of MIS theory proceeds as follows. We start from the Landau-frame constitutive relations of first-order viscous relativistic hydrodynamics in $d$-dimensional Minkowski space~\eqref{Pi_constitutive} with the equations of motion of the theory being nothing but the conservation of the full energy-momentum tensor, $\partial_a T^{a b} = 0$. The algebraic relation between $\Pi^{ab}$ and $\sigma^{ab}$ implied by \eqref{Pi_constitutive} entails that first-order relativistic viscous hydrodynamics violates causality. In MIS theory, this problem is overcome by promoting $\Pi^{ab}$ to a set of new independent degrees of freedom that obey a relaxation equation, in such a way that the original constitutive relation \eqref{Pi_constitutive} is recovered at times sufficiently larger than a new time-scale set by a relaxation time $\tau$, 
\beq
(\tau U^c \mathcal{D}_c + 1) \Pi^{ab} = - \eta \sigma^{ab}. \label{dynamical_constitutive_relation}
\eeq
The operator $\mathcal D_a$ is a Weyl-covariant derivative originally introduced in \cite{Loganayagam:2008is}. 

In this work, we consider infinitesimal fluctuations of this theory away from thermal equilibrium. We thus write 
\beq
\mathcal{E}(t,\textbf{x}) = \mathcal{E}_0 + \delta\mathcal{E}(t,\textbf{x}) \quad U^a = (1, \textbf{u}(t,\textbf{x})), 
\eeq
where $\mathcal{E}_0$ is the equilibrium energy density, and treat $\delta\mathcal{E} / \mathcal E_0$ and $\textbf{u}^2$ as infinitesimally small.  Moreover, we also make the symmetry assumption that the hydrodynamic variables are independent of $x^1,...,x^{d-2}$. Defining $x^{d-1} \equiv x$, this corresponds to $\delta\mathcal{E} = \delta\mathcal{E}(t,x), u_i = u_i(t,x)$. This ansatz can be viewed as the minimal generalization of a boost-invariant flow, for which the hydrodynamic variables would also be functions of $t$ and $x$, but only through the combination $\sqrt{t^2-x^2}$. 

As mentioned in the main text, the telegrapher's equation emerges when considering a shear channel fluctuation, for which $\delta\mathcal{E}(t,x) = 0$ and $u_i(t,x) = u_1(t,x) \delta_{i,1}$ with no loss of generality due to rotational invariance. Defining 
\begin{equation}
\rho(t,x) \equiv (\mathcal{E}_0 + P(\mathcal{E}_0)) u_1(t,x),\,\,\,J(t,x) \equiv \Pi_{1,d-1}(t,x), 
\end{equation}
and linearizing in the velocity fluctuation amplitude, the equation for energy-momentum conservation and the dynamical constitutive relation \eqref{dynamical_constitutive_relation} can be expressed as 
\begin{align}
&\partial_t \rho(t,x) + \partial_x J(t,x) = 0, \label{eq_conservation} \\
&\partial_t J(t,x) + \frac{D}{\tau} \partial_x \rho(t,x) = -\frac{1}{\tau}J(t,x). \label{eq_constitutive}
\end{align}
where the diffusion constant is $D = \eta/(sT)$. As mentioned in the Introduction, the linear PDE system \eqref{eq_conservation}-\eqref{eq_constitutive}  is well-known in the literature \cite{Romatschke:2009im, Grozdanov:2018fic}. Combining both equations, we recover \eqref{eq.telegraphers}. 

\section{The causality of the telegrapher's equation}
\label{causal}

In this appendix we show that the telegrapher's equation respects causality (see also \cite{Romatschke:2009im}). The basis of our proof is the following theorem \cite{strichartz2003guide}. 
\\\\
\noindent \textbf{Theorem} (Paley-Wiener). Let $f(x) \in L^2(\mathbb R)$ be supported in $x \in [-A,A]$. Then its Fourier transform $\hat f(k)$ belongs to $L^2(\mathbb R)$ and is an entire function of exponential type $A$.  
\\\\
\noindent We remind the reader than an entire function is a function which is analytic everywhere in the finite complex plane; an entire function of exponential type $A$ is an entire function that obeys the bound 
\beq
|f(z)| \leq C e^{A |z|},\,\,\, \forall z \in \mathbb C. 
\eeq
with $C \in \mathbb R^+$. 

Consider the most general solution to the telegrapher's equation, and suppose that the initial data are supported only between $-R$ and $R$. 
Causality demands that, at time $t$, $\rho(t,x)$ is supported at most in the interval $|x| \leq  R +  t$. Therefore, we have to show that $\hat\rho(t,k)$ is an entire function of exponential type at most $R+t$. 

The most general square-integrable solution of the telegrapher's equation can be written as 
\beq
\rho(t,x) = \int_{\mathbb R} dk\, \hat{\rho}(t,k) e^{ikx}
\eeq
with 
\beq
\hat{\rho}(t,k) = \hat{u}(k) e^{-\frac{t}{2\tau}} \left[\cosh\left(\frac{\Delta t}{2\tau}\right) + \frac{1}{\Delta} \sinh\left(\frac{\Delta t}{2 \tau} \right) \right] + \hat{v}(k) e^{-\frac{t}{2\tau}} \frac{2\tau}{\Delta} \sinh\left(\frac{\Delta t}{2 \tau} \right). \label{general_sol_trig}
\eeq 
We start by noting that an entire function of the square root of a complex number is itself entire if the original function is even. Therefore, both $\cosh\left(\frac{\Delta t}{2\tau}\right)$ and $\frac{1}{\Delta} \sinh\left( \frac{\Delta t}{2 \tau} \right)$ are entire functions of $k$. It then follows than $\hat{\rho}(t,k)$ is entire in $k$, since $\hat{u}$, $\hat{v}$ are entire by assumption, and the product of two entire functions, as well as their sum, are themselves entire. 

To show that \eqref{general_sol_trig} is of exponential type at most $R+ t$, we proceed as follows. First, we note that both $\cosh\left(\frac{\Delta t}{2\tau}\right)$ and $\frac{1}{\Delta} \sinh\left(\frac{\Delta t}{2 \tau} \right)$ are of exponential type $\sqrt{D/\tau}t$. This also applies to their sum. Next, we recall that the product of two functions of exponential types $\sigma_1$  and $\sigma_2$ is at most exponential type $\sigma_1 + \sigma_2$. Therefore, the type of each term in the sum \eqref{general_sol_trig} is at most $R + \sqrt{D/\tau}t$. Finally,  since the type of the sum of two functions of exponential types $\sigma_1$ and $\sigma_2$ is smaller or equal than $\textrm{max}(\sigma_1, \sigma_2)$, it follows that the type of $\rho(t,k)$ is at most $R + \sqrt{D/\tau}t$. As long as $D/\tau \leq 1$, we see that the system respects relativistic causality. 

It is worth pointing out that this result conforms with the expectation that, in any local quantum system, causality bounds the diffusion constant in terms of the effective light-cone speed and the local equilibration time \cite{Hartman:2017hhp} (see also \cite{Davison:2018ofp,Davison:2018nxm,Baggioli:2020ljz}). In the case at hand, the effective light-cone speed is to be identified with the speed of light, and the local equilibration time with $\tau$.

\section{The large-scale expansion and hydrodynamics \label{sec.relationtogradexp}}

In our previous work \cite{Heller:2020uuy}, we described the most general construction of linearized hydrodynamics for a neutral conformal fluid in arbitrary number of spacetime dimensions. Specializing to a shear channel fluctuation, the hydrodynamic description of the microscopic field~$\rho$ is provided by the conservation equation \eqref{eq_conservation} in combination with the purely-spatial gradient expansion of the constitutive relations. For the MIS case, and in the notation of appendix~\ref{MIS}, the hydrodynamic gradient expansion reads  
\beq
J(t,x) = - \sum_{n=0}^\infty c_n \, \partial_x^{2n+1} \rho(t,x). \label{const_rel}
\eeq
The transport coefficients $c_n$ are extracted from the microscopic shear hydrodynamic mode by a matching computation. In MIS, they are given by  
\beq
c_n = (-1)^n \,\mathcal{C}_n \, D^{n+1} \, \tau^{n},  
\eeq
where $\mathcal{C}_n$ is the $n$-th Catalan number.

As we have illustrated extensively in the main text, the perturbative series $\rho^{(H)}_\epsilon(t,x)$ corresponds to the hydrodynamic shear mode contribution to the exact $\rho(t,x)$. It is then natural to ask whether the effective description of MIS theory in terms of classical hydrodynamics contains exactly the same physical information as the perturbative sector of our transseries. 

Let us view \eqref{const_rel} as a formal series, making no assumptions about the relative size of subsequent terms, and perform the rescaling \eqref{rescaling} both at the level of the gradient expansion \eqref{const_rel} and the conservation equation. Then, it turns out that the perturbative sector of our transseries, $\rho^{(H)}_\epsilon(t,x)$, solves the resulting system order-by-order in an expansion around $\epsilon = 0$. Equivalently, if we ignored the specific values of the transport coefficients $c_n$ but somebody handed to us  $\rho^{(H)}_\epsilon(t,x)$, we could fix the former by demanding that the latter is a solution order-by-order in an expansion around $\epsilon = 0$. This procedure could then be viewed as the position space counterpart of the matching computation performed in \cite{Heller:2020uuy}. 
\\\\
\noindent It should be noted that while $\rho^{(H)}_\epsilon(t,x)$ is a solution, the initial conditions we imposed on the $a_{n}(t,x)$ coefficients in the main text are completely unnatural from the perspective of the hydrodynamic description of the system. These boundary conditions relied on the existence of the nonperturbative sector of the momentum space transseries, which allowed us to enforce simultaneously that $\hat{\rho}(0,k) = \hat{u}(k)$, $\partial_t\hat{\rho}(0,k) = \hat{v}(k)$. This nonperturbative sector is absent now, reflecting the fact that the algebraic relation between $J(t,x)$ and $\rho(t,x)$ in the  hydrodynamic description implies the loss of the nonhydrodynamic degree of freedom. 

We now proceed to explain the natural choice of initial conditions from the perspective of the hydrodynamic description. After the spacetime rescaling, our equations of motion are given by 
\begin{align}
&\epsilon^2 \partial_t \rho(t,x) + \epsilon \partial_x J(t,x) = 0, 
&J(t,x) = - \sum_{n=0}^{\infty} c_n \epsilon^{2n+1}\partial_x^{2n+1} \rho(t,x).  
\end{align}
The ansatz 
\beq 
\rho(t,x) = \sum_{n=0}^\infty a_n(t,x) \epsilon^n \label{H_ansatz}
\eeq
results in the following nested ODE system
\beq
(\partial_t - D \, \partial_x^2) \, a_{2n}(t,x) = \sum_{q=1}^n c_q \, \partial_x^{2(q+1)} a_{2(n-q)}(t,x) \label{a_n_nested_hydro}
\eeq
with an equivalent expression for the odd coefficients. Again, the $n$-th term in the series expansion \eqref{H_ansatz} is a solution of the heat equation sourced by the $n-1$ previous orders. 

Let us assume that, at a time slice $t=t_0$, $\rho(t_0,x) = \rho_0(x)$ is known. In this situation, the natural boundary conditions to impose on \eqref{H_ansatz} are that, at $t=t_0$, the leading-order term $a_0(t_0, x)$ agrees with $\rho_0(x)$, with the remaining higher-order terms vanishing. Due to the structure of \eqref{a_n_nested_hydro}, these boundary conditions result in the vanishing of the odd order terms in \eqref{H_ansatz} for all times. 

With this choice of boundary conditions, \eqref{a_n_nested_hydro} can be explicitly solved in closed-form. In Fourier space, we find that\footnote{Since the nested ODE system \eqref{a_n_nested_hydro} is time-translation invariant, we set $t_0 = 0$ with no loss of generality.}
\beq
\hat{a}_{2n}(t,k) = \delta_{n,0}\,\hat{\rho}_0(k) - (-1)^n\, c_n\, t\, k^{2(n+1)}\, {}_1 F_1(1-n, 2+n, D k^2 t) \, e^{-D k^2 t} \hat{\rho}_0(k). 
\eeq
The expansion coefficients can be explicitly computed in position space. The zeroth-order one is the solution of the heat equation given by $a_0(t,x) = (G_0 * \rho_0)(t,x)$, where $G_0$ is given in equation \eqref{heat_kernel}. The remaining ones are
\beq
a_{2n}(t,x) = c_n \sum_{q=0}^{n-1} \frac{\Gamma(n) \Gamma(n+2)}{\Gamma(n-q)\Gamma(n+2+q)\Gamma(q+1)} D^q t^{q+1} \partial_x^{2 (n+q+1)} a_0(t,x). 
\eeq
As it happened with our previous choice of initial conditions, each term is a gradient series in $\partial_x^2$ acting on a particular solution of the heat equation.

\section{Applicability of the truncated perturbative series} 
\label{comparison}

The practical usefulness of the perturbative piece of the asymptotic expansion developed in the previous section is that, when truncated to low order, it provides a good description of the exact velocity field $\rho$ in some spacetime regions. Owing to our general discussion in the Introduction, where we introduced our large-scale expansion, we expect that these regions correspond to those in which the nonhydrodynamic mode contribution has significantly decayed.

Let us illustrate this by considering the time evolution of $\delta$-function initial data of the form $u(x) = 0,\,v(x) = \delta(x)$. In this case, $\rho(t,x)$ corresponds to the propagator of the telegrapher's equation \cite{Romatschke:2009im}, 
\beq
G(t,x) = \Theta(t) \Theta\left(\frac{D}{\tau}t^2 - x^2\right) \frac{1}{\sqrt{4 D \tau}} e^{- \frac{t}{2\tau}} I_0\left(\sqrt{\frac{t^2}{4 \tau^2} - \frac{x^2}{4 D \tau}} \right).  
\eeq
The expression above clearly demonstrate that the telegrapher's equation is causal: for any $t>0$, $\rho(t,x)$ vanishes if $|x| > \sqrt{\frac{D}{\tau}} t$. 
\begin{figure}[h!]
\begin{center}
\includegraphics[width=7.5cm]{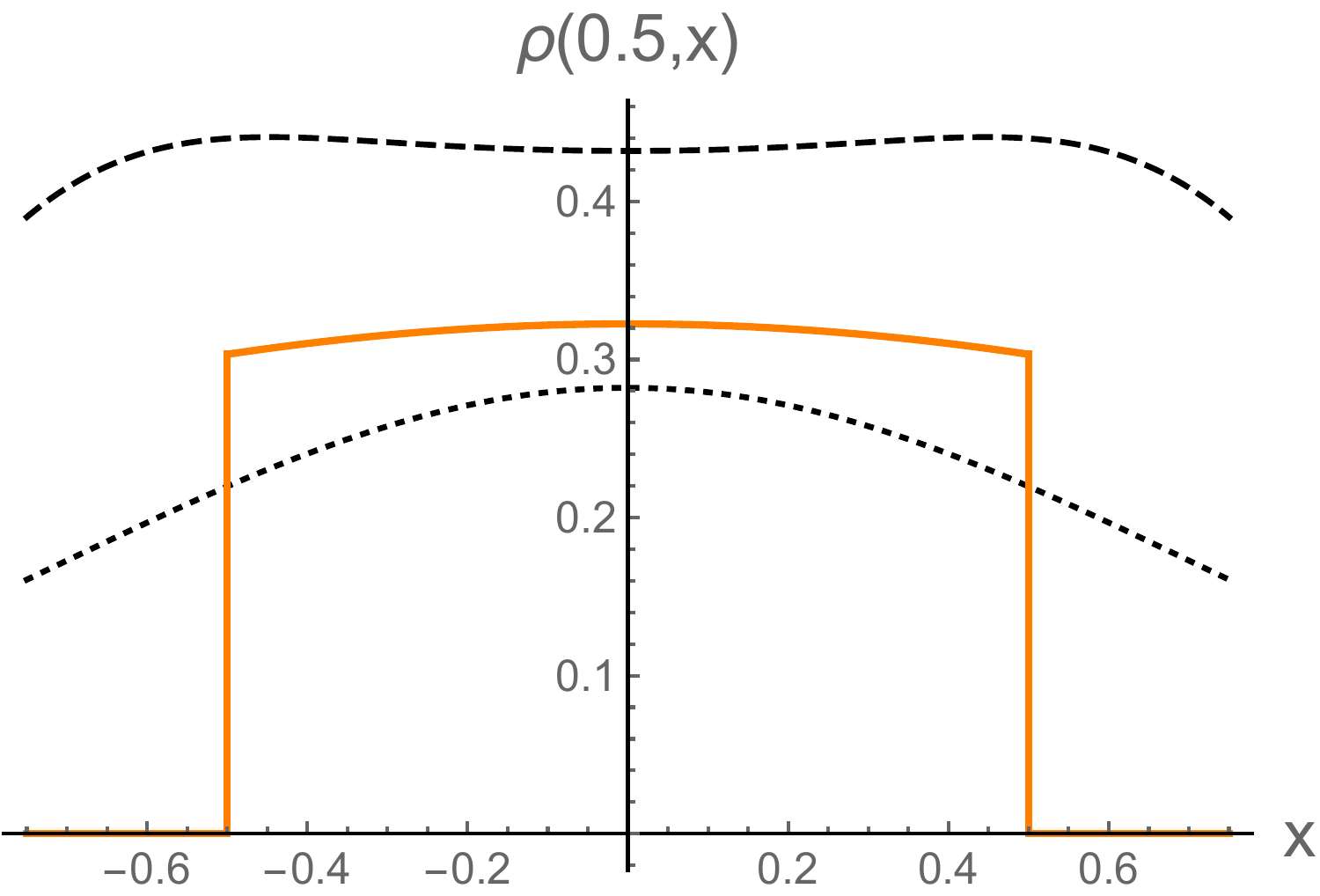}\,\,\,\includegraphics[width=7.5cm]{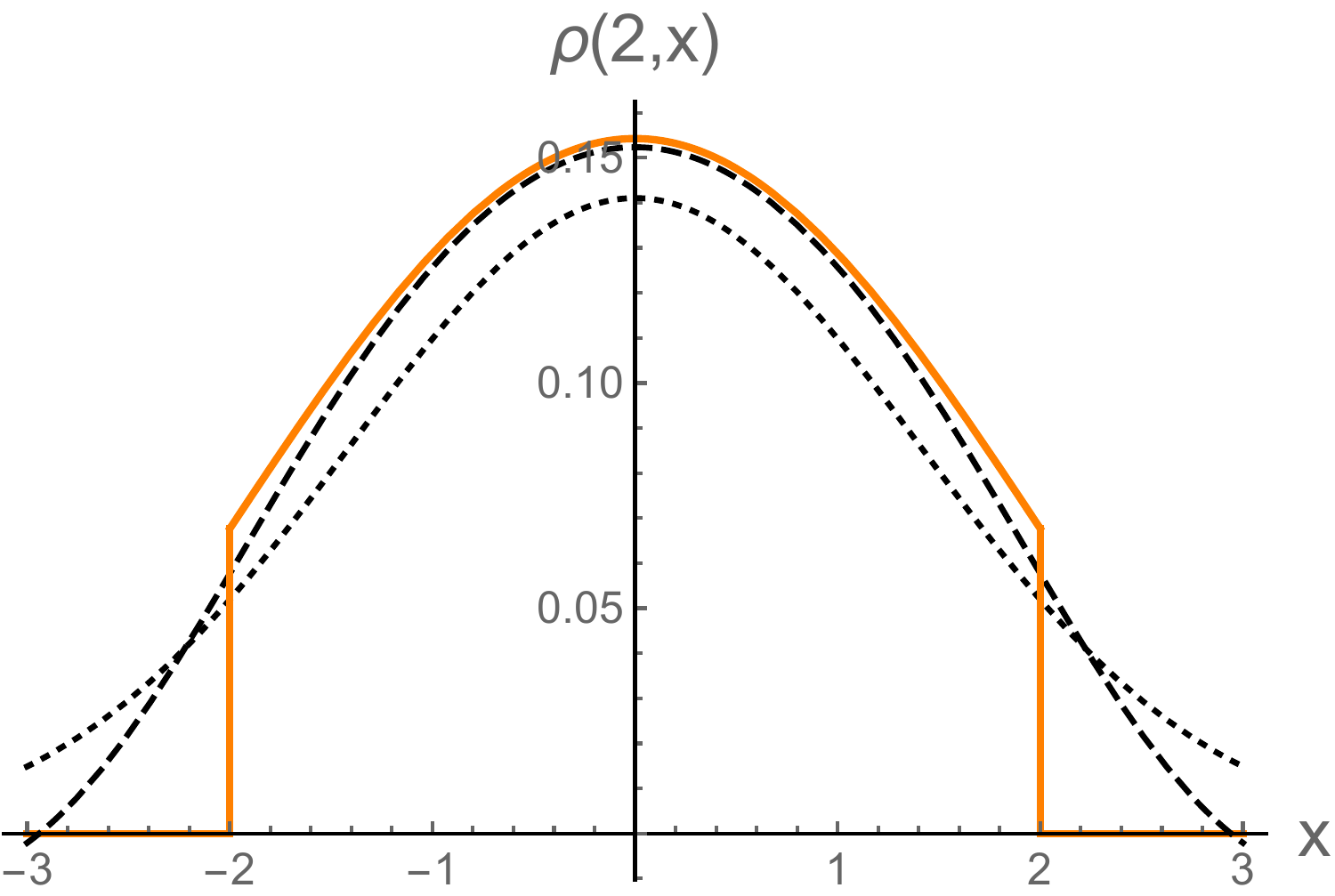}\\
\includegraphics[width=7.5cm]{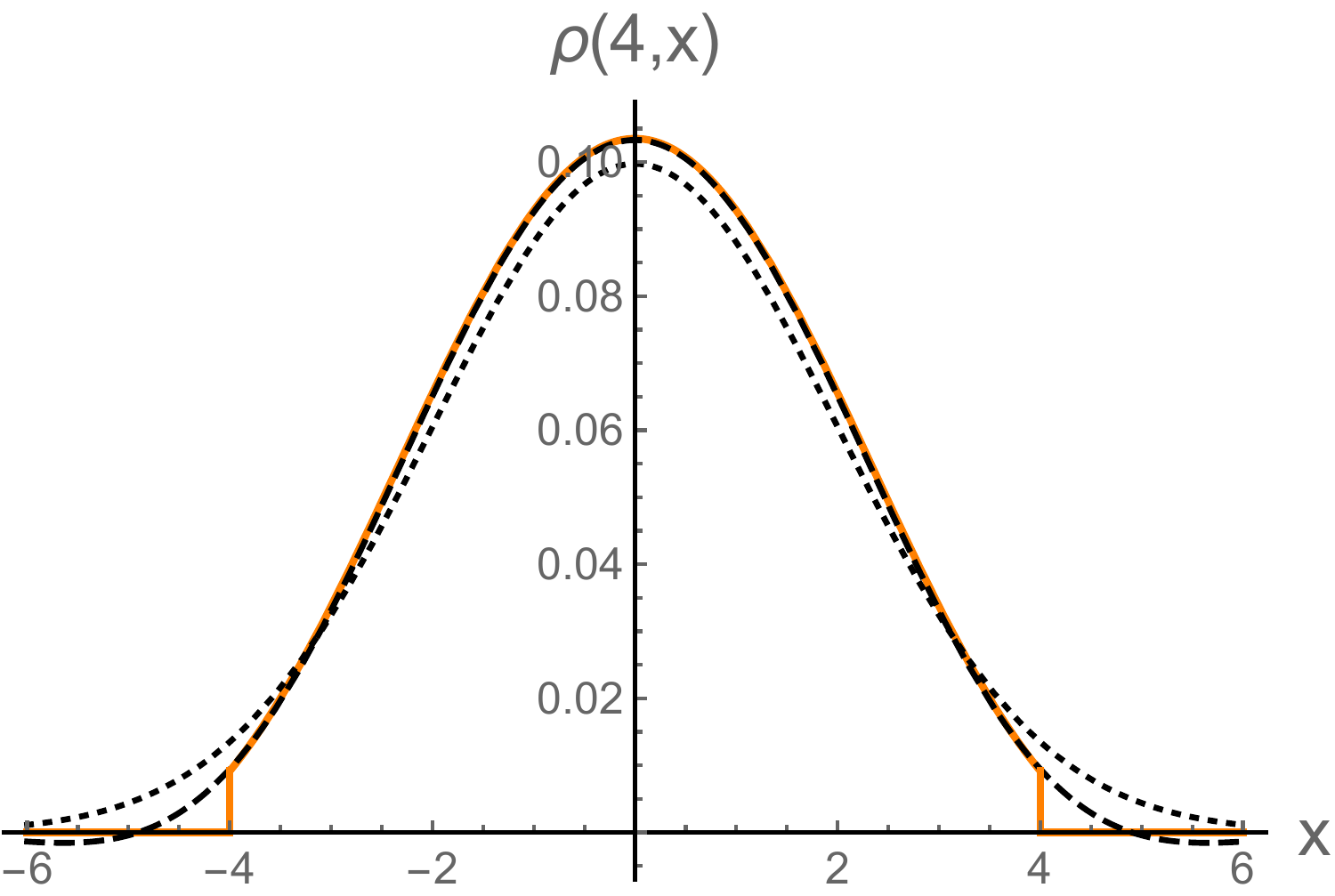}
\caption{\small Comparison between the exact $\rho$ corresponding to $\delta$-function initial data (solid orange) and the $\rho^{(H)}_{\epsilon}\big|_{\epsilon = 1}$ expansion truncated to first (dotted black) and third order (dashed black).}
\label{fig:propagator_comparison}
\end{center}
\end{figure}

We compare the exact $\rho(t,x)$ above with the with the perturbative expansion $\rho^{(H)}_{\epsilon}\big|_{\epsilon = 1}$ obtained by means of equation \eqref{rho_n_real_space_gradient} truncated to first and third nontrivial order. The results can be found in figure \ref{fig:propagator_comparison}.

We clearly see that, for any $x$, $\rho^{(H)}_{\epsilon}\big|_{\epsilon = 1}$ never provides an accurate description of $\rho$ at early times. This is due to the fact that, in this regime, the nonhydrodynamic mode contribution, which is necessary to enforce the initial condition $\rho(0,x) = 0$ we chose, is still sizable. This state of affairs changes at later times. In particular, focusing on a fixed $x$ as $t$ grows, we eventually observe a very good agreement between $\rho$ and the low-order truncated $\rho^{(H)}_{\epsilon}\big|_{\epsilon = 1}$. 

Another important point to be drawn from figure \ref{fig:propagator_comparison} is that $\rho^{(H)}_{\epsilon}\big|_{\epsilon = 1}$ is never a good approximation of the exact microscopic $\rho$ outside the causal cone of the system. While the exact $\rho$ vanishes there, $\rho^{(H)}_{\epsilon}\big|_{\epsilon = 1}$ does not. The reason behind this difference is that $\rho^{(H)}$ is solely built out of the hydrodynamic shear mode, and is therefore blind to the nonhydrodynamic contribution needed to enforce the causal response of the system. This fact provides a nice illustration of the general lesson that the nonhydrodynamic sector is essential to guarantee causality~\cite{Florkowski:2017olj}.

\section{Large-order behavior for  Gaussian initial data}
\label{sp_details}

In the main text, we showed that the large-order behavior of the perturbative series is controlled by the adjacent nonperturbative saddle points. In this appendix, we elaborate further on this connection for the case of Gaussian initial data.  

Following Berry and Howls \cite{berry1991hyperasymptotics}, we can express the $a_{n+1}(t,0)$ coefficient of the perturbative series expansion as the following contour integral, 
\begin{align}
&a_{n+1}(t,0) = \Gamma\left(n+\frac{1}{2} \right) \frac{1}{2\pi i} \oint_\Gamma dk\, \frac{\tau}{2\pi \sqrt{1-4 D \tau k^2}} \left(\frac{k^2}{-S_H(k)}\right)^{n+\frac{1}{2}} k^{-(2n+1)}, \label{a_coeff_BH} \\
&S_H(k) = -\frac{1}{2}s^2 k^2 + (-1+\sqrt{1- 4 D \tau k^2})\frac{t}{2\tau} 
\end{align}
with $\Gamma$ a positively-oriented path enclosing $k=0$ in the hydrodynamic sheet. 

It can be demonstrated that \eqref{a_coeff_BH} can be alternatively expressed as a contour integral in the nonhydrodynamic sheet,  
\begin{align}
&a_{n+1}(t,0) = \Gamma\left(n+\frac{1}{2} \right) \frac{1}{\pi i} \int_{\Gamma'} dk\, \frac{\tau}{2\pi \sqrt{1-4 D \tau k^2}} \frac{1}{\left(-S_{NH}(k)\right)^{n+\frac{1}{2}}}, \label{a_coeff_BH_deformed} \\
&S_{NH}(k) = -\frac{1}{2}s^2 k^2 - (1+\sqrt{1- 4 D \tau k^2})\frac{t}{2\tau}, 
\end{align}
where $\Gamma'$ is a path starting at $\infty - i 0$ below the right branch cut, going around the right branch point, and ending at $\infty + i 0$ above the right branch cut.\footnote{In writing \eqref{a_coeff_BH_deformed}, we have taken into account that left and right branch cuts contribute equally to the total result.} 

\begin{figure}[h!]
\begin{center}
\includegraphics[width=7.5cm]{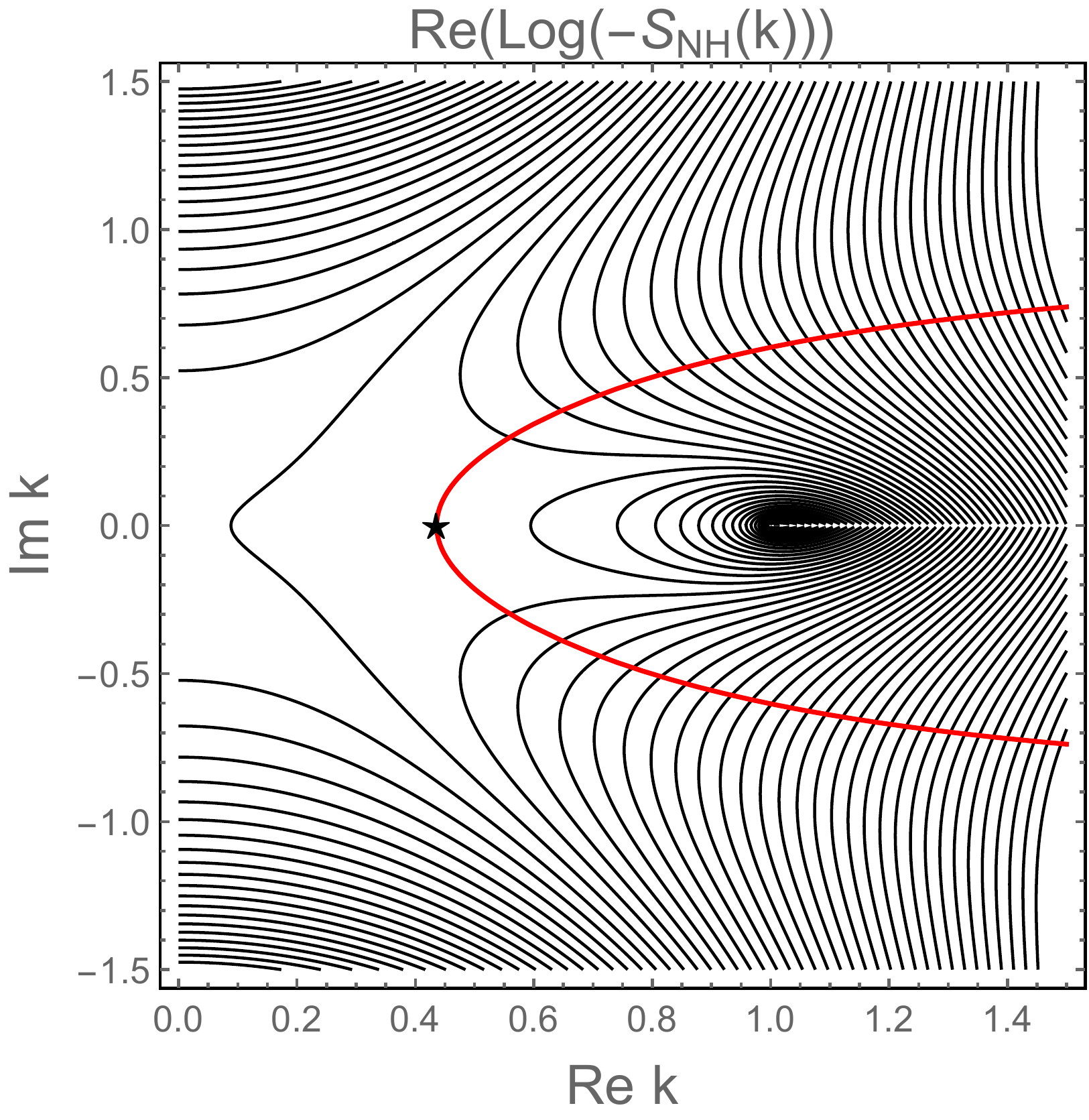}\includegraphics[width=7.5cm]{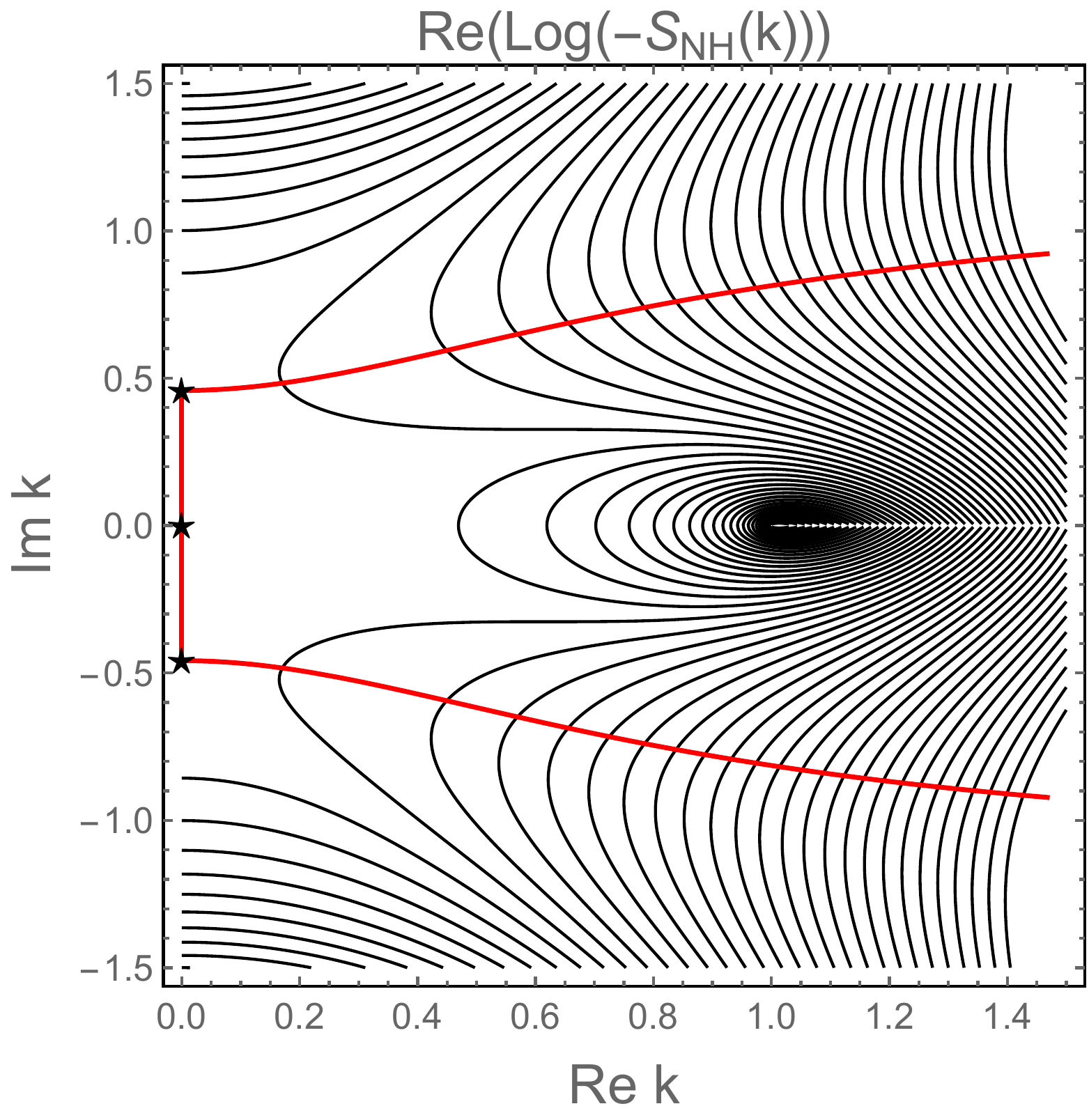}
\caption{\small Relevant steepest descent contours to perform the large-$n$ saddle point analysis (solid red), along with the adjacent saddles (black stars), for $s=1$. Left: $t = 0.9$. Right: $t=1.1$. 
}
\label{fig:adjacent_saddles_steepest_descend}
\end{center}
\end{figure}

In order to analyze the behavior of \eqref{a_coeff_BH_deformed} when $n \to \infty$, it is natural to decompose $\Gamma'$ into the steepest descent contours associated to the adjacent saddles discussed in the main text, and employ a saddle point approximation afterward.\footnote{Since at $k_* = k_+, k_-$ or $k_0$ we have that $S_{NH}'(k_*)=0$, but $S_{NH}(k_*) \neq 0$, these saddles are also stationary points of $\log(-S_{NH}(k))$.} The relevant steepest descent contours involved in computing \eqref{a_coeff_BH_deformed} in a large $n$ asymptotic expansion are depicted in figure \ref{fig:adjacent_saddles_steepest_descend}. 

\noindent For $t < t_c$, only the $k_+$ saddle contributes to $a_{n}(t,0)$. On the other hand, for $t > t_c$, we have contributions both from $k_+, k_-$ and $k_0$. 

\begin{figure}[h!]
\begin{center}
\includegraphics[width=7.5cm]{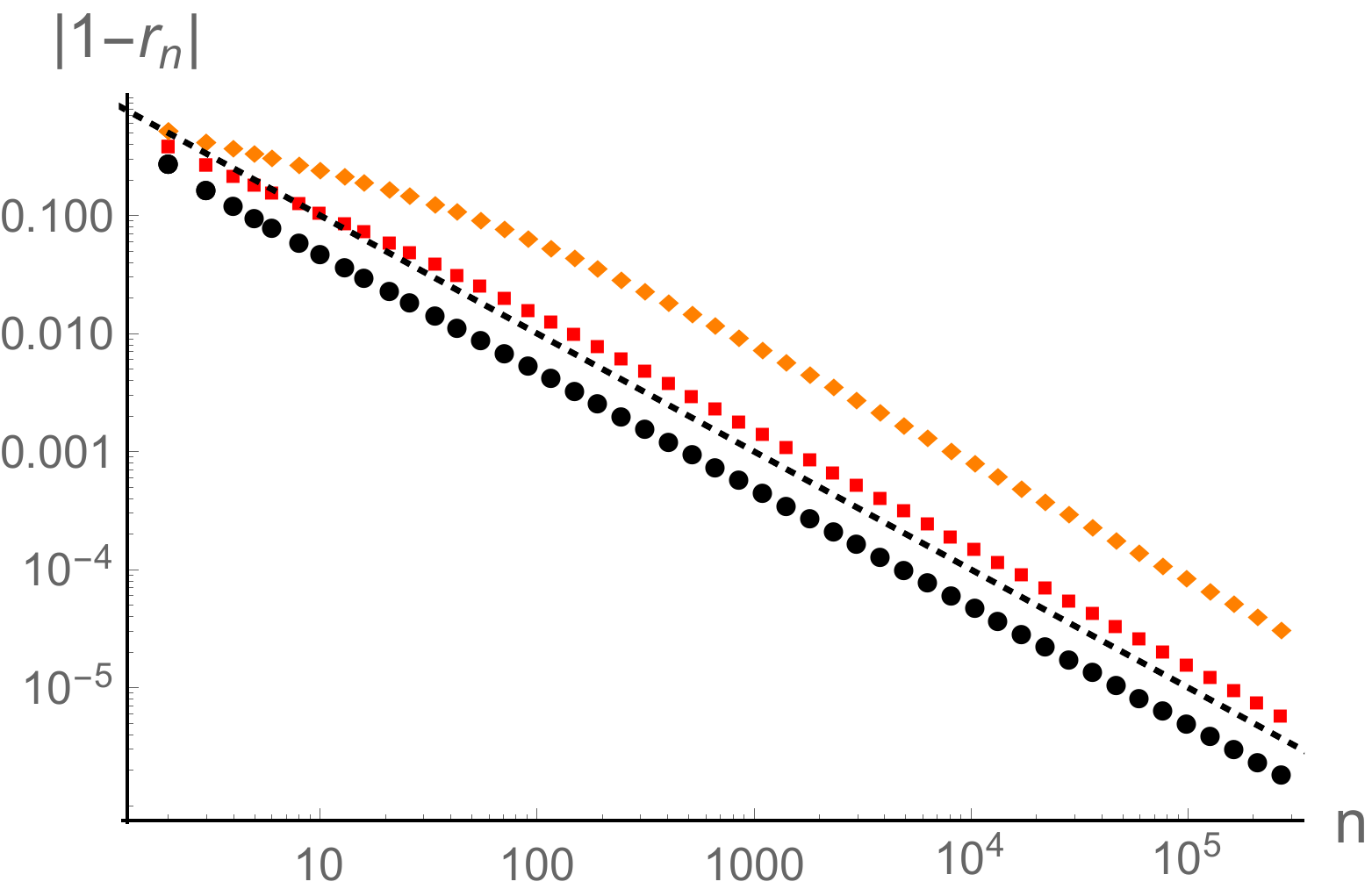}\includegraphics[width=7.5cm]{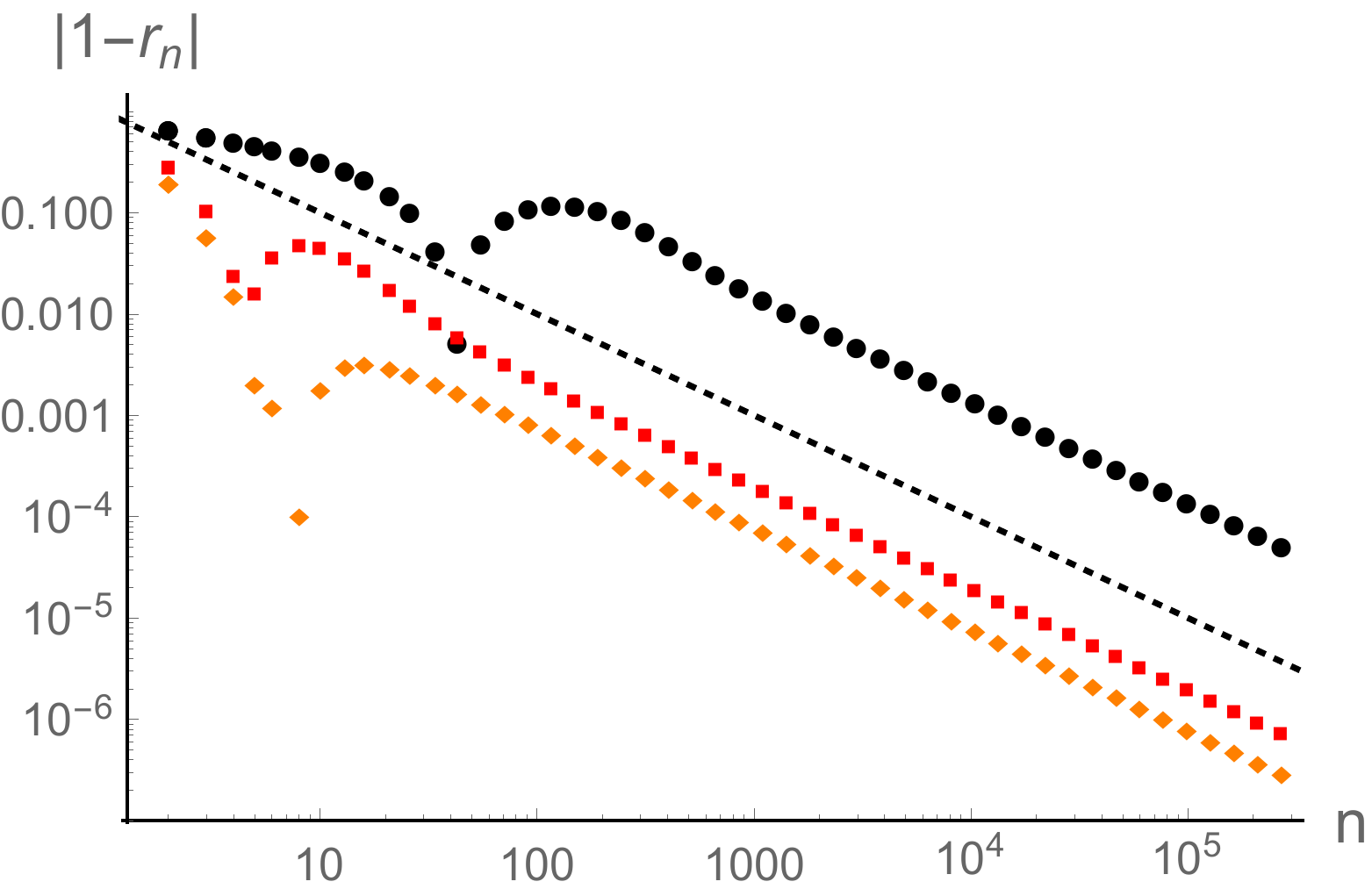}
\caption{\small Deviation ratio \eqref{r_def} between the exact $a_n(t,0)$ coefficients and the large-order prediction of the saddle point analysis, for Gaussian initial data with $s=1$. Left: $t = 0.25$ (black), $t=0.5$ (red), $t=0.75$ (orange). Right: $t=1.25$ (black), $t=3$ (red), $t = 5$ (orange). The dotted black line is a plot of the function $1/n$ to guide the reader's eye. 
}
\label{fig:sp_comparison}
\end{center}
\end{figure}

Let us focus on the former case first. At leading order, it is immediate to show that 
\beq
a_{n+1}(t,0) = \frac{\Gamma(n)}{(-S_+)^n} \left[G(k_+) \sqrt{-\frac{2}{\pi S_{NH}''(k_+)}} + O\left(\frac{1}{n}\right) \right], \label{GID_an_sp_before}
\eeq
from which it follows that $r = \lim_{n\to\infty} n \, a_n(t,0)/a_{n+1}(t,0) = - S_+$, as reported in the main text. A plot of the deviation of the ratio  
\beq
r_{n+1} = \frac{a_{n+1}^{\textrm{exact}}(t,0)}{a_{n+1}^{\textrm{s.p.}}(t,0)}, \label{r_def}
\eeq
from one, where $a_{n+1}^{\textrm{exact}}$ is given by \eqref{an_GID} and $a_{n+1}^{\textrm{s.p.}}$ by \eqref{GID_an_sp_before}, can be found in figure \ref{fig:sp_comparison} (left). We have considered several different times before $t_c$. It is readily seen that, as $n \to \infty$, $r_n \to 1$ plus a $O(1/n)$ correction, confirming the validity of equation \eqref{GID_an_sp_before}.  

On the other hand, for $t > t_c$, we have three separate contributions to consider. We find that at leading order the $k_0$ saddle contributes as
\beq
\frac{\Gamma(n)}{(-S_0)^n} \left[G(k_0) \sqrt{-\frac{2}{\pi S_{NH}''(k_0)}} + O\left(\frac{1}{n}\right) \right] \label{large_n_k0_after}
\eeq
while the first nontrivial term of the combined contribution of the $k_+$, $k_-$ saddles is given by  
\beq
\frac{\Gamma\left(n+\frac{1}{2}\right)}{(-S_+)^{n+\frac{1}{2}}}\frac{2}{i\pi}\left(G'(k_+) - \frac{S_{NH}'''(k_+)}{3 S_{NH}''(k_+)} G(k_+)\right)\frac{(-S_+)}{\left(n+\frac{1}{2}\right) S_{NH}''(k_+)} + \ldots \label{large_n_kpm_after}
\eeq
Since $|S_0| < |S_+|$, the $k_\pm$ contribution is exponentially suppressed with respect to the $k_0$ one,\footnote{Despite this, in the main text we show that Borel transform techniques were capable or unveiling it.} which now governs the divergence of the perturbative series expansion. The behavior of $r_n$ for $t > t_c$ with $a_{n}^{\textrm{s.p.}}(t,0)$ given by \eqref{large_n_k0_after} is illustrated by figure \ref{fig:sp_comparison} (right). 

\section{Large-order behavior for Lorentzian initial data}
\label{Lorentzian}

\begin{figure}[h!]
\begin{center}
\includegraphics[width=8cm]{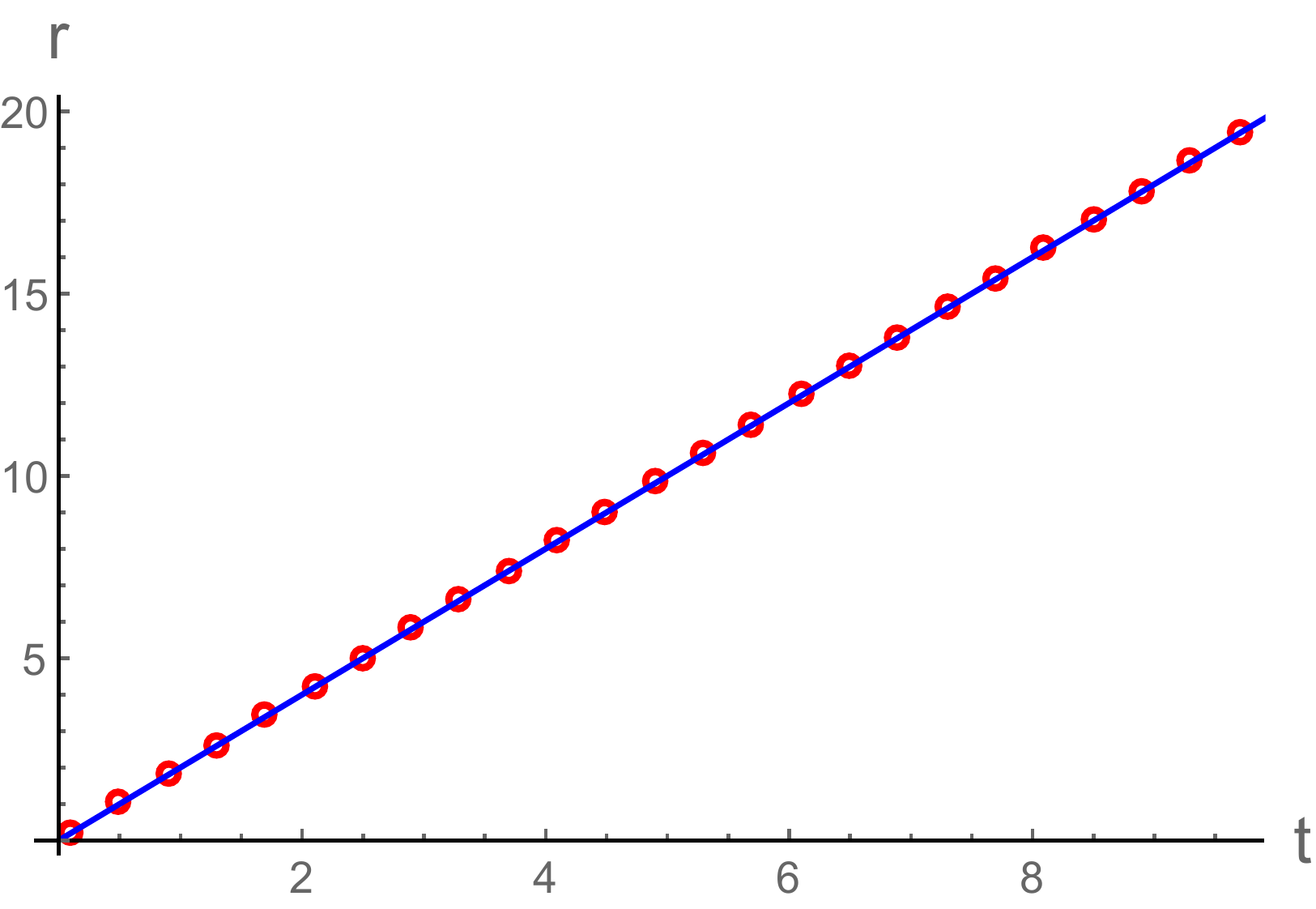}
\caption{\small Lorentzian initial data exhibits factorial growth for the series coefficients. Here we show $r$ as defined in \eqref{qn_LID} vs. $t$ for $\alpha = 1$.}
\label{fig:ratio_test_Lorentzian}
\end{center}
\end{figure}

Another family of initial data that allows to compute in closed-form the coefficients of the perturbative expansion at $x=0$ is that of Lorentzian initial data, 
\begin{equation}
v(x) = \frac{\alpha}{\pi (x^2 + \alpha^2)},\,\,\, \hat v(k) = \frac{1}{2\pi}e^{-\alpha |k|}, \label{id_Lorentzian}
\end{equation}
where we find
\begin{equation}
a_{n+1}(t,0) = \frac{\alpha^{2n+1} \tau^{n+1}}{4 \pi^2 D^{n+1}t^{2n+1}}\Gamma\left(n+\frac{1}{2}\right)^2  U\left(2n+1, n+\frac{3}{2}, \frac{\alpha^2}{ 4 D t} \right). \label{rho_n_Lorentzian}
\end{equation}
As it happened for Gaussian initial data, we find that, for $n \to \infty$, $q_n = \lim_{n\to \infty} |\frac{a_{n+1}(t,0)}{a_{n}(t,0)}|$ is linear in $n$, implying that the series is factorially divergent. The large $n$ behavior determined numerically is compatible with 
\beq
q_n \sim \frac{1}{r} n, \quad r \equiv \frac{t}{\tau}, \label{qn_LID}
\eeq
irrespectively of the value of $\alpha$. See figure \ref{fig:ratio_test_Lorentzian} for an example of this asymptotic behavior in the case $\alpha = 1$. 

\section{Reconstructing $\rho$ from the transseries: Gaussian initial data case}
\label{resummation}

In this appendix we illustrate how Borel resummation can be employed to reconstruct the exact $\rho(t,x)$ from the different transseries at our disposal. The agreement between the Borel resummation and the exact solution determined by direct numerical integration provides a nontrivial consistency check of the results presented in the main text. For the sake of brevity, we focus on Gaussian initial data of the form \eqref{GID}. 

Before starting, let us recall that the splitting of $\rho(t,x)$ into hydrodynamic and nonhydrodynamic contributions is only unequivocally defined once a particular integration path $\gamma$ is provided, due to the branch points of $\Delta(k)$.\footnote{We choose the principal branch for $\Delta(k)$.}
 To keep track of which integration path we are employing, we will denote by $\gamma_{(+,-)}$ the equivalence class of integration paths starting at $-\infty + i 0^+$ above the left branch cut and ending at $\infty - i 0^+$ below the right one, with an analogous interpretation for $\gamma_{(-,+)}$, $\gamma_{(-,-)}$ and $\gamma_{(+,+)}$.

\subsection{The perturbative sector}

\begin{figure}[h!]
\begin{center}
\includegraphics[width=7.5cm]{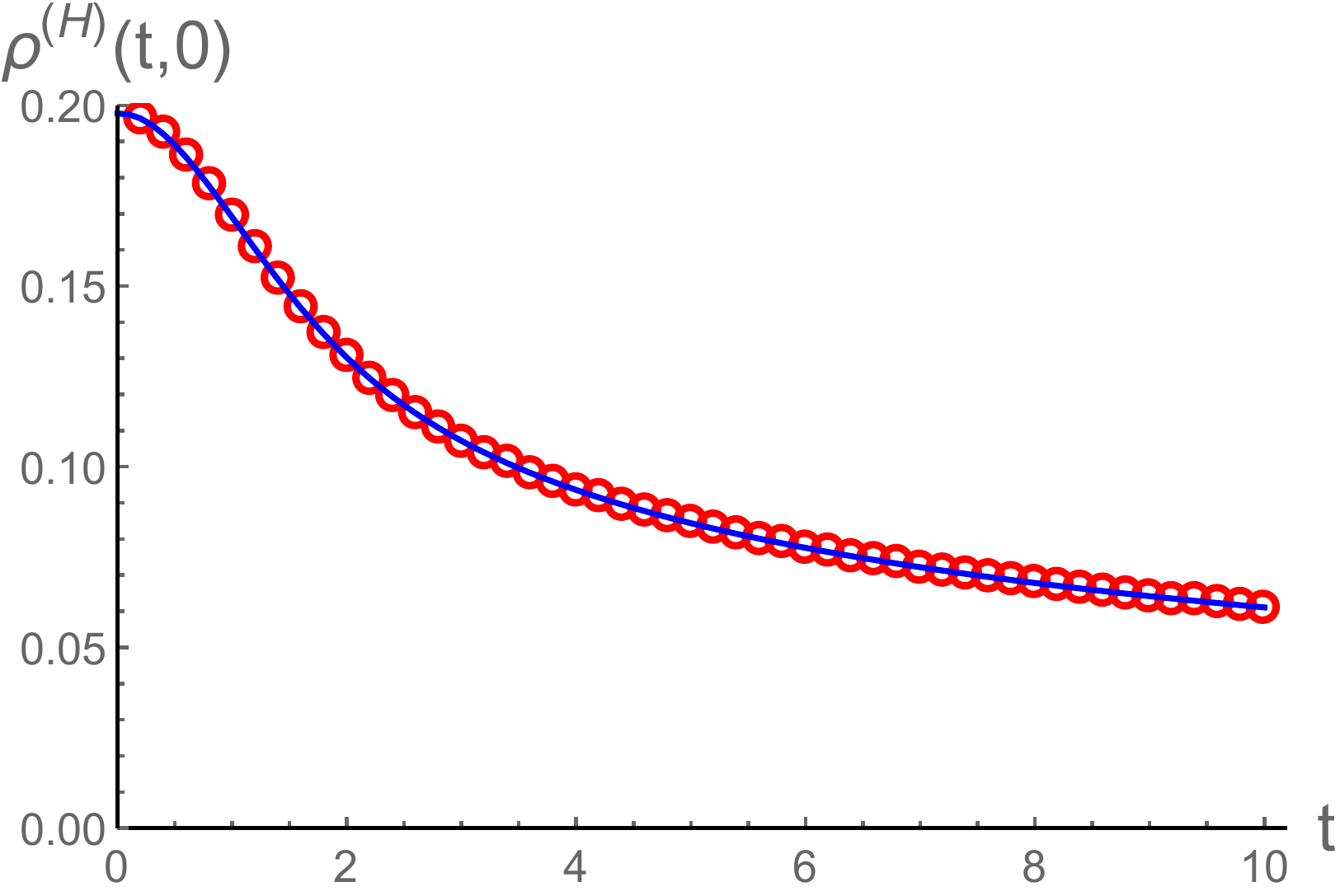}
\caption{\small Average of the lateral Borel resummations of $\rho^{(H)}_{\epsilon}(t,x)\big|_{\epsilon = 1}$ (open red circles) for Gaussian initial data with $s = 1$ compared with the $\rho^{(H)}(t,0)$ determined directly by a numerical integration along the $\gamma_{(-,-)}$ contour (solid blue curve).}
\label{fig:GID_H_resummation}
\end{center}
\end{figure}

As mentioned before, the Padé approximant of the Borel-transformed asymptotic series displays a line of pole condensation along the positive real $z$-axis starting at $z_c = -S_-$ and, as a consequence,  we have to resort to lateral Borel resummations. In turns out that the negative (positive) lateral Borel resummation corresponds to the choice of $\gamma_{(+,-)}$ ($\gamma_{(-,+)}$) integration contour in the original integral, with the average of both corresponding to the $\gamma_{(-,-)}, \gamma_{(+,+)}$ integration paths. We plot an example of this agreement in figure \ref{fig:GID_H_resummation}.

\subsection{The nonperturbative sector}

\begin{figure}[h!]
\begin{center}
\includegraphics[width=7.5cm]{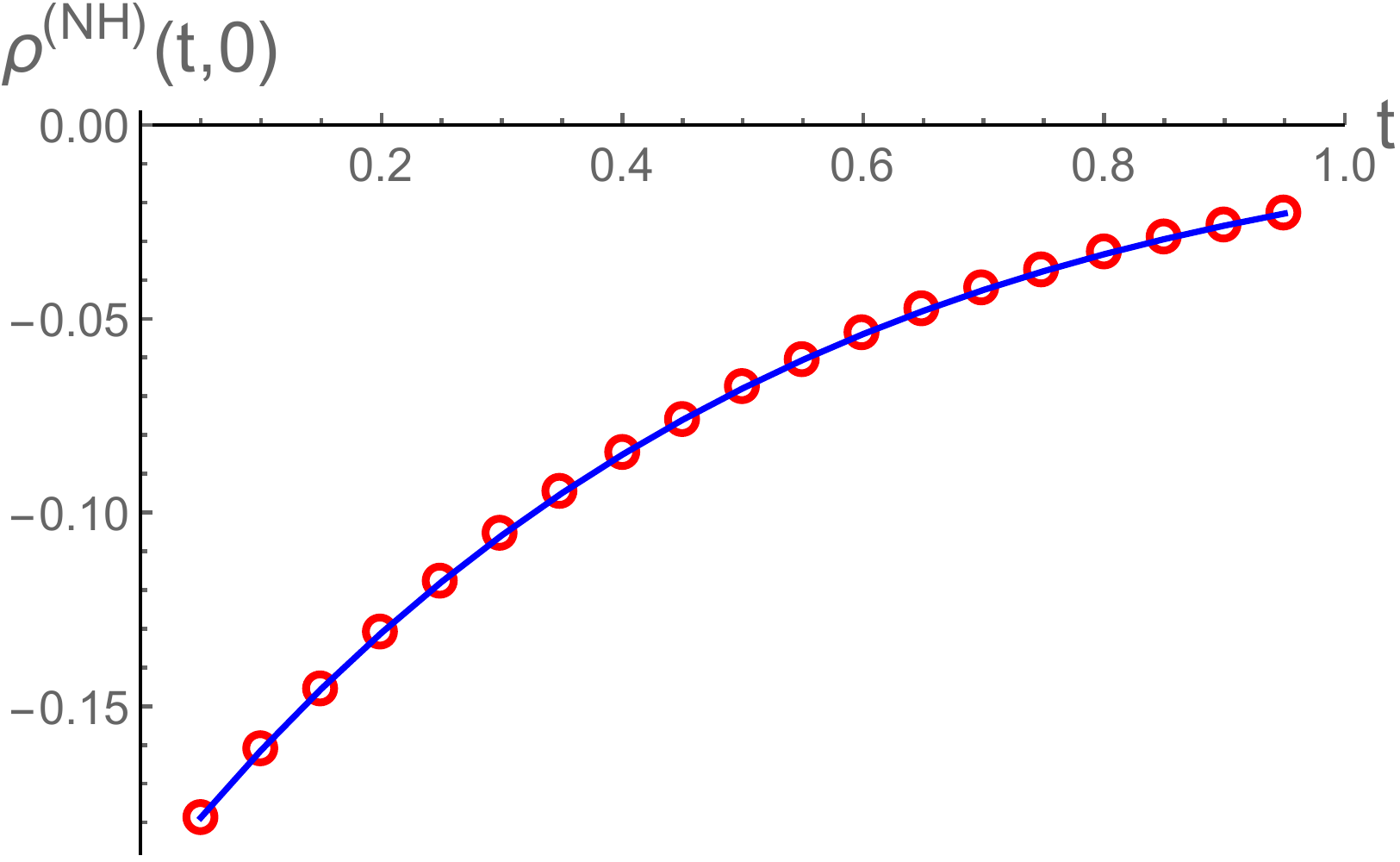}\includegraphics[width=7.5cm]{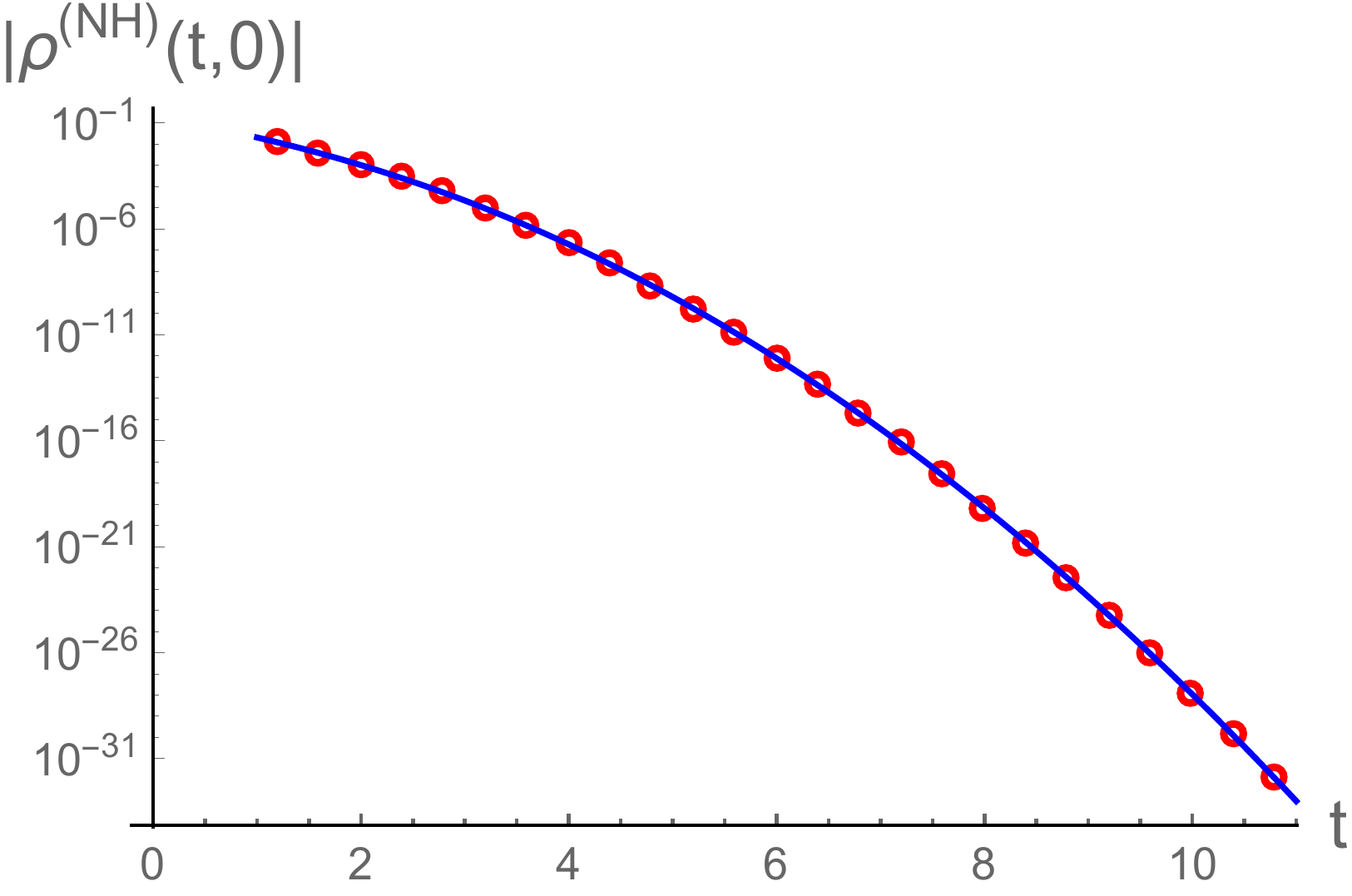}
\caption{\small Left: same as figure \ref{fig:GID_H_resummation}, but now for the nonperturbative series at $t < t_c$. Right: result of the Borel resummation of the nonperturbative series for $t > t_c$ (open circles) compared with the $\rho^{(NH)}(t,0)$ determined by a numerical integration along $\gamma_{(-,-)}$ (solid blue curve).}
\label{fig:GID_asymptotic_expansion_past_res}
\end{center}
\end{figure}

For the nonperturbative sector, we have a discontinuous change in the nonperturbative series when $t = t_c$; correspondingly, we discuss the $t < t_c$ and $t > t_c$ cases separately: 

\begin{itemize}
    \item $t < t_c$. In this regime, the Fourier integral defining $b_n(t,x)$ in terms of $b_n(t,k)$ converges. Therefore, $b_n(t,0) = -a_n(-t,0)$, with $a_n(t,0)$ given by \eqref{an_GID}. The relation between the lateral Borel resummations and the $\gamma$ integration contours is the same as for the perturbative series, with the real result being given by the average of both lateral resummations. We plot this average in figure \ref{fig:GID_asymptotic_expansion_past_res} (left), where we compare it with the $\rho^{(NH)}(t,0)$ determined by a direct numerical integration, finding very good agreement between both quantities.   
    \item $t > t_c$. In this regime, the nonperturbative sector of our transseries is given by 
     \eqref{GID_asymptotic_expansion_past_full}-\eqref{GID_asymptotic_expansion_past_coeffs}. The absence of singularities on the positive real axis in the Borel plane implies that the Borel transform of $\rho^{(NH)}_\epsilon(t,0)$ is Borel resummable; hence the integration along the real axis agrees trivially with the average of the lateral resummations. In figure \ref{fig:GID_asymptotic_expansion_past_res} (right), we compare the $\rho^{(NH)}(t,0)$ determined by a direct numerical integration with the resummation result, finding again excellent agreement. 
     
\end{itemize}

\bibliography{literature}

\providecommand{\href}[2]{#2} \providecommand{\beforedoihref}{}
  \providecommand{\afterdoihref}{}\begingroup\raggedright\begin{thebibliography}{10}

\bibitem{Florkowski:2017olj}
W.~Florkowski, M.~P. Heller and M.~Spali{\'n}ski, {\it {New theories of
  relativistic hydrodynamics in the LHC era}},
  \beforedoihref\href{http://dx.doi.org/10.1088/1361-6633/aaa091}{Rept. Prog.
  Phys.}\afterdoihref\  {\bf 81} (2018), no.~4 046001
  [\href{http://arXiv.org/abs/1707.02282}{{arXiv:1707.02282}}].

\bibitem{Romatschke:2017ejr}
P.~Romatschke and U.~Romatschke, {\em {Relativistic Fluid Dynamics In and Out
  of Equilibrium}}.
\newblock Cambridge Monographs on Mathematical Physics. Cambridge University
  Press, 5, 2019.

\bibitem{Busza:2018rrf}
W.~Busza, K.~Rajagopal and W.~van~der Schee, {\it {Heavy Ion Collisions: The
  Big Picture, and the Big Questions}},
  \beforedoihref\href{http://dx.doi.org/10.1146/annurev-nucl-101917-020852}{Ann.
  Rev. Nucl. Part. Sci.}\afterdoihref\  {\bf 68} (2018) 339--376
  [\href{http://arXiv.org/abs/1802.04801}{{arXiv:1802.04801}}].

\bibitem{Shen:2020mgh}
C.~Shen and L.~Yan, {\it {Recent development of hydrodynamic modeling in
  heavy-ion collisions}},
  \href{http://arXiv.org/abs/2010.12377}{{arXiv:2010.12377}}.

\bibitem{Hartnoll:2016apf}
S.~A. Hartnoll, A.~Lucas and S.~Sachdev, {\it {Holographic quantum matter}},
  \href{http://arXiv.org/abs/1612.07324}{{arXiv:1612.07324}}.

\bibitem{Kovtun:2004de}
P.~Kovtun, D.~T. Son and A.~O. Starinets, {\it {Viscosity in strongly
  interacting quantum field theories from black hole physics}},
  \beforedoihref\href{http://dx.doi.org/10.1103/PhysRevLett.94.111601}{Phys.
  Rev. Lett.}\afterdoihref\  {\bf 94} (2005) 111601
  [\href{http://arXiv.org/abs/hep-th/0405231}{{arXiv:hep-th/0405231}}].

\bibitem{Heller:2013fn}
M.~P. Heller, R.~A. Janik and P.~Witaszczyk, {\it {Hydrodynamic Gradient
  Expansion in Gauge Theory Plasmas}},
  \beforedoihref\href{http://dx.doi.org/10.1103/PhysRevLett.110.211602}{Phys.Rev.Lett.}\afterdoihref\
  {\bf 110} (2013), no.~21 211602
  [\href{http://arXiv.org/abs/1302.0697}{{arXiv:1302.0697}}].

\bibitem{Maldacena:1997re}
J.~M. Maldacena, {\it {The Large N limit of superconformal field theories and
  supergravity}},  Adv.Theor.Math.Phys. {\bf 2} (1998) 231--252
  [\href{http://arXiv.org/abs/hep-th/9711200}{{arXiv:hep-th/9711200}}].

\bibitem{Witten:1998qj}
E.~Witten, {\it {Anti-de Sitter space and holography}},  Adv.Theor.Math.Phys.
  {\bf 2} (1998) 253--291
  [\href{http://arXiv.org/abs/hep-th/9802150}{{arXiv:hep-th/9802150}}].

\bibitem{Gubser:1998bc}
S.~Gubser, I.~R. Klebanov and A.~M. Polyakov, {\it {Gauge theory correlators
  from noncritical string theory}},
  \beforedoihref\href{http://dx.doi.org/10.1016/S0370-2693(98)00377-3}{Phys.Lett.}\afterdoihref\
  {\bf B428} (1998) 105--114
  [\href{http://arXiv.org/abs/hep-th/9802109}{{arXiv:hep-th/9802109}}].

\bibitem{Hubeny:2010ry}
V.~E. Hubeny and M.~Rangamani, {\it {A Holographic view on physics out of
  equilibrium}},
  \beforedoihref\href{http://dx.doi.org/10.1155/2010/297916}{Adv. High Energy
  Phys.}\afterdoihref\  {\bf 2010} (2010) 297916
  [\href{http://arXiv.org/abs/1006.3675}{{arXiv:1006.3675}}].

\bibitem{Bjorken:1982qr}
J.~Bjorken, {\it {Highly Relativistic Nucleus-Nucleus Collisions: The Central
  Rapidity Region}},
  \beforedoihref\href{http://dx.doi.org/10.1103/PhysRevD.27.140}{Phys.Rev.}\afterdoihref\
  {\bf D27} (1983) 140--151.

\bibitem{Baier:2007ix}
R.~Baier, P.~Romatschke, D.~T. Son, A.~O. Starinets and M.~A. Stephanov, {\it
  {Relativistic viscous hydrodynamics, conformal invariance, and holography}},
  \beforedoihref\href{http://dx.doi.org/10.1088/1126-6708/2008/04/100}{JHEP}\afterdoihref\
  {\bf 04} (2008) 100
  [\href{http://arXiv.org/abs/0712.2451}{{arXiv:0712.2451}}].

\bibitem{Heller:2015dha}
M.~P. Heller and M.~Spali{\'n}ski, {\it {Hydrodynamics Beyond the Gradient
  Expansion: Resurgence and Resummation}},
  \beforedoihref\href{http://dx.doi.org/10.1103/PhysRevLett.115.072501}{Phys.
  Rev. Lett.}\afterdoihref\  {\bf 115} (2015), no.~7 072501
  [\href{http://arXiv.org/abs/1503.07514}{{arXiv:1503.07514}}].

\bibitem{Basar:2015ava}
G.~Ba{\c s}ar and G.~V. Dunne, {\it {Hydrodynamics, resurgence, and
  transasymptotics}},
  \beforedoihref\href{http://dx.doi.org/10.1103/PhysRevD.92.125011}{Phys.
  Rev.}\afterdoihref\  {\bf D92} (2015), no.~12 125011
  [\href{http://arXiv.org/abs/1509.05046}{{arXiv:1509.05046}}].

\bibitem{Aniceto:2015mto}
I.~Aniceto and M.~Spali{\'n}ski, {\it {Resurgence in Extended Hydrodynamics}},
  \beforedoihref\href{http://dx.doi.org/10.1103/PhysRevD.93.085008}{Phys.
  Rev.}\afterdoihref\  {\bf D93} (2016), no.~8 085008
  [\href{http://arXiv.org/abs/1511.06358}{{arXiv:1511.06358}}].

\bibitem{Florkowski:2016zsi}
W.~Florkowski, R.~Ryblewski and M.~Spali\'nski, {\it {Gradient expansion for
  anisotropic hydrodynamics}},
  \beforedoihref\href{http://dx.doi.org/10.1103/PhysRevD.94.114025}{Phys. Rev.
  D}\afterdoihref\  {\bf 94} (2016), no.~11 114025
  [\href{http://arXiv.org/abs/1608.07558}{{arXiv:1608.07558}}].

\bibitem{Heller:2016rtz}
M.~P. Heller, A.~Kurkela, M.~Spali{\'n}ski and V.~Svensson, {\it
  {Hydrodynamization in kinetic theory: Transient modes and the gradient
  expansion}},
  \beforedoihref\href{http://dx.doi.org/10.1103/PhysRevD.97.091503}{Phys.
  Rev.}\afterdoihref\  {\bf D97} (2018), no.~9 091503
  [\href{http://arXiv.org/abs/1609.04803}{{arXiv:1609.04803}}].

\bibitem{Heller:2018qvh}
M.~P. Heller and V.~Svensson, {\it {How does relativistic kinetic theory
  remember about initial conditions?}},
  \beforedoihref\href{http://dx.doi.org/10.1103/PhysRevD.98.054016}{Phys.
  Rev.}\afterdoihref\  {\bf D98} (2018), no.~5 054016
  [\href{http://arXiv.org/abs/1802.08225}{{arXiv:1802.08225}}].

\bibitem{Casalderrey-Solana:2017zyh}
J.~Casalderrey-Solana, N.~I. Gushterov and B.~Meiring, {\it {Resurgence and
  Hydrodynamic Attractors in Gauss-Bonnet Holography}},
  \href{http://arXiv.org/abs/1712.02772}{{arXiv:1712.02772}}.

\bibitem{Aniceto:2018uik}
I.~Aniceto, B.~Meiring, J.~Jankowski and M.~Spali\'nski, {\it {The large
  proper-time expansion of Yang-Mills plasma as a resurgent transseries}},
  \beforedoihref\href{http://dx.doi.org/10.1007/JHEP02(2019)073}{JHEP}\afterdoihref\
  {\bf 02} (2019) 073
  [\href{http://arXiv.org/abs/1810.07130}{{arXiv:1810.07130}}].

\bibitem{Behtash:2020vqk}
A.~Behtash, S.~Kamata, M.~Martinez, T.~Schaefer and V.~Skokov, {\it
  {Transasymptotics and hydrodynamization of the Fokker-Planck equation for
  gluons}},  \href{http://arXiv.org/abs/2011.08235}{{arXiv:2011.08235}}.

\bibitem{Denicol:2019lio}
G.~S. Denicol and J.~Noronha, {\it {Exact hydrodynamic attractor of an
  ultrarelativistic gas of hard spheres}},
  \href{http://arXiv.org/abs/1908.09957}{{arXiv:1908.09957}}.

\bibitem{Janik:2006gp}
R.~A. Janik and R.~B. Peschanski, {\it {Gauge/gravity duality and
  thermalization of a boost-invariant perfect fluid}},
  \beforedoihref\href{http://dx.doi.org/10.1103/PhysRevD.74.046007}{Phys. Rev.
  D}\afterdoihref\  {\bf 74} (2006) 046007
  [\href{http://arXiv.org/abs/hep-th/0606149}{{arXiv:hep-th/0606149}}].

\bibitem{Heller:2014wfa}
M.~P. Heller, R.~A. Janik, M.~Spali{\'n}ski and P.~Witaszczyk, {\it {Coupling
  hydrodynamics to nonequilibrium degrees of freedom in strongly interacting
  quark-gluon plasma}},
  \beforedoihref\href{http://dx.doi.org/10.1103/PhysRevLett.113.261601}{Phys.Rev.Lett.}\afterdoihref\
  {\bf 113} (2014), no.~26 261601
  [\href{http://arXiv.org/abs/1409.5087}{{arXiv:1409.5087}}].

\bibitem{Dorigoni:2014hea}
D.~Dorigoni, {\it {An Introduction to Resurgence, Trans-Series and Alien
  Calculus}},
  \beforedoihref\href{http://dx.doi.org/10.1016/j.aop.2019.167914}{Annals
  Phys.}\afterdoihref\  {\bf 409} (2019) 167914
  [\href{http://arXiv.org/abs/1411.3585}{{arXiv:1411.3585}}].

\bibitem{Aniceto:2018bis}
I.~Aniceto, G.~Basar and R.~Schiappa, {\it {A Primer on Resurgent Transseries
  and Their Asymptotics}},
  \beforedoihref\href{http://dx.doi.org/10.1016/j.physrep.2019.02.003}{Phys.
  Rept.}\afterdoihref\  {\bf 809} (2019) 1--135
  [\href{http://arXiv.org/abs/1802.10441}{{arXiv:1802.10441}}].

\bibitem{Heller:2020uuy}
M.~P. Heller, A.~Serantes, M.~Spali\'nski, V.~Svensson and B.~Withers, {\it
  {The hydrodynamic gradient expansion in linear response theory}},
  \href{http://arXiv.org/abs/2007.05524}{{arXiv:2007.05524}}.

\bibitem{Romatschke:2015gic}
P.~Romatschke, {\it {Retarded correlators in kinetic theory: branch cuts, poles
  and hydrodynamic onset transitions}},
  \beforedoihref\href{http://dx.doi.org/10.1140/epjc/s10052-016-4169-7}{Eur.
  Phys. J.}\afterdoihref\  {\bf C76} (2016), no.~6 352
  [\href{http://arXiv.org/abs/1512.02641}{{arXiv:1512.02641}}].

\bibitem{Romatschke:2009im}
P.~Romatschke, {\it {New Developments in Relativistic Viscous Hydrodynamics}},
  \beforedoihref\href{http://dx.doi.org/10.1142/S0218301310014613}{Int. J. Mod.
  Phys. E}\afterdoihref\  {\bf 19} (2010) 1--53
  [\href{http://arXiv.org/abs/0902.3663}{{arXiv:0902.3663}}].

\bibitem{Bu:2014sia}
Y.~Bu and M.~Lublinsky, {\it {All order linearized hydrodynamics from
  fluid-gravity correspondence}},
  \beforedoihref\href{http://dx.doi.org/10.1103/PhysRevD.90.086003}{Phys.Rev.}\afterdoihref\
  {\bf D90} (2014), no.~8 086003
  [\href{http://arXiv.org/abs/1406.7222}{{arXiv:1406.7222}}].

\bibitem{Bu:2014ena}
Y.~Bu and M.~Lublinsky, {\it {Linearized fluid/gravity correspondence: from
  shear viscosity to all order hydrodynamics}},
  \beforedoihref\href{http://dx.doi.org/10.1007/JHEP11(2014)064}{JHEP}\afterdoihref\
  {\bf 11} (2014) 064
  [\href{http://arXiv.org/abs/1409.3095}{{arXiv:1409.3095}}].

\bibitem{Grozdanov:2018fic}
S.~s. Grozdanov, A.~Lucas and N.~Poovuttikul, {\it {Holography and
  hydrodynamics with weakly broken symmetries}},
  \beforedoihref\href{http://dx.doi.org/10.1103/PhysRevD.99.086012}{Phys. Rev.
  D}\afterdoihref\  {\bf 99} (2019), no.~8 086012
  [\href{http://arXiv.org/abs/1810.10016}{{arXiv:1810.10016}}].

\bibitem{Davison:2016hno}
R.~A. Davison, L.~V. Delacr\'etaz, B.~Gout\'eraux and S.~A. Hartnoll, {\it
  {Hydrodynamic theory of quantum fluctuating superconductivity}},
  \beforedoihref\href{http://dx.doi.org/10.1103/PhysRevB.94.054502}{Phys. Rev.
  B}\afterdoihref\  {\bf 94} (2016), no.~5 054502
  [\href{http://arXiv.org/abs/1602.08171}{{arXiv:1602.08171}}]. [Erratum:
  Phys.Rev.B 96, 059902 (2017)].

\bibitem{Delacretaz:2017zxd}
L.~V. Delacr\'etaz, B.~Gout\'eraux, S.~A. Hartnoll and A.~Karlsson, {\it
  {Theory of hydrodynamic transport in fluctuating electronic charge density
  wave states}},
  \beforedoihref\href{http://dx.doi.org/10.1103/PhysRevB.96.195128}{Phys. Rev.
  B}\afterdoihref\  {\bf 96} (2017), no.~19 195128
  [\href{http://arXiv.org/abs/1702.05104}{{arXiv:1702.05104}}].

\bibitem{Kaminski:2009dh}
M.~Kaminski, K.~Landsteiner, J.~Mas, J.~P. Shock and J.~Tarrio, {\it
  {Holographic Operator Mixing and Quasinormal Modes on the Brane}},
  \beforedoihref\href{http://dx.doi.org/10.1007/JHEP02(2010)021}{JHEP}\afterdoihref\
  {\bf 02} (2010) 021
  [\href{http://arXiv.org/abs/0911.3610}{{arXiv:0911.3610}}].

\bibitem{Davison:2011ek}
R.~A. Davison and A.~O. Starinets, {\it {Holographic zero sound at finite
  temperature}},
  \beforedoihref\href{http://dx.doi.org/10.1103/PhysRevD.85.026004}{Phys. Rev.
  D}\afterdoihref\  {\bf 85} (2012) 026004
  [\href{http://arXiv.org/abs/1109.6343}{{arXiv:1109.6343}}].

\bibitem{Chen:2017dsy}
C.-F. Chen and A.~Lucas, {\it {Origin of the Drude peak and of zero sound in
  probe brane holography}},
  \beforedoihref\href{http://dx.doi.org/10.1016/j.physletb.2017.10.023}{Phys.
  Lett. B}\afterdoihref\  {\bf 774} (2017) 569--574
  [\href{http://arXiv.org/abs/1709.01520}{{arXiv:1709.01520}}].

\bibitem{Davison:2014lua}
R.~A. Davison and B.~Gout\'eraux, {\it {Momentum dissipation and effective
  theories of coherent and incoherent transport}},
  \beforedoihref\href{http://dx.doi.org/10.1007/JHEP01(2015)039}{JHEP}\afterdoihref\
  {\bf 01} (2015) 039
  [\href{http://arXiv.org/abs/1411.1062}{{arXiv:1411.1062}}].

\bibitem{Grozdanov:2016vgg}
S.~Grozdanov, N.~Kaplis and A.~O. Starinets, {\it {From strong to weak coupling
  in holographic models of thermalization}},
  \beforedoihref\href{http://dx.doi.org/10.1007/JHEP07(2016)151}{JHEP}\afterdoihref\
  {\bf 07} (2016) 151
  [\href{http://arXiv.org/abs/1605.02173}{{arXiv:1605.02173}}].

\bibitem{Grozdanov:2016fkt}
S.~Grozdanov and A.~O. Starinets, {\it {Second-order transport, quasinormal
  modes and zero-viscosity limit in the Gauss-Bonnet holographic fluid}},
  \beforedoihref\href{http://dx.doi.org/10.1007/JHEP03(2017)166}{JHEP}\afterdoihref\
  {\bf 03} (2017) 166
  [\href{http://arXiv.org/abs/1611.07053}{{arXiv:1611.07053}}].

\bibitem{Grozdanov:2017kyl}
S.~Grozdanov and N.~Poovuttikul, {\it {Generalised global symmetries in
  holography: magnetohydrodynamic waves in a strongly interacting plasma}},
  \beforedoihref\href{http://dx.doi.org/10.1007/JHEP04(2019)141}{JHEP}\afterdoihref\
  {\bf 04} (2019) 141
  [\href{http://arXiv.org/abs/1707.04182}{{arXiv:1707.04182}}].

\bibitem{Hofman:2017vwr}
D.~M. Hofman and N.~Iqbal, {\it {Generalized global symmetries and
  holography}},
  \beforedoihref\href{http://dx.doi.org/10.21468/SciPostPhys.4.1.005}{SciPost
  Phys.}\afterdoihref\  {\bf 4} (2018), no.~1 005
  [\href{http://arXiv.org/abs/1707.08577}{{arXiv:1707.08577}}].

\bibitem{Grozdanov:2018ewh}
S.~Grozdanov and N.~Poovuttikul, {\it {Generalized global symmetries in states
  with dynamical defects: The case of the transverse sound in field theory and
  holography}},
  \beforedoihref\href{http://dx.doi.org/10.1103/PhysRevD.97.106005}{Phys. Rev.
  D}\afterdoihref\  {\bf 97} (2018), no.~10 106005
  [\href{http://arXiv.org/abs/1801.03199}{{arXiv:1801.03199}}].

\bibitem{Baggioli:2020whu}
M.~Baggioli, M.~Vasin, V.~V. Brazhkin and K.~Trachenko, {\it {Field Theory of
  Dissipative Systems with Gapped Momentum States}},
  \beforedoihref\href{http://dx.doi.org/10.1103/PhysRevD.102.025012}{Phys. Rev.
  D}\afterdoihref\  {\bf 102} (2020), no.~2 025012
  [\href{http://arXiv.org/abs/2004.13613}{{arXiv:2004.13613}}].

\bibitem{Jimenez-Alba:2014iia}
A.~Jimenez-Alba, K.~Landsteiner and L.~Melgar, {\it {Anomalous magnetoresponse
  and the St\"uckelberg axion in holography}},
  \beforedoihref\href{http://dx.doi.org/10.1103/PhysRevD.90.126004}{Phys. Rev.
  D}\afterdoihref\  {\bf 90} (2014) 126004
  [\href{http://arXiv.org/abs/1407.8162}{{arXiv:1407.8162}}].

\bibitem{Stephanov:2014dma}
M.~Stephanov, H.-U. Yee and Y.~Yin, {\it {Collective modes of chiral kinetic
  theory in a magnetic field}},
  \beforedoihref\href{http://dx.doi.org/10.1103/PhysRevD.91.125014}{Phys. Rev.
  D}\afterdoihref\  {\bf 91} (2015), no.~12 125014
  [\href{http://arXiv.org/abs/1501.00222}{{arXiv:1501.00222}}].

\bibitem{Withers:2018srf}
B.~Withers, {\it {Short-lived modes from hydrodynamic dispersion relations}},
  \beforedoihref\href{http://dx.doi.org/10.1007/JHEP06(2018)059}{JHEP}\afterdoihref\
  {\bf 06} (2018) 059
  [\href{http://arXiv.org/abs/1803.08058}{{arXiv:1803.08058}}].

\bibitem{Grozdanov:2019kge}
S.~Grozdanov, P.~K. Kovtun, A.~O. Starinets and P.~Tadi\'c, {\it {Convergence
  of the Gradient Expansion in Hydrodynamics}},
  \beforedoihref\href{http://dx.doi.org/10.1103/PhysRevLett.122.251601}{Phys.
  Rev. Lett.}\afterdoihref\  {\bf 122} (2019), no.~25 251601
  [\href{http://arXiv.org/abs/1904.01018}{{arXiv:1904.01018}}].

\bibitem{Grozdanov:2019uhi}
S.~Grozdanov, P.~K. Kovtun, A.~O. Starinets and P.~Tadi\'c, {\it {The complex
  life of hydrodynamic modes}},
  \beforedoihref\href{http://dx.doi.org/10.1007/JHEP11(2019)097}{JHEP}\afterdoihref\
  {\bf 11} (2019) 097
  [\href{http://arXiv.org/abs/1904.12862}{{arXiv:1904.12862}}].

\bibitem{Abbasi:2020ykq}
N.~Abbasi and S.~Tahery, {\it {Complexified quasinormal modes and the
  pole-skipping in a holographic system at finite chemical potential}},
  \beforedoihref\href{http://dx.doi.org/10.1007/JHEP10(2020)076}{JHEP}\afterdoihref\
  {\bf 10} (2020) 076
  [\href{http://arXiv.org/abs/2007.10024}{{arXiv:2007.10024}}].

\bibitem{Jansen:2020hfd}
A.~Jansen and C.~Pantelidou, {\it {Quasinormal modes in charged fluids at
  complex momentum}},
  \beforedoihref\href{http://dx.doi.org/10.1007/JHEP10(2020)121}{JHEP}\afterdoihref\
  {\bf 10} (2020) 121
  [\href{http://arXiv.org/abs/2007.14418}{{arXiv:2007.14418}}].

\bibitem{Choi:2020tdj}
C.~Choi, M.~Mezei and G.~S\'arosi, {\it {Pole skipping away from maximal
  chaos}},  \href{http://arXiv.org/abs/2010.08558}{{arXiv:2010.08558}}.

\bibitem{Arean:2020eus}
D.~Arean, R.~A. Davison, B.~Gout\'eraux and K.~Suzuki, {\it {Hydrodynamic
  diffusion and its breakdown near AdS$_2$ fixed points}},
  \href{http://arXiv.org/abs/2011.12301}{{arXiv:2011.12301}}.

\bibitem{Baggioli:2019jcm}
M.~Baggioli, V.~V. Brazhkin, K.~Trachenko and M.~Vasin, {\it {Gapped momentum
  states}},
  \beforedoihref\href{http://dx.doi.org/10.1016/j.physrep.2020.04.002}{Phys.
  Rept.}\afterdoihref\  {\bf 865} (2020) 1--44
  [\href{http://arXiv.org/abs/1904.01419}{{arXiv:1904.01419}}].

\bibitem{Baggioli:2018vfc}
M.~Baggioli and K.~Trachenko, {\it {Low frequency propagating shear waves in
  holographic liquids}},
  \beforedoihref\href{http://dx.doi.org/10.1007/JHEP03(2019)093}{JHEP}\afterdoihref\
  {\bf 03} (2019) 093
  [\href{http://arXiv.org/abs/1807.10530}{{arXiv:1807.10530}}].

\bibitem{Baggioli:2018nnp}
M.~Baggioli and K.~Trachenko, {\it {Maxwell interpolation and close
  similarities between liquids and holographic models}},
  \beforedoihref\href{http://dx.doi.org/10.1103/PhysRevD.99.106002}{Phys. Rev.
  D}\afterdoihref\  {\bf 99} (2019), no.~10 106002
  [\href{http://arXiv.org/abs/1808.05391}{{arXiv:1808.05391}}].

\bibitem{Baggioli:2020loj}
M.~Baggioli, {\it {How small hydrodynamics can go}},
  \href{http://arXiv.org/abs/2010.05916}{{arXiv:2010.05916}}.

\bibitem{Arias:2014msa}
R.~E. Arias and I.~S. Landea, {\it {Hydrodynamic Modes of a holographic
  $p-$wave superfluid}},
  \beforedoihref\href{http://dx.doi.org/10.1007/JHEP11(2014)047}{JHEP}\afterdoihref\
  {\bf 11} (2014) 047
  [\href{http://arXiv.org/abs/1409.6357}{{arXiv:1409.6357}}].

\bibitem{Gran:2018vdn}
U.~Gran, M.~Torns\"o and T.~Zingg, {\it {Exotic Holographic Dispersion}},
  \beforedoihref\href{http://dx.doi.org/10.1007/JHEP02(2019)032}{JHEP}\afterdoihref\
  {\bf 02} (2019) 032
  [\href{http://arXiv.org/abs/1808.05867}{{arXiv:1808.05867}}].

\bibitem{Baggioli:2019aqf}
M.~Baggioli, U.~Gran, A.~J. Alba, M.~Torns\"o and T.~Zingg, {\it {Holographic
  Plasmon Relaxation with and without Broken Translations}},
  \beforedoihref\href{http://dx.doi.org/10.1007/JHEP09(2019)013}{JHEP}\afterdoihref\
  {\bf 09} (2019) 013
  [\href{http://arXiv.org/abs/1905.00804}{{arXiv:1905.00804}}].

\bibitem{Baggioli:2019sio}
M.~Baggioli, U.~Gran and M.~Torns\"o, {\it {Transverse Collective Modes in
  Interacting Holographic Plasmas}},
  \beforedoihref\href{http://dx.doi.org/10.1007/JHEP04(2020)106}{JHEP}\afterdoihref\
  {\bf 04} (2020) 106
  [\href{http://arXiv.org/abs/1912.07321}{{arXiv:1912.07321}}].

\bibitem{chapman2005exponential}
S.~J. Chapman and D.~B. Mortimer, {\it Exponential asymptotics and stokes lines
  in a partial differential equation},  Proceedings of the Royal Society A:
  Mathematical, Physical and Engineering Sciences {\bf 461} (2005), no.~2060
  2385--2421.

\bibitem{chapman2007shock}
S.~Chapman, C.~Howls, J.~King and A.~O. Daalhuis, {\it Why is a shock not a
  caustic? the higher-order stokes phenomenon and smoothed shock formation},
  Nonlinearity {\bf 20} (2007), no.~10 2425.

\bibitem{Howls2004Aug}
C.~Howls, P.~Langman and A.~Olde~Daalhuis, {\it On the higher--order stokes
  phenomenon},  Proceedings of the Royal Society of London. Series A:
  Mathematical, Physical and Engineering Sciences {\bf 460} (2004), no.~2048
  2285--2303.

\bibitem{kurkela2019attracts}
A.~Kurkela, W.~van~der Schee, U.~A. Wiedemann and B.~Wu, {\it {Early- and
  Late-Time Behavior of Attractors in Heavy-Ion Collisions}},
  \beforedoihref\href{http://dx.doi.org/10.1103/PhysRevLett.124.102301}{Phys.
  Rev. Lett.}\afterdoihref\  {\bf 124} (2020), no.~10 102301
  [\href{http://arXiv.org/abs/1907.08101}{{arXiv:1907.08101}}].

\bibitem{Romatschke:2017acs}
P.~Romatschke, {\it {Relativistic Hydrodynamic Attractors with Broken
  Symmetries: Non-Conformal and Non-Homogeneous}},
  \beforedoihref\href{http://dx.doi.org/10.1007/JHEP12(2017)079}{JHEP}\afterdoihref\
  {\bf 12} (2017) 079
  [\href{http://arXiv.org/abs/1710.03234}{{arXiv:1710.03234}}].

\bibitem{Kurkela2020Dec}
A.~Kurkela, S.~F. Taghavi, U.~A. Wiedemann and B.~Wu, {\it {Hydrodynamization
  in systems with detailed transverse profiles}},
  \beforedoihref\href{http://dx.doi.org/10.1016/j.physletb.2020.135901}{Phys.
  Lett. B}\afterdoihref\  {\bf 811} (2020) 135901
  [\href{http://arXiv.org/abs/2007.06851}{{arXiv:2007.06851}}].

\bibitem{Ambrus:2021sjg}
V.~E. Ambrus, S.~Busuioc, J.~A. Fotakis, K.~Gallmeister and C.~Greiner, {\it
  {Bjorken flow attractors with transverse dynamics}},
  \href{http://arXiv.org/abs/2102.11785}{{arXiv:2102.11785}}.

\bibitem{berry1991hyperasymptotics}
M.~V. Berry and C.~J. Howls, {\it Hyperasymptotics for integrals with saddles},
   Proceedings of the Royal Society of London. Series A: Mathematical and
  Physical Sciences {\bf 434} (1991), no.~1892 657--675.

\bibitem{Serone:2017nmd}
M.~Serone, G.~Spada and G.~Villadoro, {\it {The Power of Perturbation Theory}},
   \beforedoihref\href{http://dx.doi.org/10.1007/JHEP05(2017)056}{JHEP}\afterdoihref\
  {\bf 05} (2017) 056
  [\href{http://arXiv.org/abs/1702.04148}{{arXiv:1702.04148}}].

\bibitem{Costin:2019xql}
O.~Costin and G.~V. Dunne, {\it {Resurgent extrapolation: rebuilding a function
  from asymptotic data. Painlev\'e I}},
  \beforedoihref\href{http://dx.doi.org/10.1088/1751-8121/ab477b}{J. Phys.
  A}\afterdoihref\  {\bf 52} (2019), no.~44 445205
  [\href{http://arXiv.org/abs/1904.11593}{{arXiv:1904.11593}}].

\bibitem{Bhattacharyya:2008jc}
S.~Bhattacharyya, V.~E. Hubeny, S.~Minwalla and M.~Rangamani, {\it {Nonlinear
  Fluid Dynamics from Gravity}},
  \beforedoihref\href{http://dx.doi.org/10.1088/1126-6708/2008/02/045}{JHEP}\afterdoihref\
  {\bf 02} (2008) 045
  [\href{http://arXiv.org/abs/0712.2456}{{arXiv:0712.2456}}].

\bibitem{Loganayagam:2008is}
R.~Loganayagam, {\it {Entropy Current in Conformal Hydrodynamics}},
  \beforedoihref\href{http://dx.doi.org/10.1088/1126-6708/2008/05/087}{JHEP}\afterdoihref\
  {\bf 0805} (2008) 087
  [\href{http://arXiv.org/abs/0801.3701}{{arXiv:0801.3701}}].

\bibitem{strichartz2003guide}
R.~S. Strichartz, {\em A guide to distribution theory and Fourier transforms}.
\newblock World Scientific Publishing Company, 2003.

\bibitem{Hartman:2017hhp}
T.~Hartman, S.~A. Hartnoll and R.~Mahajan, {\it {Upper Bound on Diffusivity}},
  \beforedoihref\href{http://dx.doi.org/10.1103/PhysRevLett.119.141601}{Phys.
  Rev. Lett.}\afterdoihref\  {\bf 119} (2017), no.~14 141601
  [\href{http://arXiv.org/abs/1706.00019}{{arXiv:1706.00019}}].

\bibitem{Davison:2018ofp}
R.~A. Davison, S.~A. Gentle and B.~Gout\'eraux, {\it {Slow relaxation and
  diffusion in holographic quantum critical phases}},
  \beforedoihref\href{http://dx.doi.org/10.1103/PhysRevLett.123.141601}{Phys.
  Rev. Lett.}\afterdoihref\  {\bf 123} (2019), no.~14 141601
  [\href{http://arXiv.org/abs/1808.05659}{{arXiv:1808.05659}}].

\bibitem{Davison:2018nxm}
R.~A. Davison, S.~A. Gentle and B.~Gout\'eraux, {\it {Impact of irrelevant
  deformations on thermodynamics and transport in holographic quantum critical
  states}},
  \beforedoihref\href{http://dx.doi.org/10.1103/PhysRevD.100.086020}{Phys. Rev.
  D}\afterdoihref\  {\bf 100} (2019), no.~8 086020
  [\href{http://arXiv.org/abs/1812.11060}{{arXiv:1812.11060}}].

\bibitem{Baggioli:2020ljz}
M.~Baggioli and W.-J. Li, {\it {Universal Bounds on Transport in Holographic
  Systems with Broken Translations}},
  \beforedoihref\href{http://dx.doi.org/10.21468/SciPostPhys.9.1.007}{SciPost
  Phys.}\afterdoihref\  {\bf 9} (2020), no.~1 007
  [\href{http://arXiv.org/abs/2005.06482}{{arXiv:2005.06482}}].

\end{thebibliography}\endgroup
\bibliographystyle{bibstyl}

\end{document}